\documentclass[journal=acssci,manuscript=article]{achemso}
\setkeys{acs}{articletitle = true}

\usepackage[version=3]{mhchem}    
\usepackage{xr}                   
\usepackage[hidelinks]{hyperref}       
\setkeys{acs}{maxauthors=2000, etalmode=truncate} 

\author{Riccardo Alessandri}
\affiliation{Pritzker School of Molecular Engineering,
University of Chicago, Chicago, Illinois 60637, United States}
\author{Juan J. de Pablo}
\affiliation{Pritzker School of Molecular Engineering,
University of Chicago, Chicago, Illinois 60637, United States}
\email{depablo@uchicago.edu}

\title[]
  {\Large
   Prediction of Electronic Properties 
   of Radical-Containing Polymers 
   at Coarse-Grained Resolutions}

\begin{document}
\vspace{-0.5cm} 
%
\begin{abstract}
The properties of soft electronic materials depend on the coupling of electronic 
and conformational degrees of freedom over a wide range of spatiotemporal scales.
Description of such properties requires 
multiscale approaches capable of, at the same time,
accessing electronic properties and 
sampling the conformational space of soft materials.
This could in principle be realized by connecting the coarse-grained (CG) methodologies
required for adequate conformational sampling to conformationally-averaged electronic
property distributions via backmapping to atomistic-resolution level models and
repeated quantum-chemical calculations.
Computational demands of such approaches, however,
have hindered their application in high-throughput
computer-aided soft materials discovery.
Here, we present a 
method that, combining machine learning and CG techniques,
can replace traditional backmapping-based approaches without sacrificing accuracy. 
We illustrate the 
method for an emerging class of soft electronic materials, 
namely non-conjugated, radical-containing polymers,
promising materials for all-organic energy storage.
Supervised machine learning models are trained to learn the dependence
of electronic properties on polymer conformation at CG resolutions.
We then parametrize 
CG models that retain electronic structure information,
simulate CG condensed phases, and predict the electronic properties 
of such phases solely from the CG degrees of freedom.
We validate our method by comparing it against a full backmapping-based approach, 
and find good agreement between both methods.
This work demonstrates the potential
of the proposed method to accelerate multiscale workflows,
and provides a framework for the development of CG models that retain electronic structure information.
\end{abstract}


\section{\normalsize INTRODUCTION}
\label{sec:intro}
Radical-containing polymers, also known as open-shell macromolecules, 
macromolecular radicals, or simply radical polymers,
possess intriguing redox, optoelectronic, and magnetic characteristics
that make them appealing for applications ranging from energy storage and optoelectronics to spintronics and 
memory storage~\cite{2020SWang-ACSMacroLett, 2022YTan-JAmChemSoc, 2017JKim-Joule, 
2012TJanoschka-AdvMater, 2009KOyaizu-AdvMater, 2018DAWilcox-AnnRevChemBiomolEng}.
Non-conjugated, radical-containing polymers
are organic polymers that have a non-conjugated backbone bearing pendant stable radical sites.
They constitute a class of charge-carrying polymers that do not rely on
$\pi$-conjugation to transport charges successfully.
The rational design of radical polymers 
with enhanced characteristics could be greatly advanced by deriving relationships that connect their
molecular structure, morphology, and electronic properties.
These relationships are inherently multiscale, involving the coupling of electronic and
conformational degrees of freedom over a wide range of spatiotemporal scales. 
New modeling approaches capable of describing such coupling are needed.

Recent work has shown that coarse-grained (CG) models can be used to probe polymeric material length- 
and timescales reaching the mesoscale.~\cite{2011JJdePablo-AnnuRevPhysChem, 2013MGSaunders-AnnRevBiophys,
2020TTFoley-PNAS, 2021PCTSouza-NatMethods}
In contrast, explicit quantum-chemical calculations---usually using density functional theory 
(DFT)---, which are necessary to access electronic properties~\cite{2018YOlivier-JPhysChemLett, 
2019PFriederich-AdvMater, 2021YTan-JAmChemSoc}, 
are extraordinarily demanding and can only capture picosecond and Angstrom-level processes.
To bridge these two scales, that is, 
to take into account large-scale morphological features 
generated via self-assembly processes (possibly as a function of processing conditions) 
when computing electronic properties,
recent efforts have sought to introduce new multiscale modeling approaches~\cite{2019PFriederich-AdvMater, 2021NEJackson-JPhysChemB}.
In such approaches, the soft material morphology generated 
via CG simulations is backmapped to the atomistic resolution required for 
the quantum-chemical calculations, and such calculations are then performed 
on conformations drawn from the backmapped morphologies. 
These multiscale approaches have been 
primarily developed in the context of organic 
semiconductors,~\cite{2015PGemunden-MacromolRapidCommun,
                      2019PFriederich-AdvMater,         
                      2020RAlessandri-AdvFunctMater,    
                      2020NRolland-ComputMaterSci,      
                      2021NEJackson-JPhysChemB}         
and provide a means to access the conformational dependence of electronic properties. 
However, considerable computational demands and workflow complexity of such have limited their applicability for 
high-throughput computer-aided materials discovery.

Computational studies of radical-containing polymers
have been limited~\cite{2020SWang-ACSMacroLett, 2022YTan-JAmChemSoc, 
2014TWKemper-JPhysChemC, 2015TWKemper-JPhysChemC, 2016TWKemper-JPhysChemC,
2018YJoo-Science, 2021YTan-JAmChemSoc,
2018KSato-JAmChemSoc, 
2020LBello-Macromolecules,
2022CHLi-JMaterChemA}.  
\citeauthor{2014TWKemper-JPhysChemC} performed what appear to be the first~\cite{2014TWKemper-JPhysChemC} 
and one of the few molecular dynamics (MD) investigations of radical polymers. 
They simulated the widely-used
p-type radical polymer poly(2,2,6,6-tetramethylpiperidinyloxy-4-yl methacrylate) 
(PTMA), a polymethacrylate with pendant 
nitroxide radical TEMPO (2,2,6,6-tetramethylpiperidin-1-oxyl) groups,
and provided early insights into some of PTMA's molecular-scale dynamics 
and its coupling to electronic properties~\cite{2014TWKemper-JPhysChemC, 2015TWKemper-JPhysChemC, 2016TWKemper-JPhysChemC}.
More recent efforts that relied on coarser models have sought to predict the conductivity
of radical polymers~\cite{2018KSato-JAmChemSoc, 2020LBello-Macromolecules}. 
Such models require as input molecular-level information about the specific redox-active unit 
in order to be predictive and to differentiate between the 
performance of different polymers.
A different type of effort has focused on molecule-specific characteristics by exploring sets 
of different redox-active units with DFT-based 
methods~\cite{2021YTan-JAmChemSoc, 2022CHLi-JMaterChemA}.
A limitation of these studies has been the fact that they do not account for condensed-phase conditions, 
which may affect the resulting electronic properties.

In this work we rely on machine learning (ML) to connect conformational properties 
to electronic structure properties in condensed phases. 
The use of ML techniques to study polymers has been limited by a variety of inherent 
challenges.~\cite{2016TBereau-APLMaterials, 2018ALFerguson-JPhysCondensMatter, 
2019NEJackson-CurrOpinChemEng, 2021LChen-MaterSciEngRRep} 
These include 
the complexity of data representation in systems that are disordered, 
the dependence of properties on an ensemble of configurations rather than on a single one,
and the dependence of processes on interactions that span multiple length- and timescales. 
To the best of our knowledge,
previous work aimed at predicting the \textit{electronic properties of polymers} using ML has focused 
exclusively on polythiophene-based conjugated polymers.~\cite{2019NEJackson-SciAdv, 
2019EDMiller-AIChE, 2020NEJackson-Macromolecules, 2020LSimine-PNAS, 2021CKLee-JChemPhys}
\citeauthor{2019NEJackson-SciAdv}, in particular,
introduced the concept of electronic coarse-graining (ECG),~\cite{2019NEJackson-SciAdv}
a ML-based methodology aimed at the prediction of electronic properties 
based on a molecule's CG representation,
and applied it mostly to prototypical conjugated polymers and
single molecules in vacuum.~\cite{2019NEJackson-SciAdv, 
2020NEJackson-Macromolecules, 2022GSivaraman-JChemTheoryComput, 2022JCMaier-JChemPhys}
However, no actual CG models---that require not only a CG mapping but also bonded and
nonbonded interaction parameters---have been developed so far in the ECG framework;
therefore, no validation against traditional backmapping-based approaches has been made.
A comparison against backmapping-based approaches has been made by \citeauthor{2020LSimine-PNAS},
where a similar approach to ECG is used to infer the absorption spectra of a polythiophene system
from CG degrees of freedom. However, significant discrepancies between the two approaches
were reported and ascribed to the backmapping protocol.~\cite{2020LSimine-PNAS}
Building on the ECG work, in what follows
we present an efficient ML-enabled method that can replace 
state-of-the-art but demanding backmapping-based multiscale approaches. 
We illustrate the method in the context of
nonconjugated, radical-containing polymers, which represent an emerging class of conducting materials.
We begin by
(1) training supervised ML models to learn the dependence of several
electronic properties on molecular conformations at CG resolution, 
and identifying CG mappings that
retain electronic structure information.
For this first part, we take inspiration from the ECG
methodology.~\cite{2019NEJackson-SciAdv}
Next, we (2) parametrize
CG models that retain electronic structure information
using established~\cite{2003DReith-JComputChem, 2021PCTSouza-NatMethods} coarse-graining techniques.
Finally, we (3) use such CG models to simulate condensed phases
and the trained ML models to predict electronic properties 
solely from the CG degrees of freedom.
We validate our method by comparing its results to those obtained 
via a standard backmapping-based approach,
and show that the proposed method offers a $10^6$ speedup 
with respect to backmapping-based approaches without loss of accuracy.
By exploring different coarse-graining strategies, we are able to delineate general 
guidelines for development of CG models that retain electronic structure information, 
and outline a viable pathway to
accelerate multiscale workflows aimed at predicting
electronic properties while encompassing CG-level spatiotemporal scales.

\newpage 

\section{\normalsize RESULTS}
\label{sec:results}
\begin{figure}[htbp]
  \centering
  \includegraphics[width=1.00\textwidth]{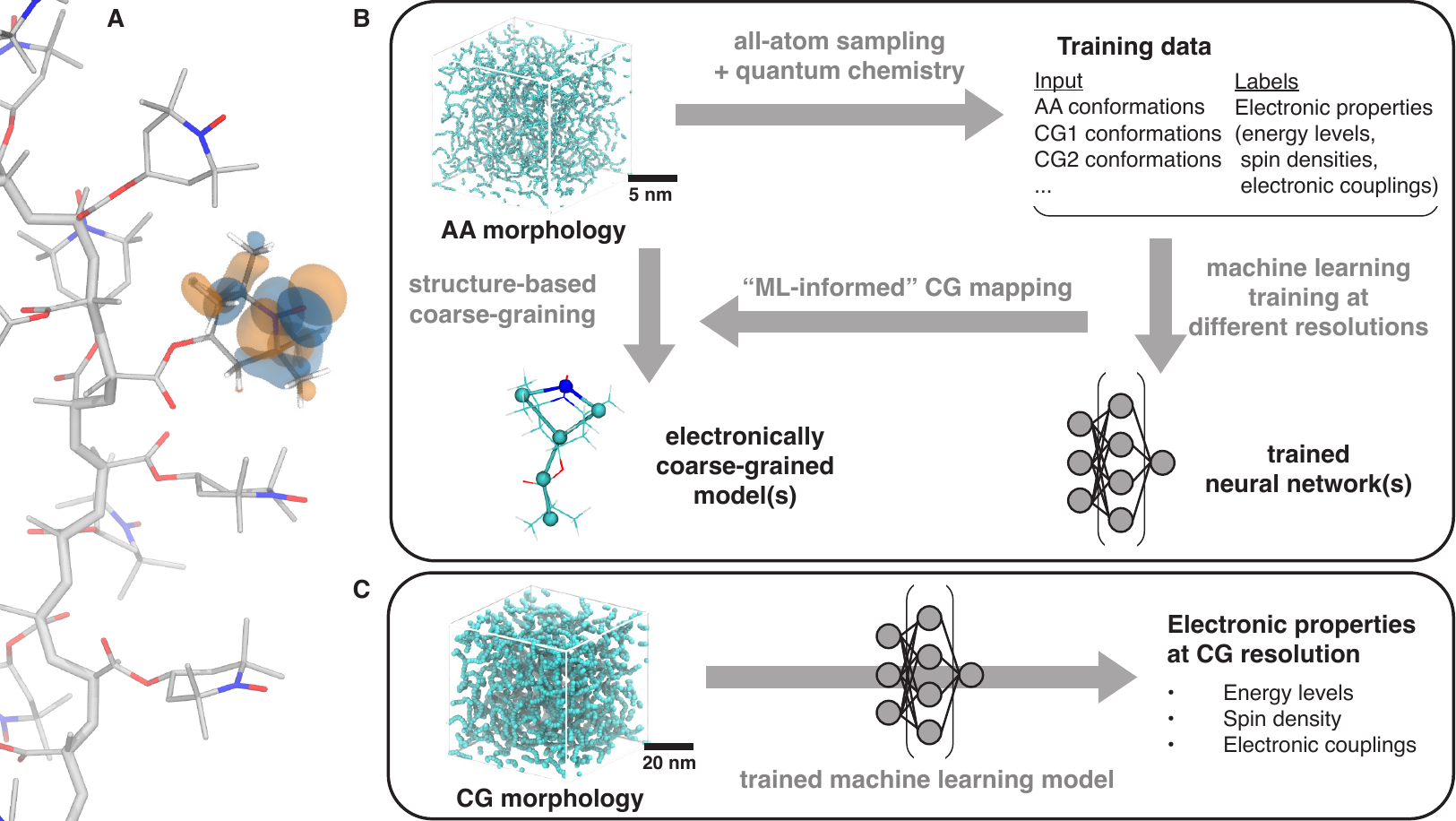}
    \caption{\textbf{Schematic representation of the proposed method for non-conjugated radical polymers}.
             (\textbf{A}) PTMA radical polymer: its non-conjugated backbone
             localizes electronic structure features---such as the SOMO orbital represented---to
             the pendant radical groups. 
             Hydrogen atoms are only shown on the radical group on which the SOMO orbital is localized.
             (\textbf{B}) Development:
             starting from an all-atom (AA) model, 
             conformations are sampled via MD simulations. Quantum-chemical
             calculations are performed on those conformations to generate the training dataset; the dataset serves to
             train machine learning models---in this case NNs---at different levels of CG
             resolution. The information gained during training is used to inform the
             development of CG models that are built on reference AA data 
             via structure-based coarse-graining techniques.
             (\textbf{C}) Application:
             the CG models permit study of structures over larger spatiotemporal scales;
             the trained machine learning models are used to predict electronic properties 
             based on the conformations taken from the CG morphology.
             The renderings of the morphologies show only polymer backbones for clarity.
            }
  \label{fig:schematic-of-method-for-RAPs}
\end{figure}

\textbf{Method overview.}
Figure~\ref{fig:schematic-of-method-for-RAPs} shows a schematic of 
the proposed method applied to non-conjugated, radical polymers.
Only the main aspects of the method are outlined here, and readers are referred to the
Materials and Methods section for additional details. 
To illustrate the method, we use the well-studied, p-type 
radical polymer PTMA, which bears pendant TEMPO units 
(Figure~\ref{fig:schematic-of-method-for-RAPs}A). 
An all-atom (AA) model is used to generate condensed-phase conformations of PTMA using MD simulations. 
The electronic properties of the resulting configurations are computed
via DFT calculations (Figure~\ref{fig:schematic-of-method-for-RAPs}B).
Given the nonconjugated backbone of this class of radical-containing polymers, 
the monomers can be treated as electronically independent (see, e.g.,
the singly-occupied molecular orbital (SOMO) distribution in 
Figure~\ref{fig:schematic-of-method-for-RAPs}A).
The data generated in this way---all-atom molecular conformations with associated electronic
properties---are used as training data to develop supervised ML models.
More specifically, we train feed-forward artificial neural networks (NNs) using the conformations 
(in the form of a reciprocal distance matrix) as input, and the electronic properties as labels. 
The all-atom molecular conformations can be mapped onto the corresponding CG configurations 
and NNs can therefore be trained \textit{at different CG levels of resolution} 
(Figure~\ref{fig:schematic-of-method-for-RAPs}B).
We build two different datasets: 
one containing single monomer conformations and associated 
monomer conformation-dependent properties (SOMO energy level,
spin density), and a second containing dimer conformations and 
associated dimer conformation-dependent properties (electronic couplings).
After training, we generate two main outputs: \textit{(1)} trained NNs 
for the different electronic properties (one for each different resolution) and 
\textit{(2)} information on the CG mapping required
for the CG model to retain sufficient information about specific electronic properties. 
The latter information, together with established coarse-graining
techniques, are used
to develop CG models for simulation of polymer 
condensed phases over larger spatial and temporal scales; 
the trained NNs allow one to retrieve electronic
properties at the CG level with only negligible computational demands
(Figure~\ref{fig:schematic-of-method-for-RAPs}C),
thereby permitting the investigation of
electronic properties over large ensembles of CG morphologies.

\newpage 

\textbf{Monomer conformation-dependent electronic properties.} 
We first examine the performance of the NN models for prediction
of electronic properties that depend on the conformation of a single PTMA \textit{monomer},
using the \textit{SOMO energy level} and the \textit{spin density} 
as examples of the target properties to be learned.
The SOMO energy level determines the position of the transport energy level and
hence the charge transport type (p-type vs. n-type), while knowledge of the spin density
distribution gives access to the degree of (de)localization of the unpaired electron(s).
The latter property contributes to the stability of the radical center.
Both energy levels and spin densities are conformation-dependent;
the extent to which they depend on conformation contributes to the degree of energetic disorder in the system.

\begin{figure}[htbp]
  \centering
  \includegraphics[width=1.00\textwidth]{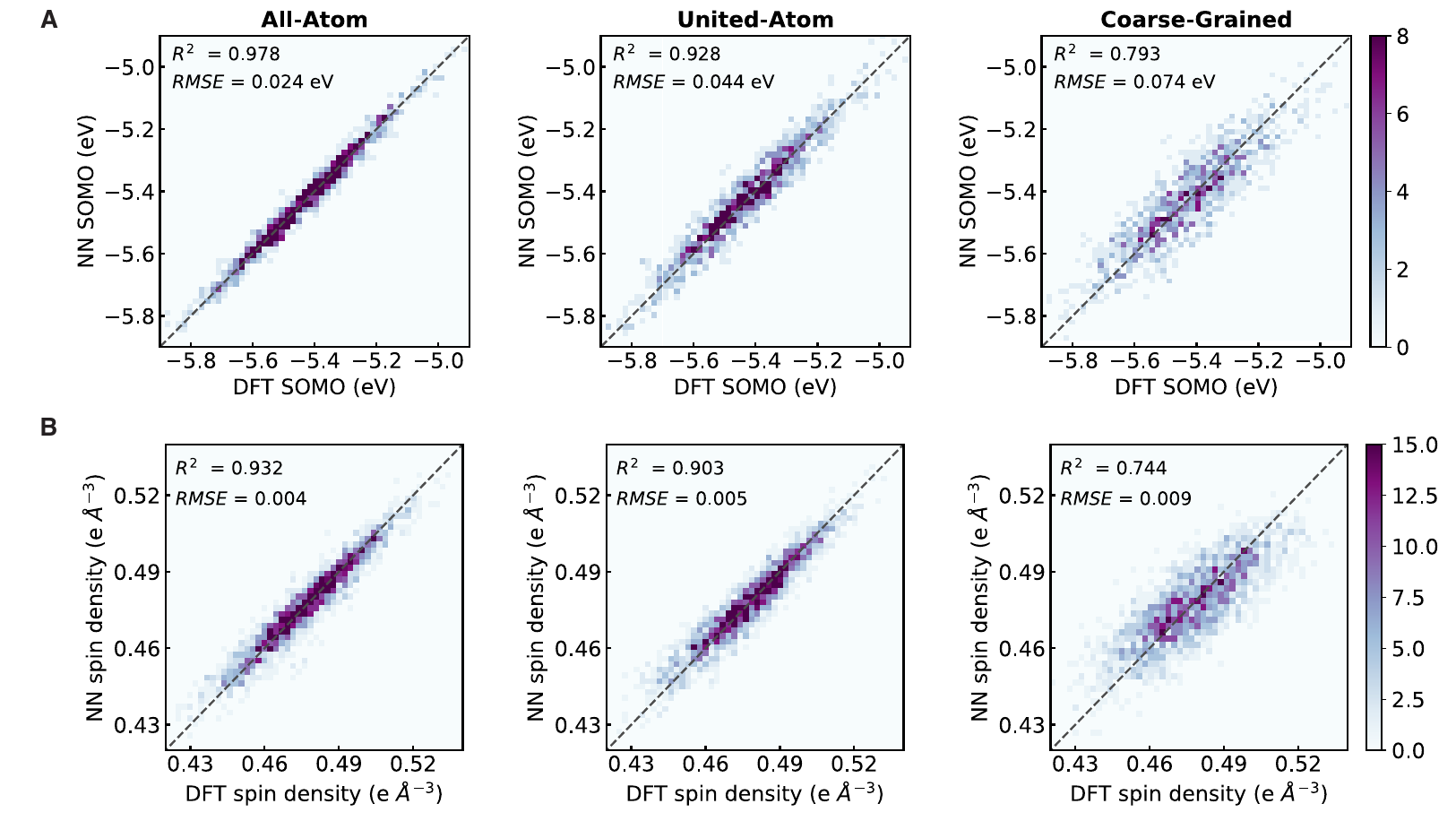}
    \caption{\textbf{Predictive performance of NN models for monomer conformation-dependent 
             electronic properties at all-atom and coarse-grained resolutions.}
             Two-dimensional histograms showing the best accuracy achieved by the NNs
             for different resolutions---the reference all-atom resolution, 
             united-atom, and CG---when learning 
             (\textbf{A}) SOMO energy levels and 
             (\textbf{B}) spin densities.
             In the CG case, the mapping used is GBNO1 (see Figure~\ref{fig:results-mappings-and-NO}).
             The $R^2$ scores and RMSE are indicated on each plot and
             represent the results on the test set.
             All correlations fit to a linear model with slope $=1$.
             The NN's hyperparameters are optimized for each resolution.
            }
  \label{fig:results-energies-spins}
\end{figure}

\begin{figure}[htbp]
  \centering
  \includegraphics[width=1.00\textwidth]{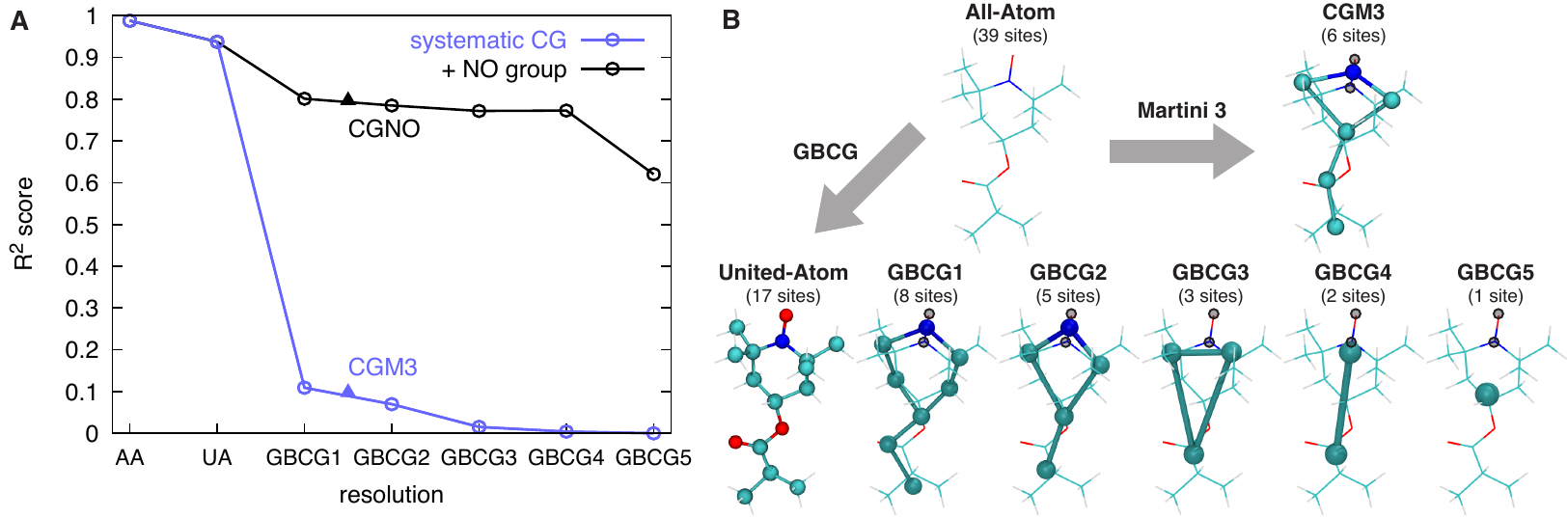}
    \caption{\textbf{Dependence of the predictive performance of NN models on CG resolution.}
             (\textbf{A}) $R^2$ score on the test set as a function of CG
             resolution for prediction of the SOMO energy level: 
             all-atom (AA), united-atom (UA), and CG mappings obtained with 
             the systematic graph-based coarse-graining approach (GBCGX, with X$=[1..5]$;
             blue line, hollow circles), and CG Martini 3 mapping (CGM3, blue triangle). 
             The black data points 
             (black line, hollow black circles for the GBCG-based mappings,
             black triangle for the CGM3-based mapping, CGNO)
             represent mappings where the nitrogen and oxygen atoms
             of the nitroxide group are described by an explicit CG site each
             (represented by semitransparent black circles 
             in the renderings of Panel \textbf{B}).
             (\textbf{B}) Renderings of the AA, UA, and CG mappings and the underlying
             atomistic structure. The number of sites that each resolution entails is also reported.
             For the detailed atom-to-bead correspondence, see Figures S10 and S12.
             The NN's hyperparameters are optimized for each resolution.
            }
  \label{fig:results-mappings-and-NO}
\end{figure}

Figure~\ref{fig:results-energies-spins} shows that SOMO energies and spin densities
can be predicted at CG resolution.
Figure~\ref{fig:results-energies-spins} shows correlation plots between
the DFT reference SOMO energies and spin densities---computed at the B3LYP/6-311G(d,p) 
level of DFT (Materials and Methods)---and the same quantities as predicted by the NN model
at AA, united-atom (UA), and CG resolutions.
The performance of the NN on the AA configurations
($R^2=0.978$, $RMSE=0.024$ eV for SOMO energies;
 $R^2=0.932$, $RMSE=0.004$ for spin densities)
represents the maximum achievable performance for a given molecular representation,
given that all other resolutions 
involve a loss of information for the NN model to 
perform the regression task on.
At UA resolution,
the NN also achieves very good predictive performance
($R^2=0.928$, $RMSE=0.044$ eV for SOMO energies;
 $R^2=0.903$, $RMSE=0.005$ for spin densities),
just below that obtained at AA resolution.
At CG resolution, 
there is a noticeable drop of performance 
with respect to the UA and AA cases, but
the NN still provides
$R^2=0.793$, $RMSE=0.074$ eV for SOMO energies
and $R^2=0.744$, $RMSE=0.009$ for spin densities.
To which degree these increased errors would eventually impact
the calculation of charge transport rates~\cite{2018KSato-JAmChemSoc}
is difficult to estimate and will require a separate study.
The CG mapping used here is ``GBNO1'' (see Figure~\ref{fig:results-mappings-and-NO}B);
a detailed investigation of the NN model accuracy
for different CG mapping choices and resolutions is presented in the next section.

It is useful to investigate the minimum data requirements 
needed by the NN to learn monomer conformation-dependent electronic properties.
Figure S1
shows that after $\approx$4,000 datapoints, the performance
of the NN starts to plateau at all resolutions.
This represents a relatively small number of datapoints, and calculations at higher
levels of theory could therefore be easily used to train such NNs.
We also note that the required NNs 
are rather small---e.g., 4 hidden layers with 3 neurons each
already maximize the performance at CG resolutions---, and 
the same performance is achieved for a relatively wide range of hyperparameters (see Figure S8).


\textbf{Impact of CG resolution and specific CG mapping.} 
The CG level of resolution and the specific CG mapping have considerable influence 
on the ability of the NN to predict electronic properties. 
In this section we provide an analysis of resolution that is useful 
for identifying design principles for CG models 
that represent the best compromise between computational efficiency and accuracy.

To design CG mappings, we resort to two different strategies.
First, we apply the Graph-Based Coarse-Graining (GBCG) algorithm of 
\citeauthor{2019MAWebb-JChemTheoryComput} to arrive at a series of increasingly 
coarser mappings in a systematic and automated
manner with minimal human intervention~\cite{2019MAWebb-JChemTheoryComput}. 
Second, we devise a mapping according to the 
Martini 3 building-block coarse-graining strategy~\cite{2021PCTSouza-NatMethods}. 
All the mappings are shown in Figure~\ref{fig:results-mappings-and-NO}B; 
the atom-to-bead correspondence is given in detail in Figures 
S10
and S12.
Additional details are provided in the Materials and Methods section and the Supplementary Material.

Figure~\ref{fig:results-mappings-and-NO}A shows how the predictive performance of the 
NN model---as quantified by the R$^2$ coefficient---changes 
as a function of the resolution of the molecular representation 
for the prediction of the SOMO energy level.
We see that GBCG mappings lead to poor predictive performance (R$^2<0.2$), 
even at the finest CG resolution level, GBCG1. 
The same is true for the Martini 3 CG mapping 
(CGM3, the resolution of which is between GBCG1 and GBCG2). 
Given the high predictive performance of the UA resolution, these results indicate that 
these ``standard'' CG mappings are missing key degrees of freedom. 
Notably, they do not include the nitroxide group explicitly.
As can be seen from Figure~\ref{fig:results-mappings-and-NO}A (black line), 
the explicit inclusion of the nitroxide group
(i.e., the positions of the nitrogen and oxygen atoms are included 
in the distance matrix used as the NN input vector)
improves considerably the performance across the different CG mappings, 
with R$^2$ remaining around 0.8 for GBCG2--GBCG4.
Even for resolution GBCG5, which represents the PTMA monomer with only 1 site, when 
expanded to include the explicit description of the nitroxide group one can achieve a R$^2$ above 0.6.
This result is expected, 
and very recently observed also in Ref.~\citenum{2022JCMaier-JChemPhys}:
the SOMO wavefunction is localized around the 
nitroxide group (see Figure~\ref{fig:schematic-of-method-for-RAPs}A) and, consequently, 
the conformation of this group 
relative to the other atoms in the TEMPO unit
governs the SOMO wavefunction coefficients and associated energy.

Further confirmation of the importance of the nitroxide group for prediction of SOMO energies
comes from a complementary test, 
where we exclude either the nitrogen or the oxygen atom of this group from the UA and AA resolutions
(Figure S2). Such exclusions lead to a dramatic decrease 
of predictive performance at the AA and UA levels, with R$^2$ values $<0.5$ 
(Figure S2). 
Figure S2 shows that excluding the nitrogen from the UA and AA representations  
leads to R$^2$ values of $\approx0.42$ while excluding the oxygen from the same 
representations degrades R$^2$ down to $\approx0.1$,
indicating that the oxygen position is more informative than that of the nitrogen for SOMO energy predictions.
In the case of strongly localized radical groups such as the nitroxide in PTMA,
the inclusion of the nitroxide atoms is essential for the NN model to be
able to accurately infer the SOMO energy.


\textbf{Dimer conformation-dependent electronic properties.} 
We now examine the accuracy of NNs for the prediction of electronic
properties that depend on the conformation of a \textit{dimer},
using the \textit{electronic couplings} as the target property to be learned.
Electronic couplings are a key molecular-scale property that determines charge transport,
the latter being in turn critical for many applications
of radical-containing polymers~\cite{2020SWang-ACSMacroLett, 2022YTan-JAmChemSoc}.
Charge transport in these materials is generally believed~\cite{2020SWang-ACSMacroLett, 
2022YTan-JAmChemSoc} to occur via a hopping mechanism,
where a charge jumps between a radical site and an adjacent ionized 
site---in the case of PTMA, and of p-type transport in general, a cation site.
We approximate the electronic coupling as the orbital overlap between 
the SOMO of the neutral radical and the lowest unoccupied molecular orbital (LUMO) 
of the cation, $\langle \phi_{SOMO} | \phi_{LUMO} \rangle$ (Materials and Methods),
and we train NNs to learn its base 10 logarithm, $\log_{10}(\langle \phi_{SOMO} | \phi_{LUMO} \rangle)$.

\begin{figure}[htbp]
  \centering
  \includegraphics[width=1.00\textwidth]{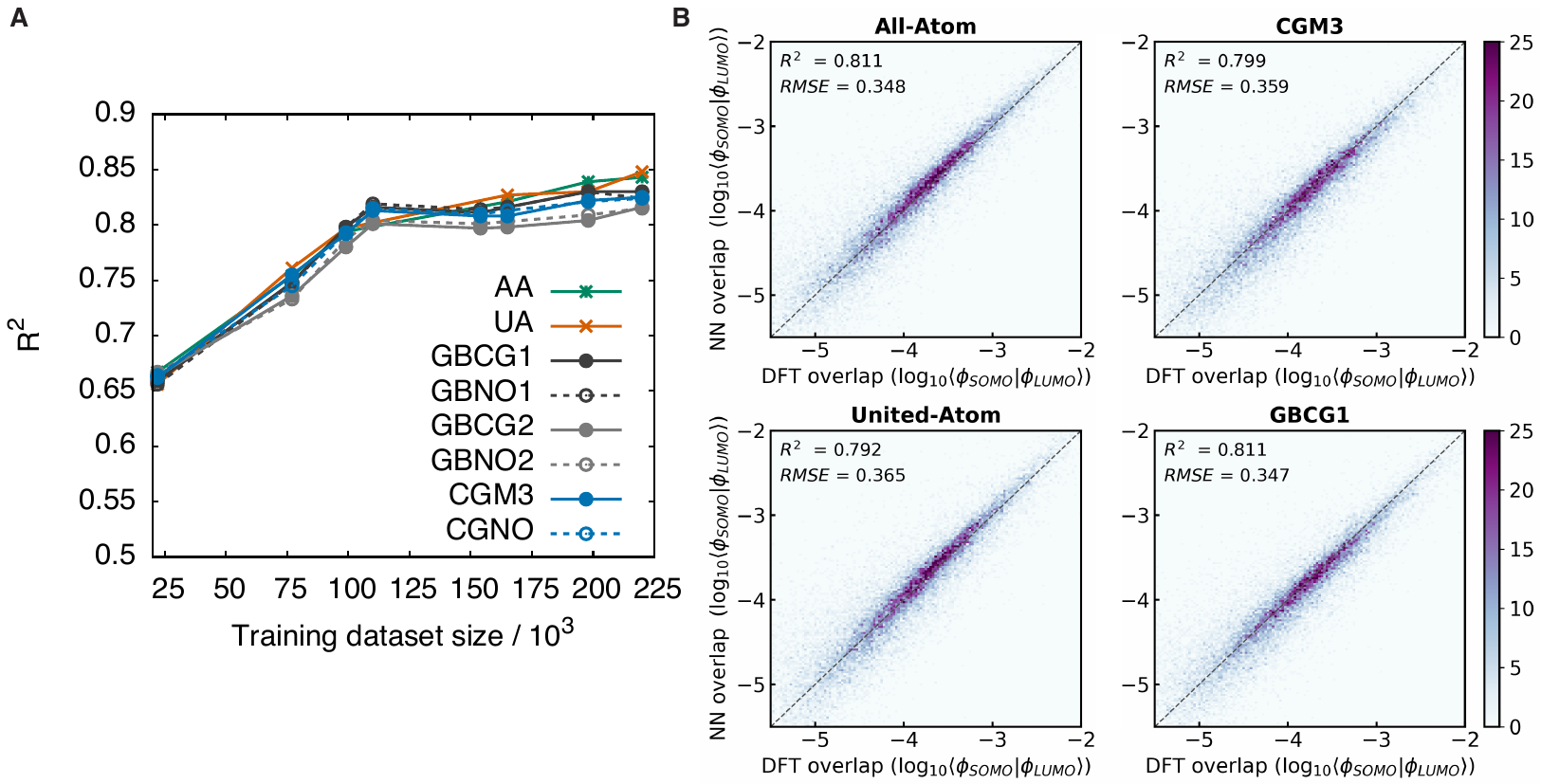}
    \caption{\textbf{Predictive performance of NN models for dimer conformation-dependent 
             electronic properties at all-atom and coarse-grained resolutions.}
             (\textbf{A}) NN performance as a function of training dataset size for different 
             resolutions: all-atom (AA), united-atom (UA), GBCG1-2, GBNO1-2, and CG Martini (CGM3 and CGNO). 
             See Figure~\ref{fig:results-mappings-and-NO}B for a schematic representation of the mappings.
             (\textbf{B}) Two-dimensional histograms showing the best accuracy achieved by the 
             NNs for the different resolutions. 
             The $R^2$ scores and RMSE are indicated on each plot and
             represent the results on the test set.
             The NN's hyperparameters are optimized for each resolution and dataset size.
            }
  \label{fig:results-couplings}
\end{figure}

As shown in Figure~\ref{fig:results-couplings}, the logarithm of the electronic coupling
can be learned with good and comparable accuracy at AA, UA, GBCG1-GBCG2, and CGM3 resolutions. 
The predictive performances (on the test set)---as quantified by the R$^2$ coefficient---are around 
0.8-0.81 at AA, UA, GBCG1, and CGM3 resolutions.
The performance gap between AA and UA resolutions and the best CG resolutions are small, 
particularly when compared to the monomer conformation-dependent
properties depicted in Figure~\ref{fig:results-energies-spins}.
Electronic couplings, however, are significantly more complex---i.e., data-intensive and 
requiring larger NNs---and difficult to learn than energy levels or spin densities
(compare Figure~\ref{fig:results-couplings}A to Figure S1),
across all resolutions.
In particular, if a training dataset of $\approx$4,000 data points 
was sufficient to train good-performing NNs for the prediction of SOMO energies 
(Figure S1),
at least $\approx$100,000 data points are necessary for the analogous task
on electronic couplings (Figure~\ref{fig:results-couplings}A).
After $\approx$100,000 data points, the predictive performance starts
to plateau across the different CG resolutions.
Moreover, NNs with 400 neurons per layer are required
(see Table S1).
This complexity---both in the dataset size required, and NN architecture---is
consistent with previous reports~\cite{2019NEJackson-SciAdv, 2020C-IWang-JChemPhys}.
Overall, while more demanding, our results confirm that
the task of predicting electronic couplings at CG resolutions can be accomplished. 
This is particularly important, given their role in determining charge transport.

Another aspect that differentiates electronic couplings from the monomer-dependent 
electronic properties is the degree of chemical detail required by the CG mapping resolution.
In particular, resolving the nitroxide group is not critical in the case of electronic couplings:
adding an explicit description of the nitroxide group does not improve the results 
within the statistical uncertainty of our predictions
(compare GBCG1-2 (solid) to GBNO1-2 (dashed) and CGM3 (solid) to CGNO (dashed) 
in Figure~\ref{fig:results-couplings}A).
This observation can be understood in terms of the electronic coupling
being dominated by the relative position between the two monomers. 
The nitroxide groups do not contribute to the definition of this relative position, and 
hence their explicit representation does not improve predictions of electronic couplings significantly.


\textbf{Electronic properties of condensed-phase CG simulations.}
We now develop CG models that retain electronic structure information and apply 
the trained NNs to predict electronic properties from condensed-phase CG simulations.
We validate the NN predictions by comparing them against the 
electronic properties obtained via the reference, state-of-the-art
approach that requires backmapping and explicit quantum-chemical calculations.

To investigate what the most suitable coarse-graining strategies are
when developing CG models that retain electronic structure information, 
we first build two CG models with the same mapping and 
bonded interactions but different nonbonded interactions. 
To realize this, we choose the CGM3 mapping that, as we have seen,
allows for good accuracy in predicting electronic couplings. At the same time, the CGM3 mapping
allows us to test two sets of nonbonded interactions derived via two different CG philosophies,
namely the building-block CG Martini 3 approach~\cite{2021PCTSouza-NatMethods}
and the structure-based CG IBI method.~\cite{2003DReith-JComputChem}
For details on the CG models, see Materials and Methods and the Supplementary Material.

\begin{figure}[htbp]
  \centering
  \includegraphics[width=0.95\textwidth]{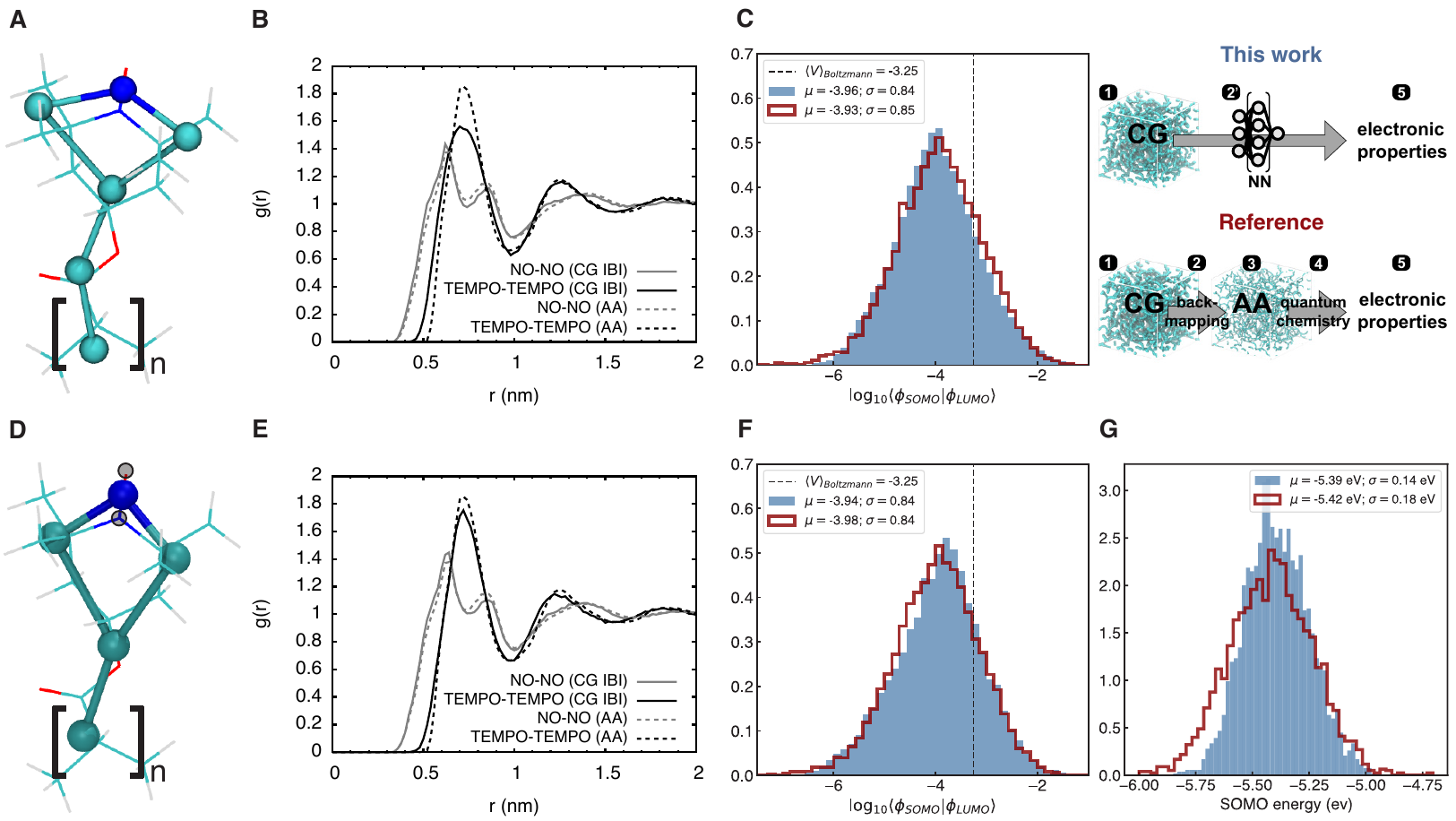}
    \caption{\textbf{Application and validation of the proposed ML-based
             method to PTMA, a prototypical radical polymer.}
             (\textbf{A}) Schematic representation of the CG model (CGM3 mapping)
             and the underlying AA structure.
             (\textbf{B}) Representative AA vs. CG RDFs for the CGM3 model 
             with nonbonded interactions obtained via IBI. 
             (\textbf{C}) Electronic couplings computed with 
             the reference backmapping-based approach (red) and 
             the proposed ML-based method (blue)
             starting from a CG morphology generated via CG MD using the CGM3 model.
             (\textbf{D}) Schematic representation of the CG model (GBNO2 mapping) 
             and the underlying AA structure;
             the nitroxide atoms are described by two virtual dummy sites 
             (transparent gray circles; see text for details). 
             (\textbf{E}) Representative AA vs. CG RDFs for the GBNO2 model.
             (\textbf{F}) Electronic couplings and 
             (\textbf{G}) SOMO energies computed with
             the reference backmapping-based approach (red) and 
             the proposed ML-based method (blue)
             starting from a CG morphology generated via CG MD using the GBNO2 model.
             Mean ($\mu$) and standard deviation ($\sigma$) of the 
             electronic property distributions are reported in the legends.
             }
  \label{fig:CGvsNN-results}
\end{figure}

We first consider the results generated with the CGM3 IBI model (Figure~\ref{fig:CGvsNN-results}A-C). 
The model is used to produce a morphology via CG MD, 
and we then use two methods to retrieve
the electronic couplings of such a CG morphology:
the reference approach that involves backmapping to atomistic resolution,
and explicit DFT calculations 
(taken to be the ground truth; in red in Figure~\ref{fig:CGvsNN-results}C)
and the proposed ML-based method (in blue in Figure~\ref{fig:CGvsNN-results}C).
The agreement between the two electronic coupling distributions is 
excellent,
with both the NN-predicted mean ($\mu$)
and standard deviation ($\sigma$) being around $1$\%
of the corresponding reference values.
We stress that the trained NN is able to predict electronic couplings of a CG simulation
with an accuracy comparable to that of the reference approach
but at a fraction of the computational cost
(see next section for a discussion on the computational efficiency).
Moreover, the NN was trained on CG conformations mapped from an AA MD simulation
and had not seen conformations drawn from a CG simulation before.
However, the structural accuracy of the CG model (Figure~\ref{fig:CGvsNN-results}B) 
is such that the CG model spans a conformational space that is consistent with 
the underlying AA structure, and therefore the NN is able to make predictions on conformations drawn from
previously unseen CG morphology realizations with excellent accuracy.

In contrast to the CGM3 IBI model, the CGM3 Martini 3 model leads to unsatisfactory electronic property predictions (Figure S4).
In fact, not only does the electronic coupling distribution obtained 
with the Martini model exhibit a considerably larger standard deviation 
($+15$\%, Figure S4C)
than the reference distribution, but that
distribution also shows qualitative discrepancies with respect to 
the reference distribution (i.e., it is not a unimodal distribution).
These discrepancies can be rationalized by looking at the structure of the Martini CG morphology,
for example by analyzing TEMPO-TEMPO and nitroxide-nitroxide radial distribution functions (RDFs)
(Figure S4B). 
We see that the RDFs are qualitatively different, 
showing, for example, an extra peak at $\approx0.4$ nm in the case of the nitroxide-nitroxide RDF. 
These discrepancies are indicative of molecular conformations occurring in the Martini CG morphology 
that have no AA counterpart, and which are hence non-physical.
In contrast, structural accuracy is the parametrization target of structure-based 
coarse-graining techniques such as IBI and, therefore,
as shown in the previous paragraph, the resulting CG models represent 
the underlying AA structure with high fidelity (Figure~\ref{fig:CGvsNN-results}B).
This structural accuracy appears to be an essential requirement when developing 
CG models that retain electronic structure information.

Having settled on IBI as the strategy of choice for developing 
CG models that retain electronic structure information, 
we now turn to the development of a CG model that is able to retain information on 
not only electronic couplings but also energy levels.
Informed by the results of the previous sections, 
we know that an explicit description of the nitroxide group is required for energy level prediction
and we therefore develop a corresponding CG model.
We choose the GBNO2 mapping (Figure~\ref{fig:CGvsNN-results}D), 
as it represents a good compromise between NN predictive accuracy---showing 
accurate electronic coupling (Figure~\ref{fig:results-mappings-and-NO}) 
and energy level (Figure~\ref{fig:results-couplings}) predictions---and coarse-graining degree.
We derive nonbonded interactions via IBI for all the GBCG2 beads (Supplementary Material).
The nitrogen and oxygen atoms of the nitroxide group are instead described by two virtual dummy sites
that do not interact via nonbonded interactions with any of the other CG sites.
As such, the GBNO2 model is expected to allow for predictions of both electronic
couplings and energy levels.

As done for CGM3-based CG models, 
we generate a morphology via CG MD using the GBNO2 model 
just described and predict this time both energy levels and electronic couplings 
using the trained NNs (Figure~\ref{fig:CGvsNN-results}D-F).
Again, we compare the NN predictions to the ground truth taken to be 
the energy levels and electronic couplings obtained with the 
reference, backmapping-based approach (Figure~\ref{fig:CGvsNN-results}F-\ref{fig:CGvsNN-results}G).
The agreement between the two electronic coupling distributions is
also excellent in this case,
with the NN-predicted $\mu$ being within $1$\% of the reference one
and $\sigma$ matching the reference value.
In contrast, the agreement between the two SOMO energy distributions is 
only satisfactory,
with a very good agreement between the NN-predicted and reference $\mu$ ($<1$\%) 
but a NN-predicted $\sigma$ that is 22\% lower than the reference.
We note that in this case, we are looking at an electronic property that depends 
strongly on the intra-monomer degrees of freedom. 
The degeneracy of the CG representation, i.e., the fact that multiple
AA conformations correspond to the same CG conformation, is 
likely responsible~\cite{2022GSivaraman-JChemTheoryComput}
for the narrowing of the SOMO distribution predicted at CG resolution.
As discussed later, a recently proposed~\cite{2022GSivaraman-JChemTheoryComput}
Deep Kernel Learning approach may remedy this shortcoming and improve the prediction.
Albeit with this limitation to keep in mind for the SOMO energy predictions,
the GBNO2 model retains information on both energy levels and couplings, 
and can be used to explore large ensembles of morphologies at CG resolution and 
retrieve electronic structure information with negligible computational costs.

\textbf{Comparison to quantum chemistry approaches.}
To highlight the importance of considering condensed-phase effects in soft materials 
when predicting electronic properties, including electronic couplings, 
we compare our results to some of the Boltzmann-averaged electronic couplings 
that have been used in recent, state-of-the-art quantum-chemical modeling 
studies~\cite{2021YTan-JAmChemSoc, 2022CHLi-JMaterChemA}
aimed at understanding transport in radical polymers.
In such studies, the electronic coupling for a given molecular species is evaluated
by first generating gas-phase dimer structures 
either randomly~\cite{2022CHLi-JMaterChemA} or with more elaborate dimer 
surface sampling algorithms~\cite{2021YTan-JAmChemSoc}.
For each of the generated dimer structures, 
the electronic coupling is computed with the level of theory of choice, 
along with either a total or binding energy for that structure. 
The energies, which are indicative of the relative stability of each structure 
\textit{in the gas phase}, are then used to obtain a Boltzmann-averaged electronic coupling according to:
\begin{equation}
\langle V \rangle = \frac{ \sum_{i=1}^{N} V_i \exp(\frac{-E_i}{k_B T}) }{ \sum_{i=1}^{N} \exp(\frac{-E_i}{k_B T}) }
\label{eq:Boltzmann-weighting}
\end{equation}
where $V_i$ is the electronic coupling of the $i$-th dimer, 
$E_i$ its binding energy 
(calculated as the single point energy difference between the dimer ($E^{dimer}_i$) 
and the isolated species ($E^{radical}_i$, $E^{cation}_i$), according to:
$E_i = E^{dimer}_i - E^{radical}_i - E^{cation}_i$),
$k_B$ the Boltzmann constant, $T$ the temperature (300 K),
and $N$ the total number of dimers.

For a comparison to the method introduced here,
we must compute the Boltzmann-averaged electronic coupling 
at the same level of theory used in the present work.
Hence, we randomly generate 10,000 dimer structures 
following the procedure by \citeauthor{2022CHLi-JMaterChemA}
(Supplementary Material for details)~\cite{2022CHLi-JMaterChemA}
and compute the $\langle \phi_{SOMO} | \phi_{LUMO} \rangle$ orbital overlap for each structure.
For the same dimer structure,
we also compute the binding energy (at the $\omega$B97X-D/6-311G(d,p) level of DFT) 
and calculate the Boltzmann-averaged electronic coupling according to 
Eq.~\ref{eq:Boltzmann-weighting}. The resulting coupling is shown with a vertical 
dashed line in Figure~\ref{fig:CGvsNN-results}C and Figure~\ref{fig:CGvsNN-results}F.
We can see that the Boltzmann-averaged coupling overestimates the magnitude
of the mean electronic coupling that arises in condensed phases. 
The discrepancy between the two is not surprising, given that the gas-phase 
sampling of the dimer structures 
neglects any solid-state packing preference dictated by \textit{(1)} steric constraints 
(including the fact that the monomers are attached to a polymer backbone) and 
\textit{(2)} (un)favorable intra- and inter-chain interactions.
A second intrinsic limitation of computing Boltzmann-averaged electronic couplings
is the lack of an estimate of the so-called structural~\cite{2020ASBondarenko-ChemSci}
disorder present in the system, i.e., the width of the electronic coupling distribution.
This disorder may affect the charge transport landscape and may therefore 
impact the conductivity of the material.
In contrast, the proposed method,
whose accuracy depends on the structural accuracy of the CG model,
gives access to not only 
\textit{(1)} electronic couplings that account for the packing in the condensed phase,
but also 
\textit{(2)} the full distributions of electronic couplings, thereby allowing for
quantification of the structural disorder in the system.


\section{\normalsize DISCUSSION}
\label{sec:discussion}
Having demonstrated the feasibility of designing and
developing CG models that retain electronic structure information,
and using such models as a more efficient substitute for
current multiscale approaches that involve backmapping procedures,
we conclude by briefly outlining a few opportunities for future extensions 
of the presented method.

The computational efficiency is a key advantage of the proposed method
with respect to backmapping-based approaches, and is expected to enable
innovative high-throughput investigations of electronic properties
over ensembles of soft material morphologies (see below).
The upfront cost of the dataset generation and ML model training is
(considering both electronic couplings and SOMO energies):
$\approx375\cdot10^3$ CPU hours for training data generation 
and $\approx4.2\cdot10^3$ CPU hours for training.
Note that this estimation takes into account both the exploration of suitable mappings
and the hyperparameter grid search performed for all the NNs.
The training data generation dominates the upfront cost.
Computing electronic properties with the backmapping-based approach requires 
a total of $\approx27\cdot10^3$ CPU hours for a single snapshot 
($\approx2.5\cdot10^3$ hours for backmapping and relaxation, and
 $\approx24.5\cdot10^3$ hours for the quantum-chemical calculations).
Instead, the NN predictions take $\approx1.5\cdot10^{-1}$ CPU hours per snapshot.
See Table S4 for a detailed breakdown. 
Therefore, if electronic properties are predicted
for 13 CG frames, we break even with the upfront training cost
(see Figure S25).
For any further frame, the proposed method will provide a speedup of 
$10^6$ times.
Further efficiency gains may be realized if more generic ML models, i.e.,
models trained on a larger fraction of the (CG) chemical space, will 
be developed---see also discussion below.

The fact that neither the GBCG nor the Martini CG
mappings describe the nitroxide group explicitly
makes these mappings unable to predict the SOMO energy level.
Although it is unsurprising that such fine detail is being averaged out by the CG mappings,
this result highlights the limitation of current methods to devise CG mappings 
that retain specific electronic structure information (e.g., SOMO energy).
While less strongly localized radical units might be less affected by this---i.e.,
we speculate that if a radical is more delocalized, a coarser representation may
still be able to capture the conformational-dependence of the SOMO energy---a
systematic way to identify mappings that preserve specific electronic properties 
would be greatly beneficial 
for automating the CG mapping step of the proposed method.

We have shown that a structurally-accurate CG model is
essential for the proposed method.
Hence, in general, structure-based coarse-graining techniques---such as 
IBI~\cite{2003DReith-JComputChem}---should be used when deriving interactions 
for CG models that retain electronic structure information.
However, other kinds of CG models such as 
chemically-specific, building-block CG force fields 
(e.g., Martini~\cite{2021PCTSouza-NatMethods}) 
that do not use the structure as a target for parametrization,
may still be valuable.
First,
while structural inaccuracies cannot be excluded for Martini-like
CG approaches, they are also not necessarily present 
in every CG model developed within such frameworks.
In fact, the latest Martini force field update (version 3)~\cite{2021PCTSouza-NatMethods}
has taken structural accuracy more into account 
via bonded parameter optimization that targets the AA molecular volume and shape
and a wider range of systematically pre-calibrated nonbonded interactions;
for example, structural accuracy is expected to improve for conjugated 
structures~\cite{2021PCTSouza-NatMethods, 2022RAlessandri-AdvTheorySimul}.
Hence, more generally, regardless of how the CG model potentials are obtained, 
as long as structural accuracy is preserved---which can be inspected, e.g., by comparing
AA and CG RDFs---we expect a CG model to be suitable for the proposed method.
Second, given the computational efficiency of Lennard-Jones based models such
as Martini as compared to models that use (IBI-derived) 
custom nonbonded potentials (Table S5),
and the transferability of CG models developed within building-block CG frameworks,
it is still valuable to investigate how such CG strategies can contribute
to the development of CG models that retain electronic structure information.

With respect to the ML model and featurization schemes used here,
there is ample room for improvement. Here we used NNs 
due to their simplicity of implementation, and application flexibility.
However, recently Gaussian process regression has been combined with NNs
to realize a Deep Kernel Learning (DKL) approach to 
ECG~\cite{2022GSivaraman-JChemTheoryComput, 2022JCMaier-JChemPhys}.
The approach allows to incorporate
the distributional nature of the electronic property predictions 
resulting from the CG mapping degeneracy, which, as we have seen in the 
present work, may be needed for electronic properties that strongly depends
on detailed intramolecular features (Figure~\ref{fig:CGvsNN-results}G).
DKL moreover provides prediction uncertainties and it is hence
suitable for active learning strategies.
Regarding the input featurization, the (reciprocal) distance matrices used
here constitute one of the simplest molecular representations that guarantees 
rotational and translation invariance. 
However, a host of more elaborated featurization schemes, 
such as representations based on symmetry functions or graph convolutions, 
are available.~\cite{2018ZWu-ChemSci-MoleculeNet} 
In particular, input representations containing chemical identity and possibly
other physical descriptors are expected to play a key role in order
to explore the question of whether ML models 
able to predict electronic properties at CG resolution
and that are \textit{transferable} across CG chemical space can be trained.
Finally, 
applications to conjugated polymers, in contrast to the nonconjugated
polymers studied here, need to consider that electronic states
can delocalize along the polymer backbone. To tackle this problem,
long short-term memory networks combined with a $\Delta$-ML approach
have been shown~\cite{2020NEJackson-Macromolecules}
to be a promising strategy to achieve the molecular-weight
transferability necessary to consider conjugation along polymer backbones.

The proposed method opens up interesting avenues for applications that require efficient 
modeling of radical-containing polymers, and more generally, in soft electronic materials.
For example, electronic properties could be studied as a function
of morphology processing conditions in a high-throughput fashion.
In particular, charge transport networks and fluctuations
of electronic couplings over CG spatiotemporal scales can be 
studied by the proposed approach.
Electronic couplings thus gathered could be fed into, for example,
kinetic Monte Carlo simulations aimed at computing conductivities.
As discussed before, such high-throughput studies are made possible by the 
$10^6$ increase in efficiency of the proposed method
with respect to current backmapping-based approaches.

In conclusion, an efficient ML-enabled method
has been presented that connects electronic properties of soft materials with the 
CG spatiotemporal scales required to sample their phase space.
The method is 
able to replace current backmapping-based multiscale approaches and
directly connect electronic structure information to CG degrees of freedom.
The method has been demonstrated 
for nonconjugated, radical-containing polymers 
but it is suitable for any soft material with electronic properties.
Overall, the findings reported underscore the potential of the 
proposed strategy to propel the investigation of soft materials' 
electronic properties that depend on conformational degrees of freedom
over a wide range of spatiotemporal scales,
a critical step towards the bottom-up design of soft electronic materials.

\section{\normalsize MATERIALS AND METHODS} 
\label{sec:methods}
\textbf{AA models.} 
Initial parameters were obtained for a OPLS-AA/CM1A force field 
from the LigParGen server~\cite{2005WLJorgensen-PNAS, 2017LSDodda-NucleicAcidsRes}.
QUBEKit~\cite{2018AEAAllen-JChemTheoryComput, 2019JTHorton-JChemInfModel} was subsequently
used to derive parameters for bond and angle potentials from DFT calculations 
(B3LYP/6-311++G(d,p)) that are tailored to PTMA. 
Charges were obtained with Gaussian via the CHELPG method.~\cite{Gaussian16-A03}
The resulting force field led to a density for methyl methacrylate and
TEMPO---molecular fragments that make up PTMA and for which experimental 
reference densities are available---in good agreement (within $2.5$\%) 
with experiments (Table S2).

\textbf{Datasets.}
Two datasets were built: one with single PTMA monomer conformations
and associated monomer conformation-dependent electronic properties 
(namely, SOMO energy level and spin density)
and a second one with dimer conformations and associated
dimer conformation-dependent electronic properties (electronic couplings).
We refer to these as ``monomer dataset'' and ``dimer dataset'', respectively.
The following protocol was used to ensure good conformational sampling
for the training data generation. We ran a condensed-phase 
AA MD simulation at 680 K (i.e., $\approx1.3 \cdot T_g$;
see Figure S9 for the estimation of the glass transition 
temperature, $T_g$, of the AA model)
and, after 30 ns of equilibration, we gathered snapshots every 10 ns.
Each snapshot was relaxed at 300K for at least 20 ns.
The relaxed snapshots were then used to extract the conformations for which
quantum-chemical calculations were performed.
Two different procedures were used to extract conformations 
for the two datasets, both leveraging the MDAnalysis python 
library~\cite{2011NMichaud-Agrawal-JComputChem, 2016RJGowers-ProceedingsScipy}
and Open Babel~\cite{2011NMOBoyle-JCheminformatics}.
For the monomer dataset,
4 snapshots containing 3,000 conformations each were used. 
For each conformation, coordinates were extracted up to the first
carbon of the backbone (``C01'' of Figure S10)
and the carbon atom capped with 3 hydrogen atoms. 
For each such conformation, a reciprocal distance matrices at the 
different resolutions were computed and stored and an input file for 
a single point calculation (B3LYP/6-311++G(d,p)) was written.
The ($4\cdot3,000$) 12,000 conformations were filtered to increase 
the diversity of the dataset by removing all conformations that had
a RMSD of less than 0.8 {\AA} with any other conformation.
Gaussian~\cite{Gaussian16-A03} was used to obtain 
the SOMO energy and Mulliken spin density (B3LYP/6-311++G(d,p)).
A final dataset of 10,778 datapoints was obtained. 
For the dimer dataset,
24 snapshots 
containing about 9,200 dimer conformations each were used. 
A dimer was selected if the distance between the respective nitrogen atoms
was within a cutoff of 10 {\AA}. For each selected dimer,
the DFT-optimized geometry of the monomer was aligned 
(by minimizing the RMSD~\cite{Theobald2005,Liu2010})
to each MD monomer structure.
In this way, intramonomer vibrations were neglected.
Intermonomer vibrations are much faster than the intermonomer motions and
affect negligibly intermonomer couplings~\cite{2002ATroisi-JPhysChemB}.
The electronic couplings were approximated as the orbital overlap~\cite{2002ATroisi-JPhysChemB} 
between the SOMO of the neutral radical and the LUMO of the cation, 
$\langle \phi_{SOMO} | \phi_{LUMO} \rangle$.
For each dimer, Gaussian~\cite{Gaussian16-A03} was used to compute 
\textit{(1)} the orbitals of monomer $i$ and 
\textit{(2)} the orbitals of monomer $j$ of the of $i$-$j$ dimer both 
by performing a single point calculation using as starting point the 
converged orbitals of the DFT-optimized geometry;
\textit{(3)} the overlap matrix 
by performing a calculation 
that produces only the overlap matrix in the MO basis set of the dimer;
finally, Multiwfn~\cite{2012TLu-JComputChem} was used to evaluate 
$\langle \phi_{SOMO} | \phi_{LUMO} \rangle$ by providing the Gaussian
checkpoint files from the three calculations described above. 
Note that the average overlap between the 
$\langle \phi_{SOMO,i} | \phi_{LUMO,j} \rangle$ and $\langle \phi_{LUMO,i} | \phi_{SOMO,j} \rangle$
values was taken, as the two values are in general different (although, at least for the 
present system, they are very similar).
A final dataset of 
221,406 datapoints 
was obtained.

\textbf{Coarse-graining.} 
GBCG mappings were generated using the spectral grouping 
variant of the GBCG algorithm~\cite{2019MAWebb-JChemTheoryComput},
as implemented at \url{https://github.com/xmwebb/GBCG}.
The Martini CG mapping was devised following the Martini 3 
guidelines~\cite{2021PCTSouza-NatMethods, 2022RAlessandri-AdvTheorySimul}.
Bonded parameters for the CG models were obtained by fitting standard bond, angle, and 
dihedral potentials to reference distributions obtained from mapped atomistic simulation 
(Supplementary Material).
Nonbonded parameters for the structure-based CG models were derived by
iterative Boltzmann inversion (IBI)~\cite{2003DReith-JComputChem} as implemented in 
the VOTCA package~\cite{2009VRuhle-JChemTheoryComput}. 
15 and 10 pair interactions were parametrized for the CGM3 and GBNO2 CG models,
respectively. For more details, 
see the Supplementary Material.
Nonbonded parameters for the Martini model---that is, Martini bead types---were 
assigned based on Martini 3 
guidelines~\cite{2021PCTSouza-NatMethods, 2022RAlessandri-AdvTheorySimul}
(and validated by computing octanol/water free energies of transfer,
see Table S3)
and available models for similar polymers~\cite{2022FGrunewald-NatCommun}
(Supplementary Material).

\textbf{Backmapping.} 
Backmapping was performed with Backward~\cite{2014TAWassenaar-JChemTheoryComput}, 
which relaxes the AA structure
with the atomistic force field after an initial geometrical reconstruction
of the AA structure based on the CG particle positions. The geometrical
reconstruction is specified via CG-to-AA mapping files.

\textbf{Molecular dynamics simulations.} 
Starting configurations for all polymers simulations were set up with 
Polyply~\cite{2022FGrunewald-NatCommun}, and MD simulations run with 
Gromacs versions 2021.x or more recent~\cite{2015MJAbraham-GMX5}.
AA simulations used a time step of 1 fs and the Verlet scheme with a
nonbonded cutoff of 1.1 nm, dispersion correction, and the 
particle mesh Ewald (PME) method for long-range electrostatic interactions. 
Bonds involving hydrogen atoms were constrained. Temperature and pressure
were controlled by a Nos{\'e}-Hoover thermostat (coupling parameter, $\tau_T$, of 1.0 ps) 
and a Parrinello-Rahman barostat (coupling parameter, $\tau_P$ of 5.0 ps), respectively.
The Berendsen barostat was used for equilibration purposes ($\tau_P=0.5$ ps).
For the Martini CG simulations, the Verlet scheme with a straight nonbonded
cutoff of 1.1 nm. Temperature and pressure were controlled by
a velocity-rescaling thermostat ($\tau_T=1.0$ ps) 
and a Parrinello-Rahman barostat ($\tau_P=12$ ps), respectively.
For the IBI-based CG simulations, the group cutoff scheme was used in order
to use tabulated potentials, with a nonbonded cutoff of 1.5 nm.
Note that CG simulations with IBI-derived nonbonded potentials
were run with Gromacs version 2019.5 because tabulated potentials are
not available in more recent Gromacs versions.

\textbf{ML model details.} 
For each monomer or dimer conformation, a reciprocal distance matrix 
$\mathbf{D}$ between all particles of a particular resolution was computed. 
Its elements are $D^{(i,j)}_{kl}=|\mathbf{r}^{(i)}_k-\mathbf{r}^{(j)}_l|^{-1}$ 
where $\mathbf{r}$ is the position vector,
$i$ and $j$ are the monomer indices, 
and $k$ and $l$ are the atom indices.
For the monomer dataset, $j=i$ and hence 
$D^{(i,i)}_{kl}=|\mathbf{r}^{(i)}_k-\mathbf{r}^{(i)}_l|^{-1}$.
Each matrix was flattened and the resulting one-dimensional vector (of dimension $N^2$) 
used as the input feature for the NN.
The electronic properties (SOMO energy level, spin density, electronic coupling)
are the labels associated to the input vectors for the supervised ML regression task.
The SOMO energy and spin density (only the values on the nitroxide nitrogen and oxygen 
atoms were considered, given that all other atoms have a spin density $<0.07$;
hence, two values were passed to the NN)
were used as obtained from the DFT calculations.
The base 10 logarithm of $\langle \phi_{SOMO} | \phi_{LUMO} \rangle$ was
used in the case of the couplings.
A fully-connected, feed-forward NN with a $M$-dimensional 
input layer followed by 4 batch-normalized hidden layers with the
same number of neurons was used so as to have a NN flexible enough.
$M$ is the dimension of the (flattened) input
vector and it therefore depends on the molecular resolution ($M=N^2$).
Hyperparameters that were optimized include: 
the number of neurons in the hidden layers, 
the batch size, 
and the number of training epochs. 
The default learning rate of the NAdam optimizer (0.001) was used for training.
Visualizations of the hyperparameter grid search for SOMO energy prediction
for some of the resolutions can be found in 
Figure S5-S8.
Standard scaling was applied to the input and output features.
10\% of the datasets (1,077 datapoints for the monomer dataset;
22,140
for the dimer dataset)
was held out and used as the test set.
The remaining datapoints were used to build training and validation
sets by using 5-fold cross-validation. 
Hyperparameters were optimized by grid search based on the 5-fold
cross-validated performance. 
For the best performing NN models for each property and molecular resolution,
see Table S1.
The final model performance was measured by applying the best model chosen
based on the 5-fold cross-validation to the held-out test set.
When predicting electronic couplings, as done when generating the dimer dataset:
\textit{(1)} couplings were inferred for dimers for which the distance
between the nitrogen virtual sites (or the beads representing the nitroxide group
in the case of the CGM3 mapping) was within a 10 {\AA} cutoff;
\textit{(2)} the CG-mapped DFT-optimized structure were
aligned to each monomer of the CG MD simulation snapshot structure
and electronic couplings were inferred based on the reciprocal distance
matrix computed for the dimer conformation after alignment.
All the ML methods were implemented using the Keras~\cite{2015FChollet-Keras}
and scikit-learn~\cite{2011FPedregosa-Scikit-learn} libraries.

\textbf{Data availability.} 
Data and code to reproduce the findings of this work
are available at
\url{https://github.com/ricalessandri/ECG-RedoxPoly-Cathode}.
The developed AA and CG polymer models 
are also implemented in the Polyply~\cite{2022FGrunewald-NatCommun} library
(\url{https://github.com/marrink-lab/polyply_1.0}).
The TEMPO and PTMA monomer Martini models
are also available on the 
Martini 3 small molecule library~\cite{2022RAlessandri-AdvTheorySimul}
(\url{https://github.com/ricalessandri/Martini3-small-molecules}).

\begin{acknowledgement}

R.A. is supported by the Dutch Research Council (NWO Rubicon 019.202EN.028).
The work was partly supported by grant NSF-DMR-2119673 funded by the National Science Foundation.
The authors acknowledge the Research Computing Center of the University of Chicago 
for computational resources.

\end{acknowledgement}

\begin{suppinfo}

Additional results: 
NN performance vs. dataset size for SOMO energy prediction,
extended Figure~\ref{fig:results-mappings-and-NO},
extended Figure~\ref{fig:results-couplings}, 
extended Figure~\ref{fig:CGvsNN-results} (CGM3 Martini model);
ML model hyperparameters;
density and $T_g$ data of the AA model;
details on the development of the CG models: mappings,
bonded parameters, 
IBI nonbonded parameters, 
Martini nonbonded parameters;
random dimer structure generation;
computational performance data.

\end{suppinfo}


\bibliography{redox}


\begin{tocentry}

    \includegraphics[width=\textwidth]{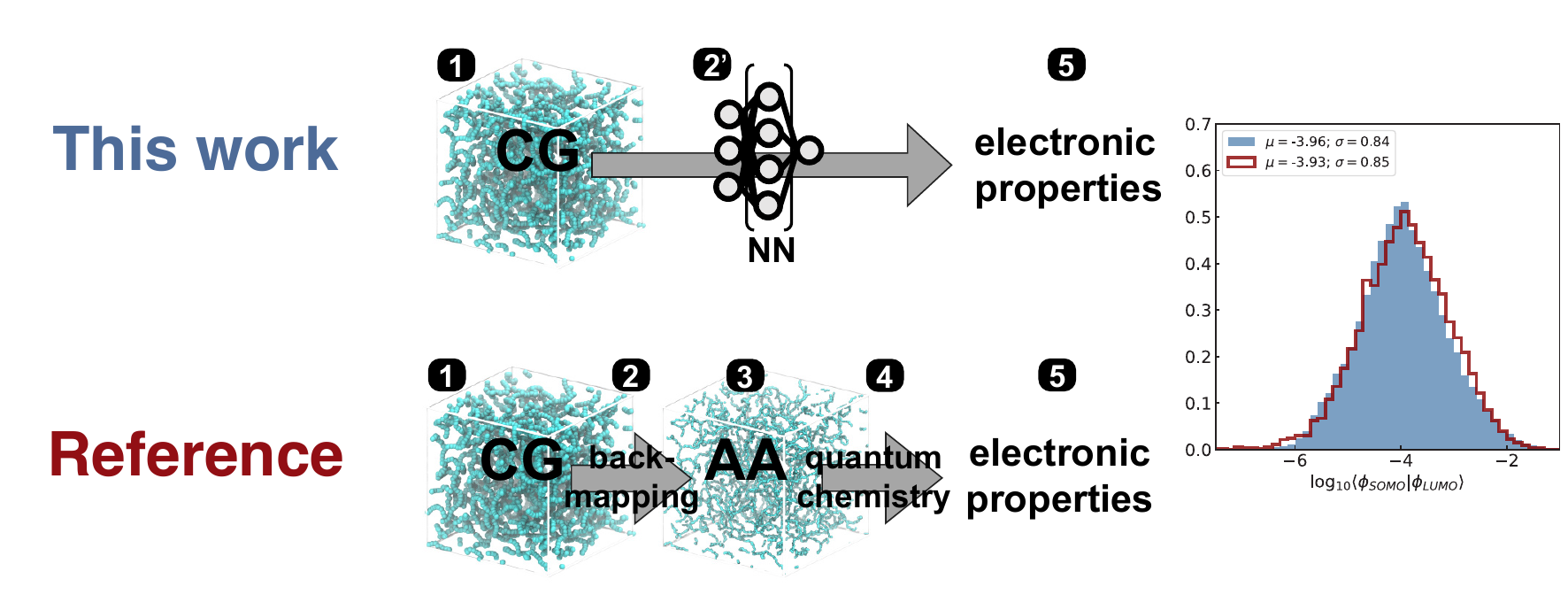}

%
%
%


\end{tocentry}

\end{document}



\newpage

\tableofcontents

\newpage

\section{Additional results}
\label{SI:sec:additional-results}
\begin{figure}[H]
  \centering
  \includegraphics[width=0.49\textwidth]{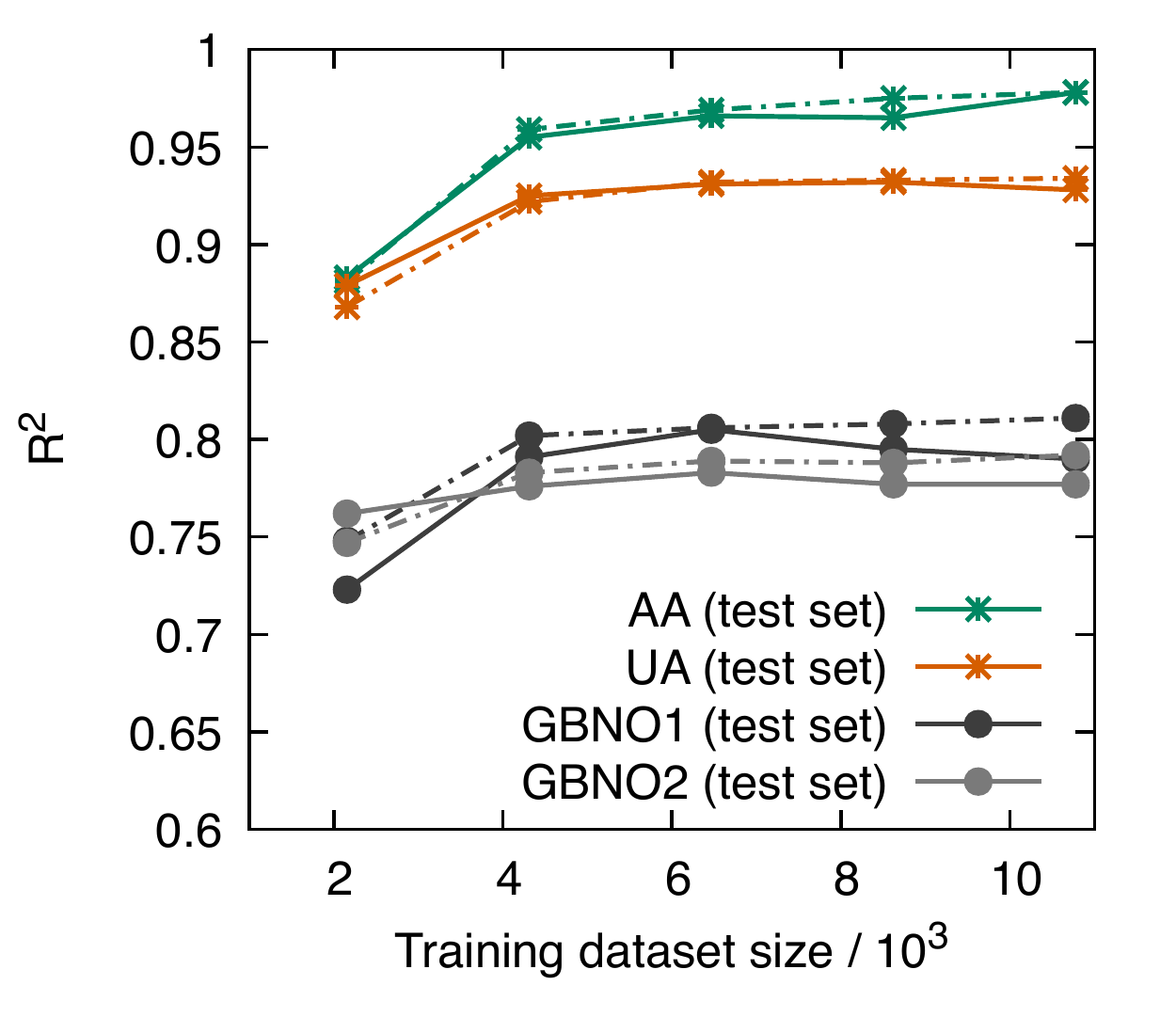}
    \caption{\textbf{NN performance vs. training dataset size for the prediction of SOMO energies.}
             NN performance as a function of training dataset size for different 
             resolutions: all-atom (AA), united-atom (UA), GBNO1, GBNO2. 
             Solid lines are results on the test set, while dash-dotted lines
             are results the fivefold cross-validated results.
             }
  \label{fig:results-SOMO-r2-vs-datasize}
\end{figure}
\begin{figure}[H]
  \centering
  \includegraphics[width=0.49\textwidth]{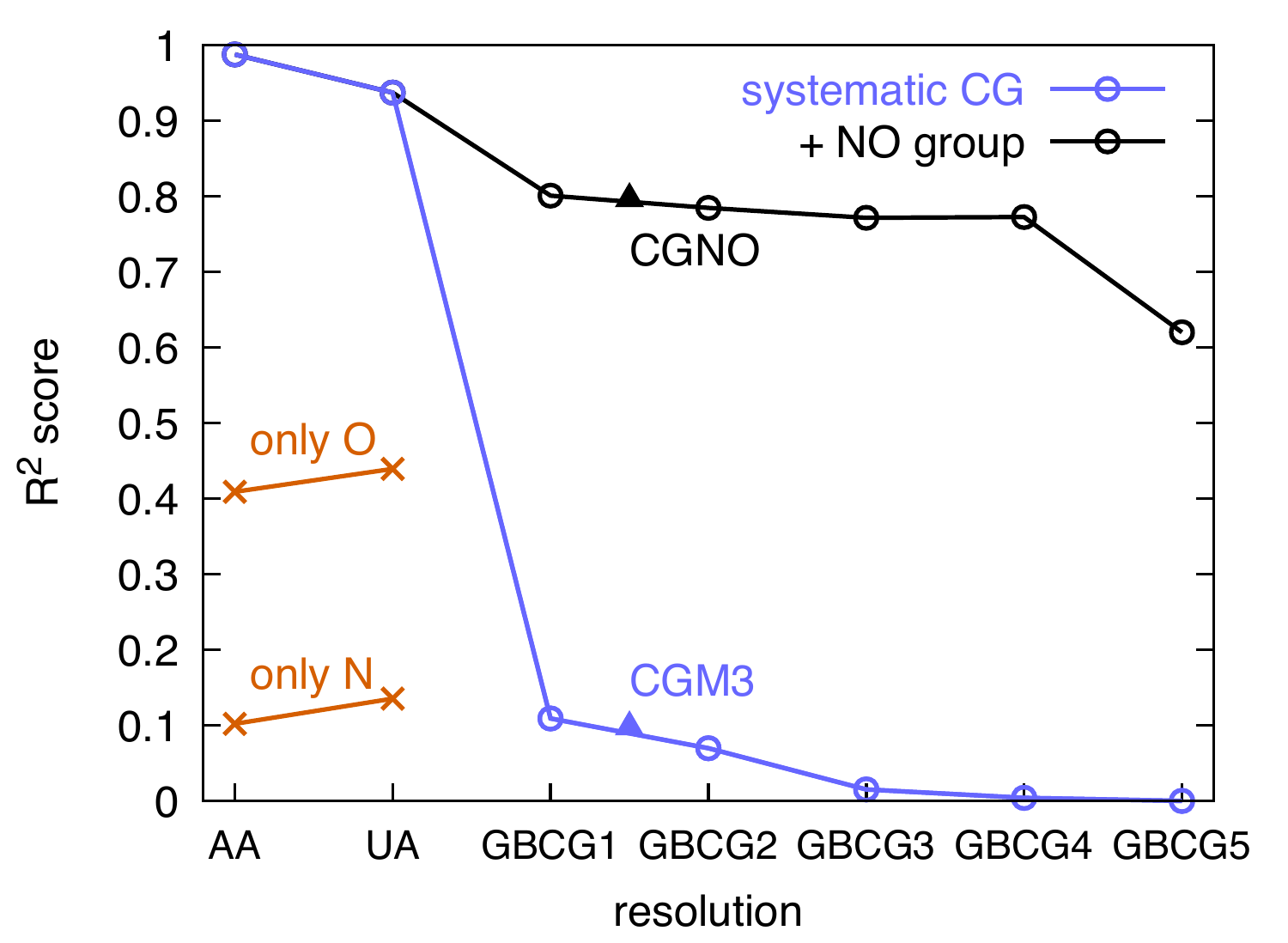}
    \caption{\textbf{Effect of excluding either N or O from the molecular representation
             on the predictive performance of NN models for SOMO energy prediction.}
             $R^2$ coefficients obtained for AA and UA
             representations without the nitrogen (``only O'', orange crosses) 
             or without the oxygen (``only N'', orange crosses). For comparison, data reproduced from 
             Figure~\ref{M-fig:results-mappings-and-NO}A of the main text are also shown.
             }
  \label{fig:results-mappings-and-NO-extra}
\end{figure}
\begin{figure}[H]
  \centering
  \includegraphics[width=0.49\textwidth]{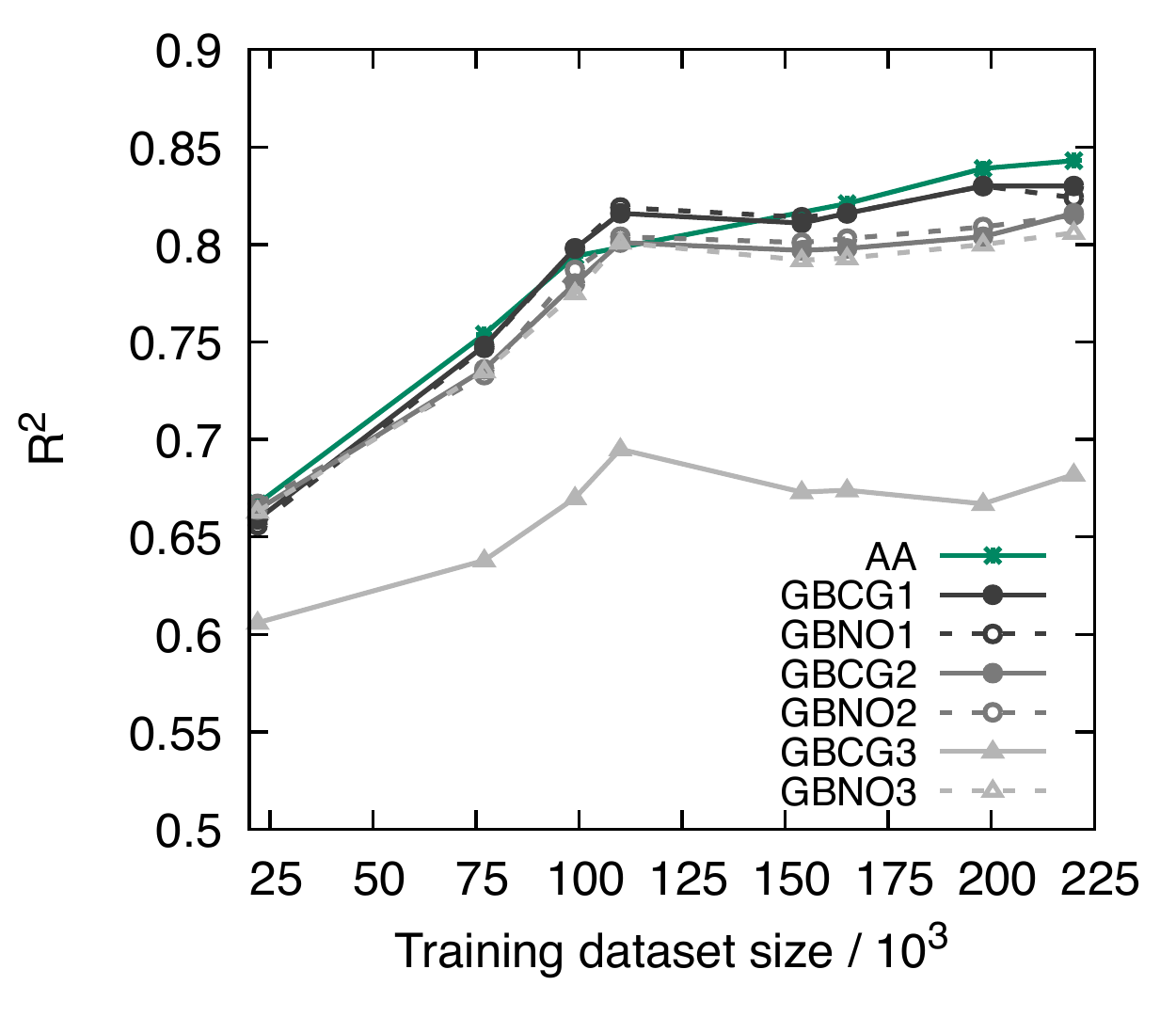}
    \caption{\textbf{Predictive performance for electronic couplings: GBCG mappings.}
             (\textbf{A}) NN performance as a function of training dataset size for different 
             resolutions: all-atom (AA), GBCG1-3, and GBNO1-3. 
             See Figure~\ref{M-fig:results-mappings-and-NO}B for a schematic representation of the mappings.
             The $R^2$ scores is for the test set.
             }
  \label{fig:results-couplings-extra}
\end{figure}

\begin{figure}[htbp]
  \centering
  \includegraphics[width=\textwidth]{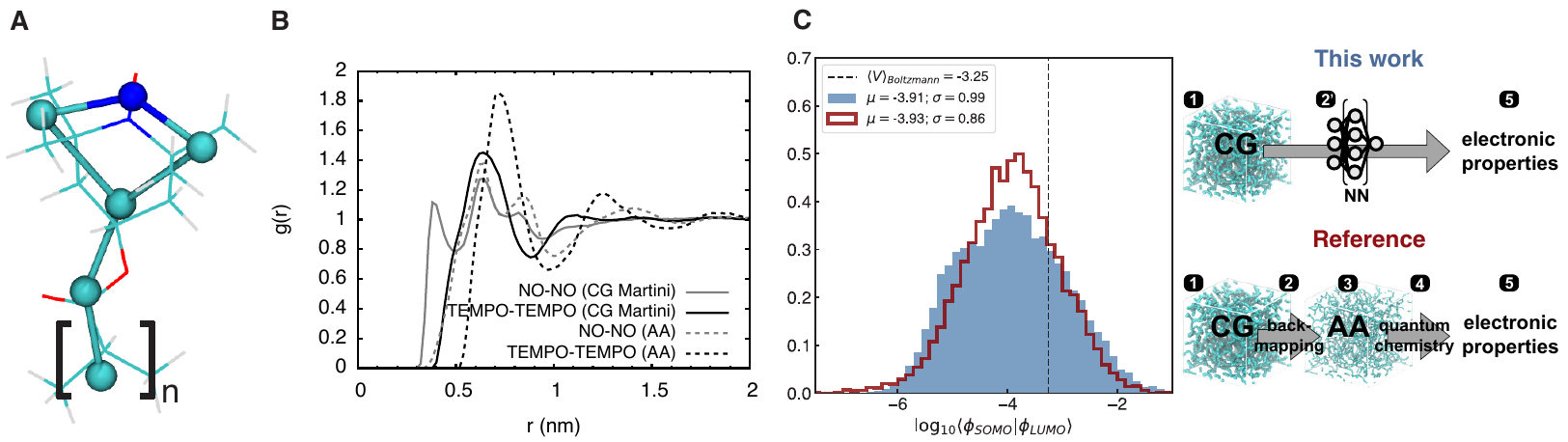}
    \caption{\textbf{Application of the method using the Martini 3 CG model.} 
             (\textbf{A}) Schematic representation of the CG model (CGM3 mapping)
             and the underlying AA structure.
             (\textbf{B}) Representative AA vs. CG RDFs for the CGM3 model 
             with Martini 3 nonbonded interactions. 
             (\textbf{C}) Electronic couplings computed with 
             the reference backmapping-based approach (red) and 
             the proposed ML-based method (blue)
             starting from a CG morphology generated via CG MD.
             Mean ($\mu$) and standard deviation ($\sigma$) of the 
             electronic property distributions are reported in the legends.
             }
  \label{fig:CGvsNN-results-M3}
\end{figure}

\newpage 

\section{Machine learning models}
\label{SI:sec:ML}
\begin{table}[htbp]
  \caption{\textbf{Hyperparameters of the highest-performing NN models.}
           For the different properties and molecular resolutions,
           the hyperparameters---optimized via grid search---for the 
           highest-performing NN models (as judged from the fivefold
           cross-validated $R^2$ score) are reported.
          }
  \centering
  \label{tbl:best-NNs}
  \begin{tabular}{lrrrr}
\hline
property, resolution & learning rate & batch size & no. epochs & no. neurons / layer  \\
\hline            
couplings, AA        & 0.002         &     1024   & 3000       & 400  \\
couplings, UA        & 0.002         &     1024   & 2000       & 400  \\
couplings, GBCG1     & 0.004         &     1024   & 2000       & 400  \\
couplings, GBNO1     & 0.001         &      256   & 2000       & 400  \\
couplings, GBCG2     & 0.004         &     1024   & 2000       & 400  \\
couplings, GBNO2     & 0.001         &      256   & 2000       & 400  \\
couplings, GBCG3     & 0.001         &      256   & 2000       & 400  \\
couplings, GBNO3     & 0.001         &      256   & 2000       & 400  \\
couplings, GBCG4     & 0.001         &      256   & 2000       & 400  \\
couplings, GBNO4     & 0.001         &      256   & 2000       & 400  \\
couplings, GBCG5     & 0.001         &      256   & 2000       & 400  \\
couplings, GBNO5     & 0.001         &      256   & 2000       & 400  \\
couplings, CGM3      & 0.001         &      256   & 2000       & 400  \\
couplings, CGNO      & 0.001         &      256   & 2000       & 400  \\
\hline            
SOMOs, AA        & 0.001         &      128   & 1000       &   3  \\
SOMOs, UA        & 0.001         &      128   & 2000       &   3  \\
SOMOs, GBNO1     & 0.001         &      256   & 2000       &   3  \\
SOMOs, GBNO2     & 0.001         &      256   & 1000       &   3  \\
SOMOs, GBNO3     & 0.001         &      128   & 1000       &  12  \\
SOMOs, GBNO4     & 0.001         &      128   & 0100       &  12  \\
SOMOs, GBNO5     & 0.001         &      256   & 2000       &  12  \\
SOMOs, CGNO      & 0.001         &      256   & 2000       &   3  \\
\hline
  \end{tabular}
\end{table}

\begin{figure}[H]
  \centering
  \includegraphics[width=0.31\textwidth]{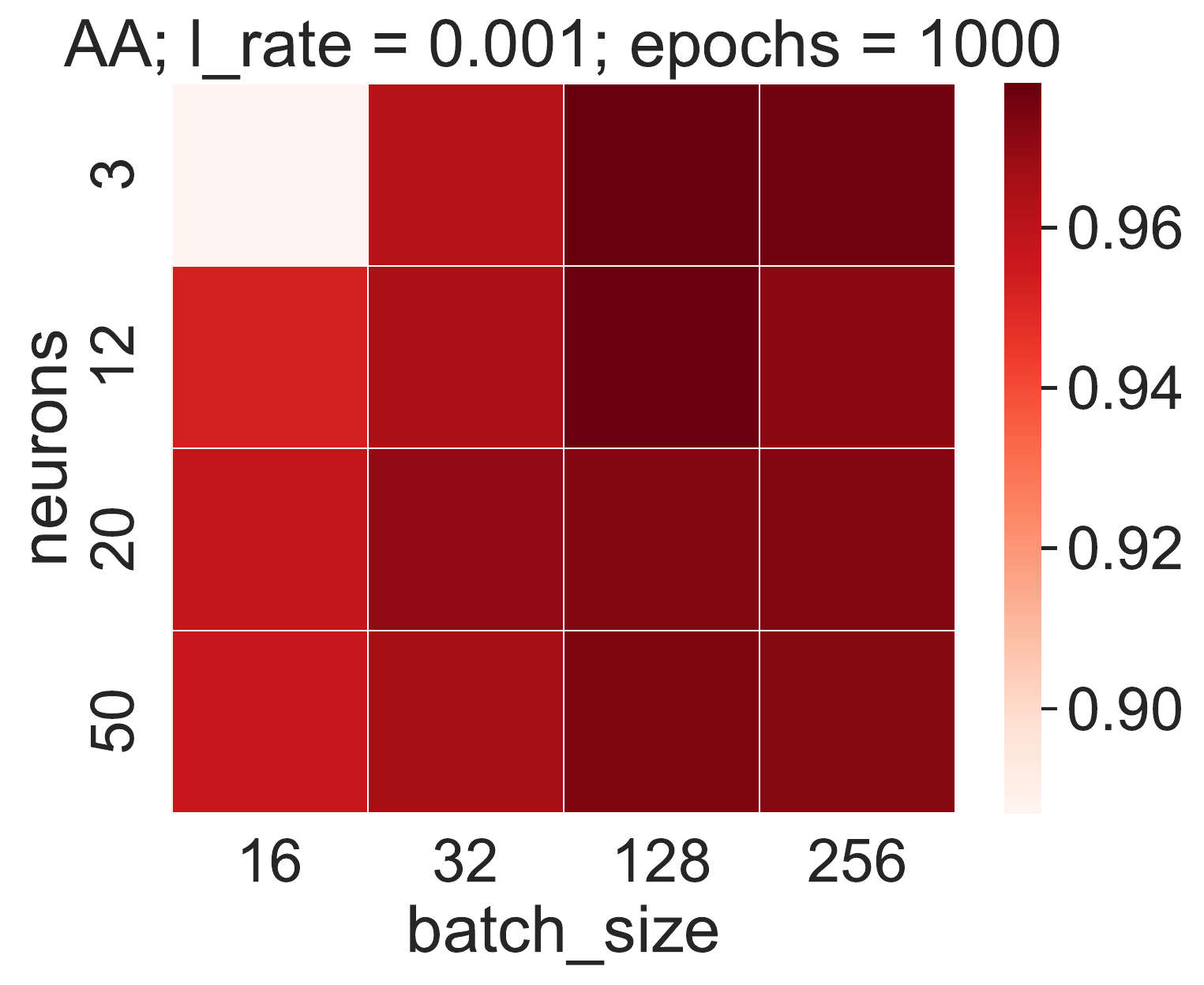}
  \includegraphics[width=0.31\textwidth]{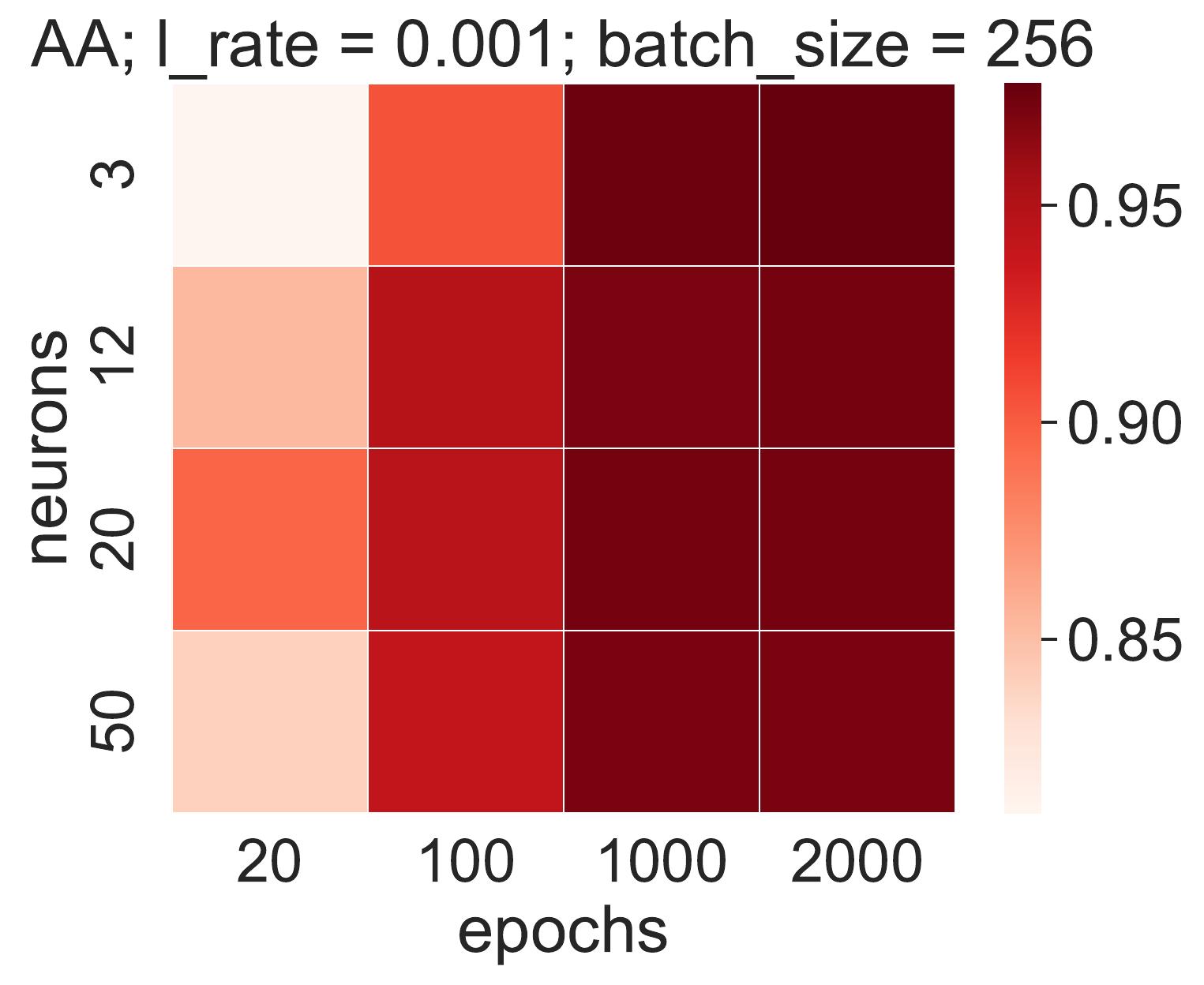}
  \includegraphics[width=0.31\textwidth]{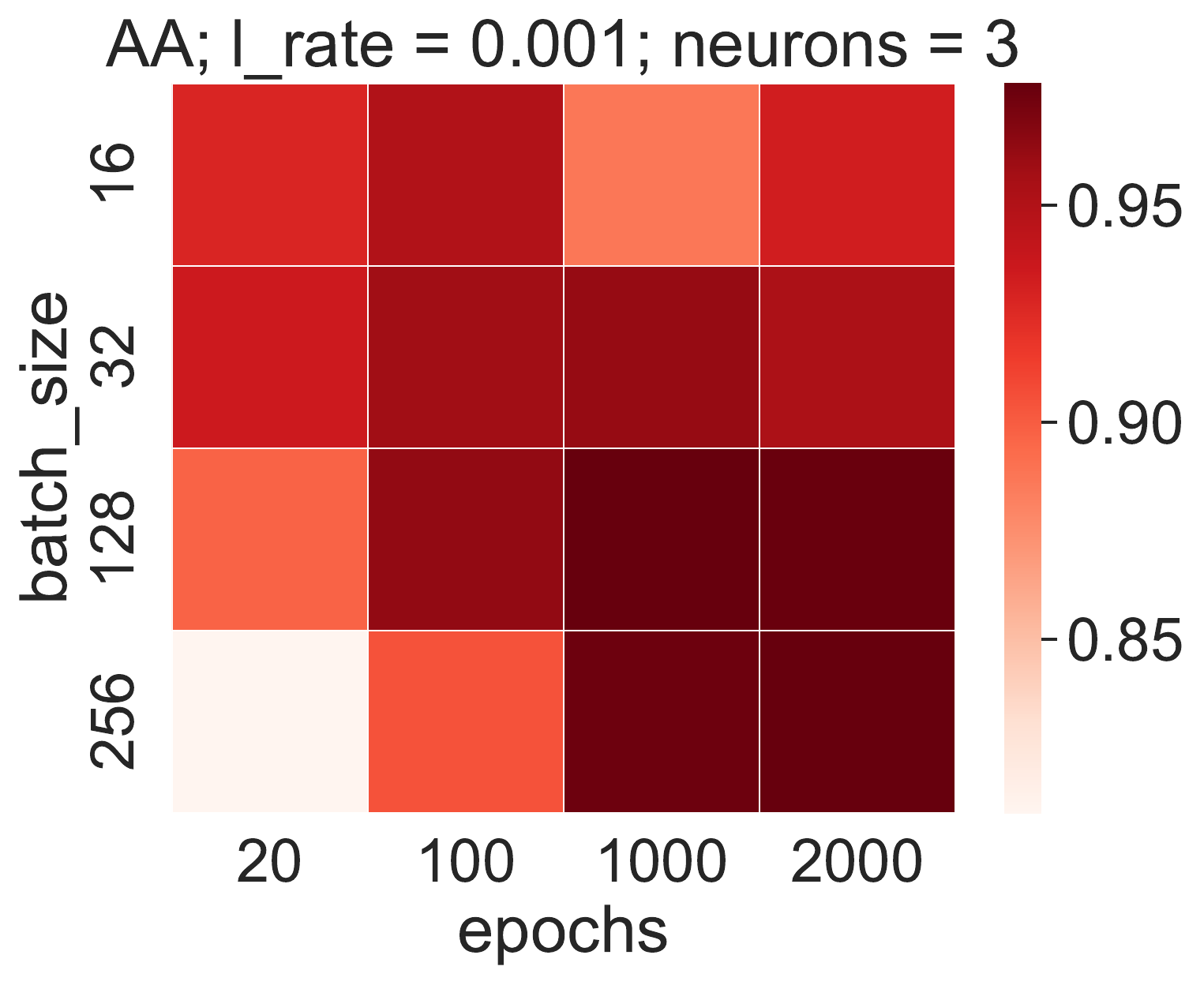}

  \includegraphics[width=0.31\textwidth]{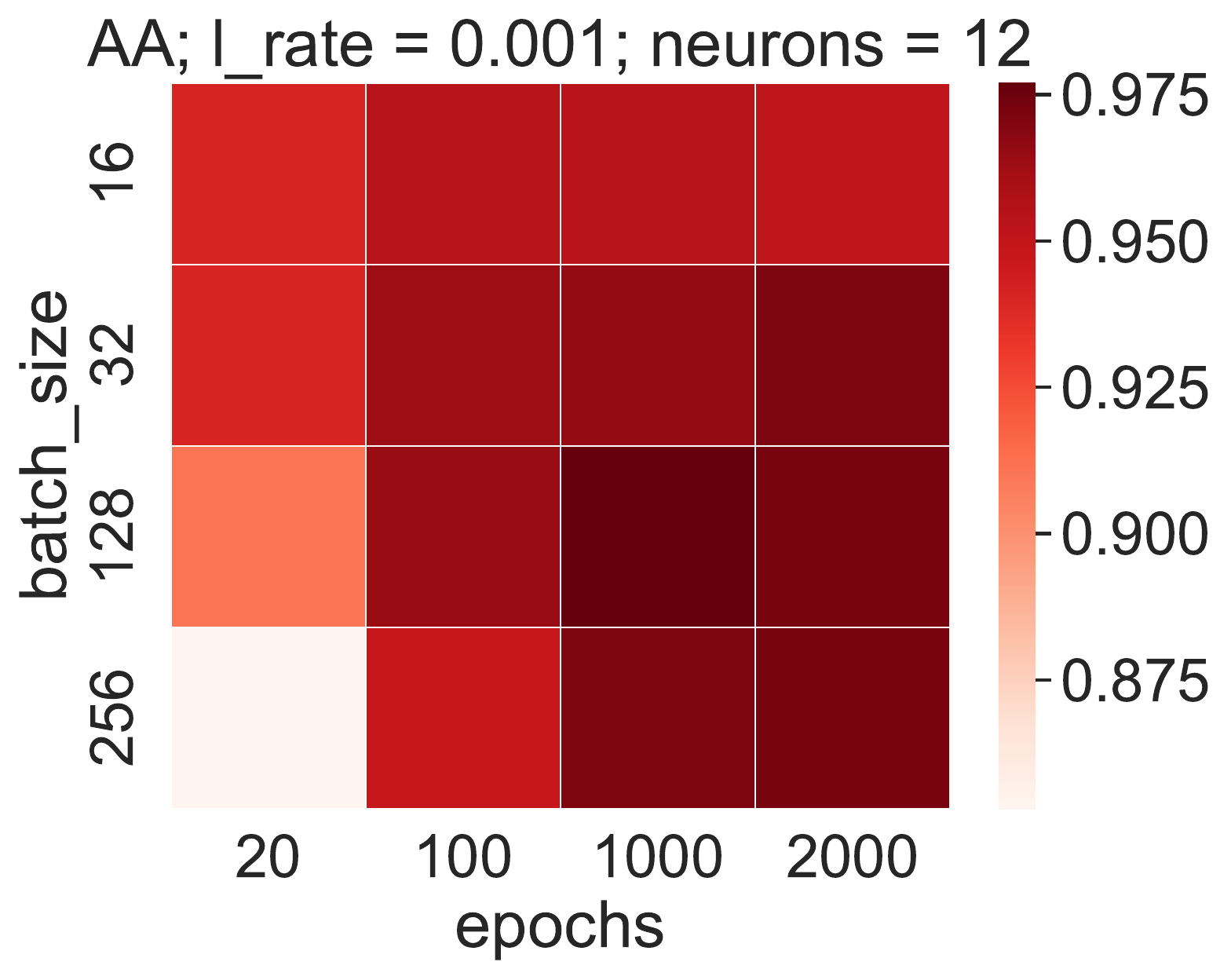}
  \includegraphics[width=0.31\textwidth]{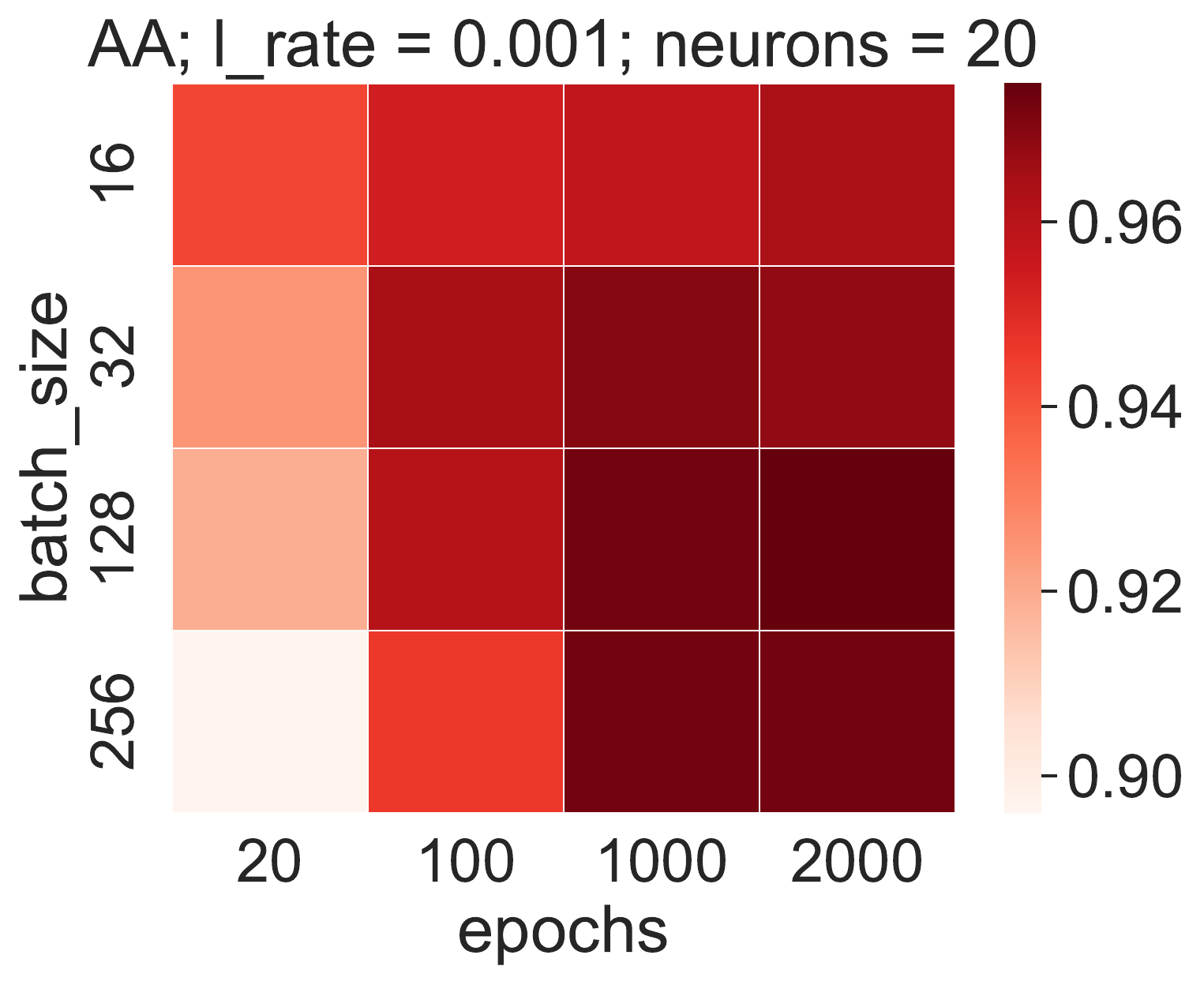}
  \includegraphics[width=0.31\textwidth]{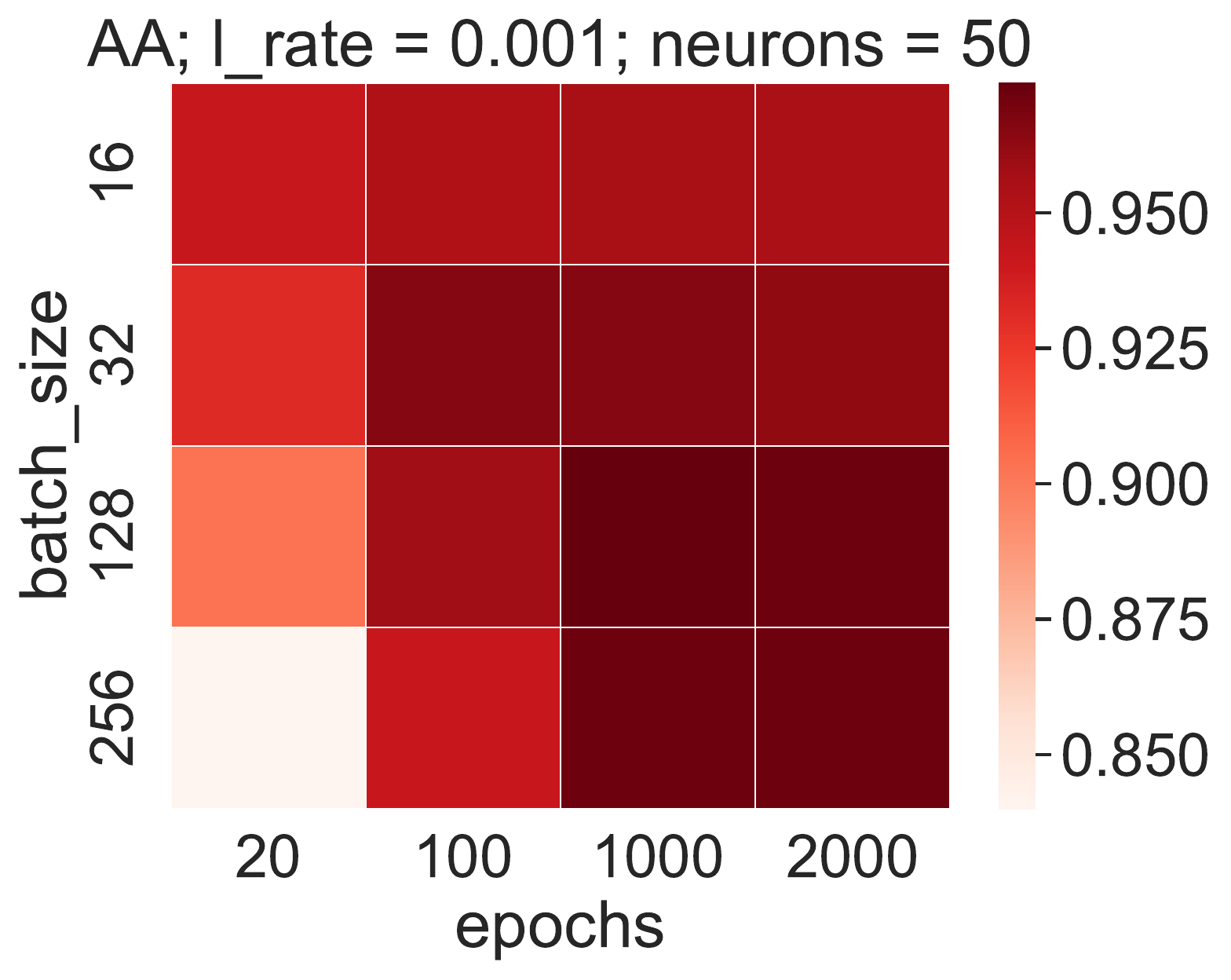}
  \caption{\textbf{Hyperparameter grid search for SOMO energy prediction: AA resolution.}
           Hyperparameters tested via grid search 
           are reported on the heatmap axes, that are colored according to the corresponding 
           fivefold cross-validated $R^2$ obtained after training.
          }
  \label{fig:hyperparameter-heatmaps-AA}
\end{figure}
\begin{figure}[H]
  \centering
  \includegraphics[width=0.31\textwidth]{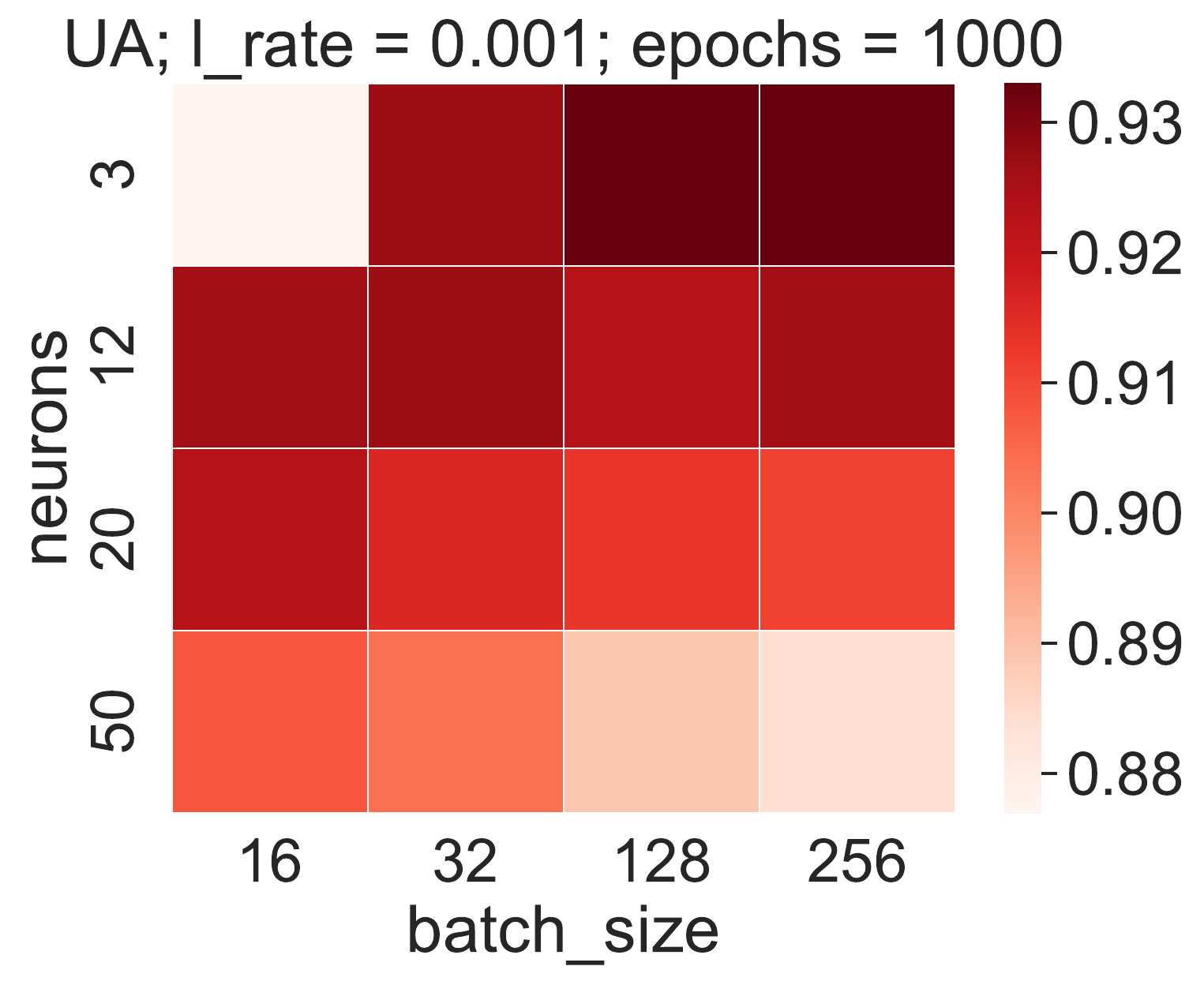}
  \includegraphics[width=0.31\textwidth]{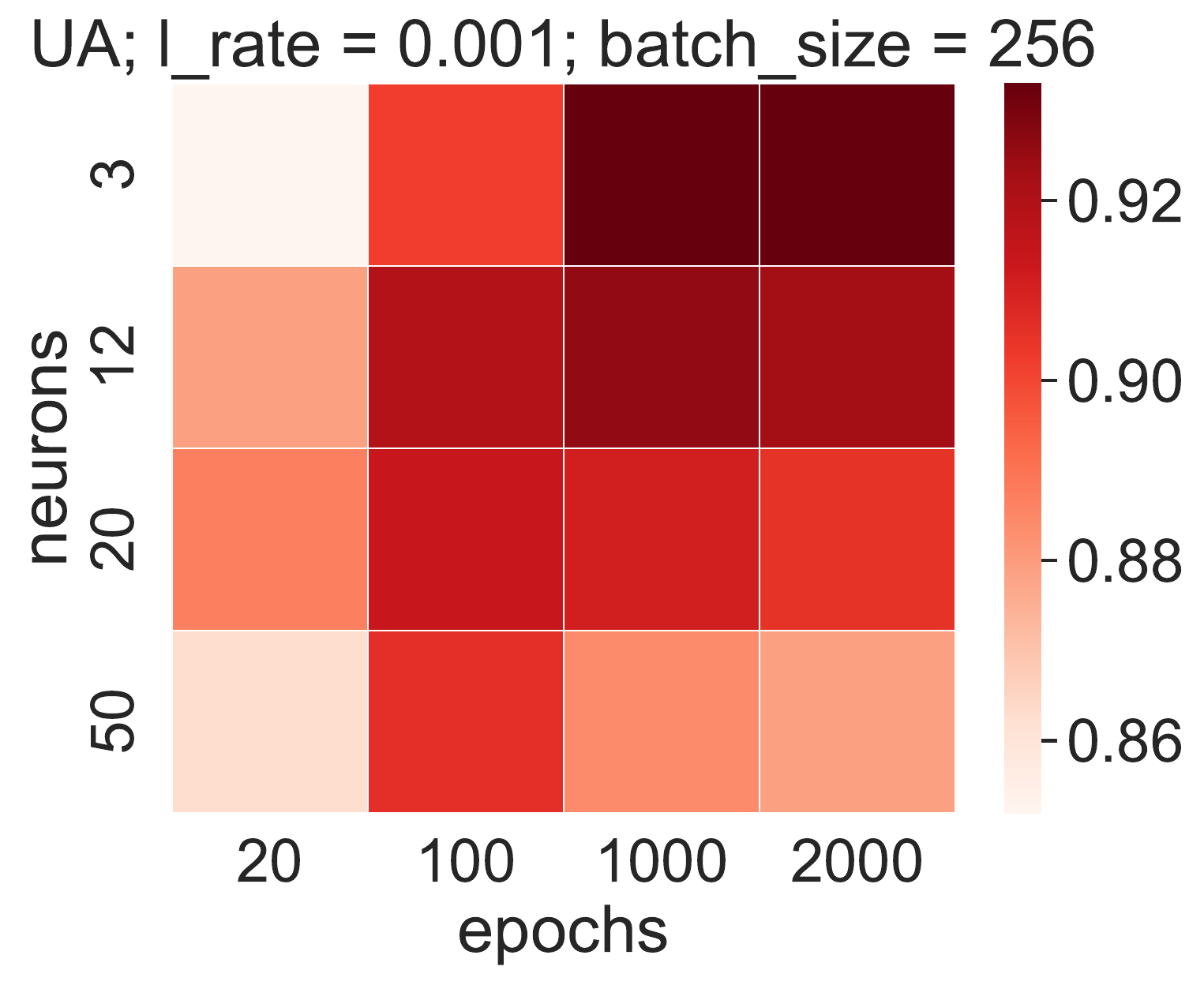}
  \includegraphics[width=0.31\textwidth]{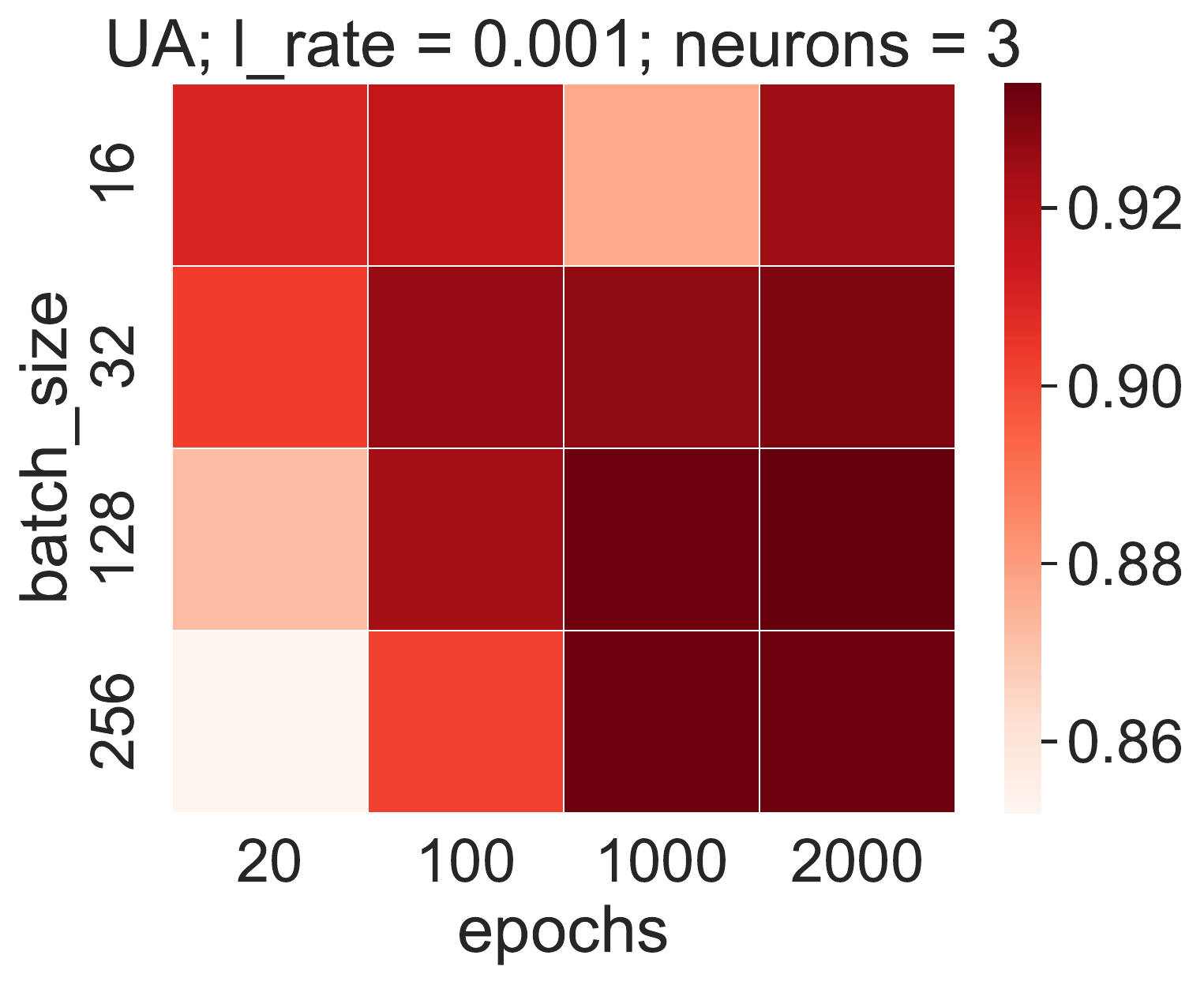}

  \includegraphics[width=0.31\textwidth]{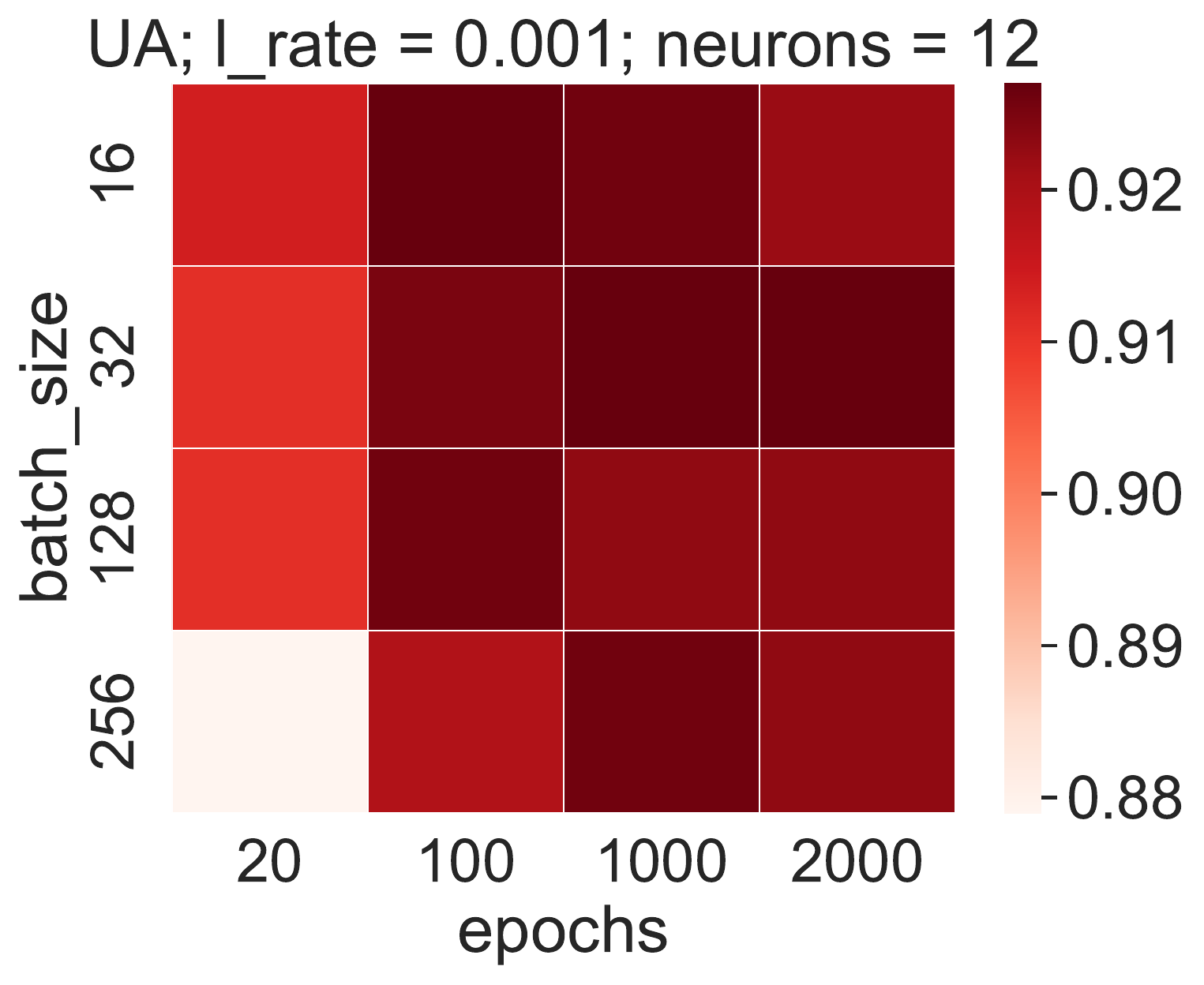}
  \includegraphics[width=0.31\textwidth]{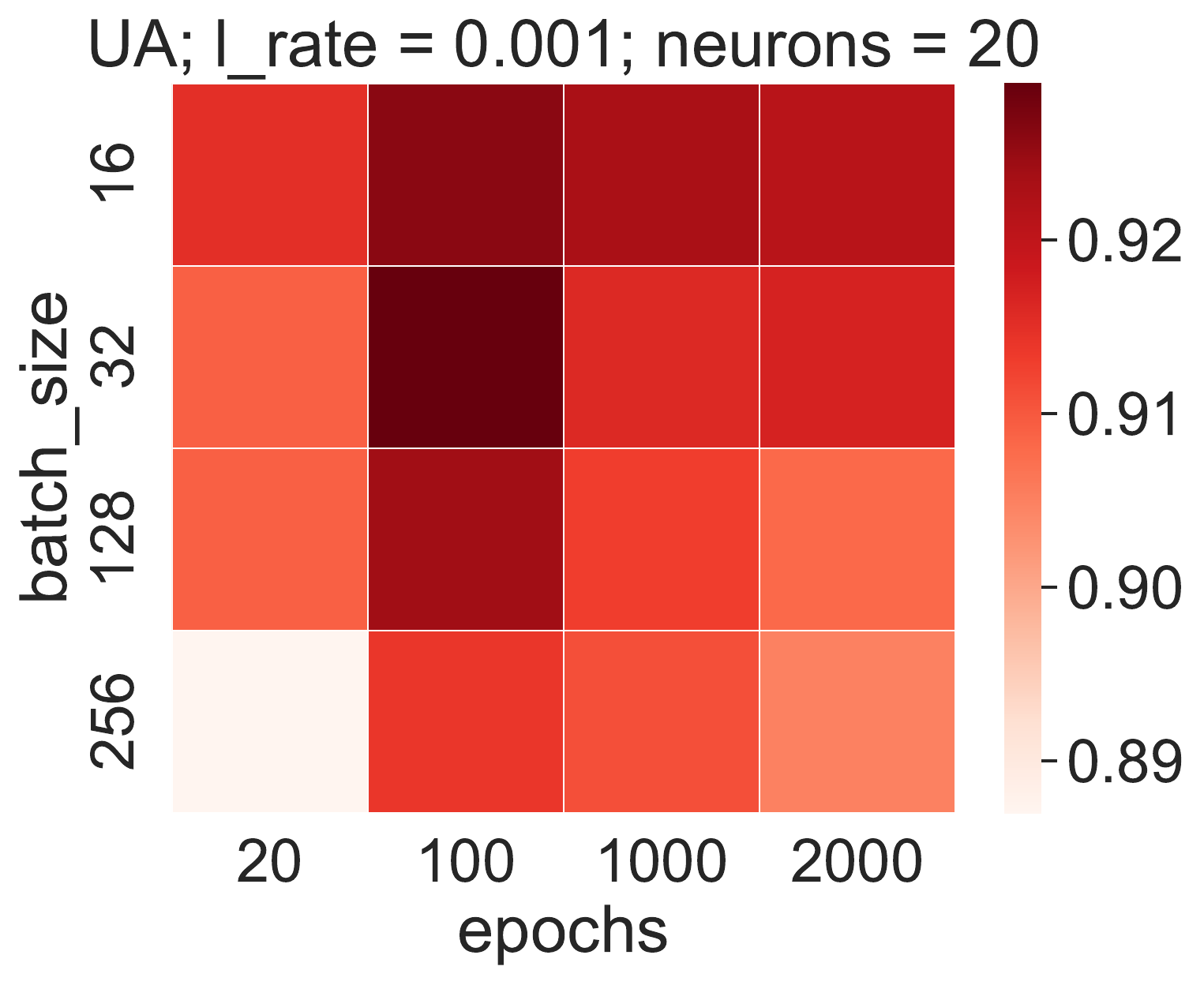}
  \includegraphics[width=0.31\textwidth]{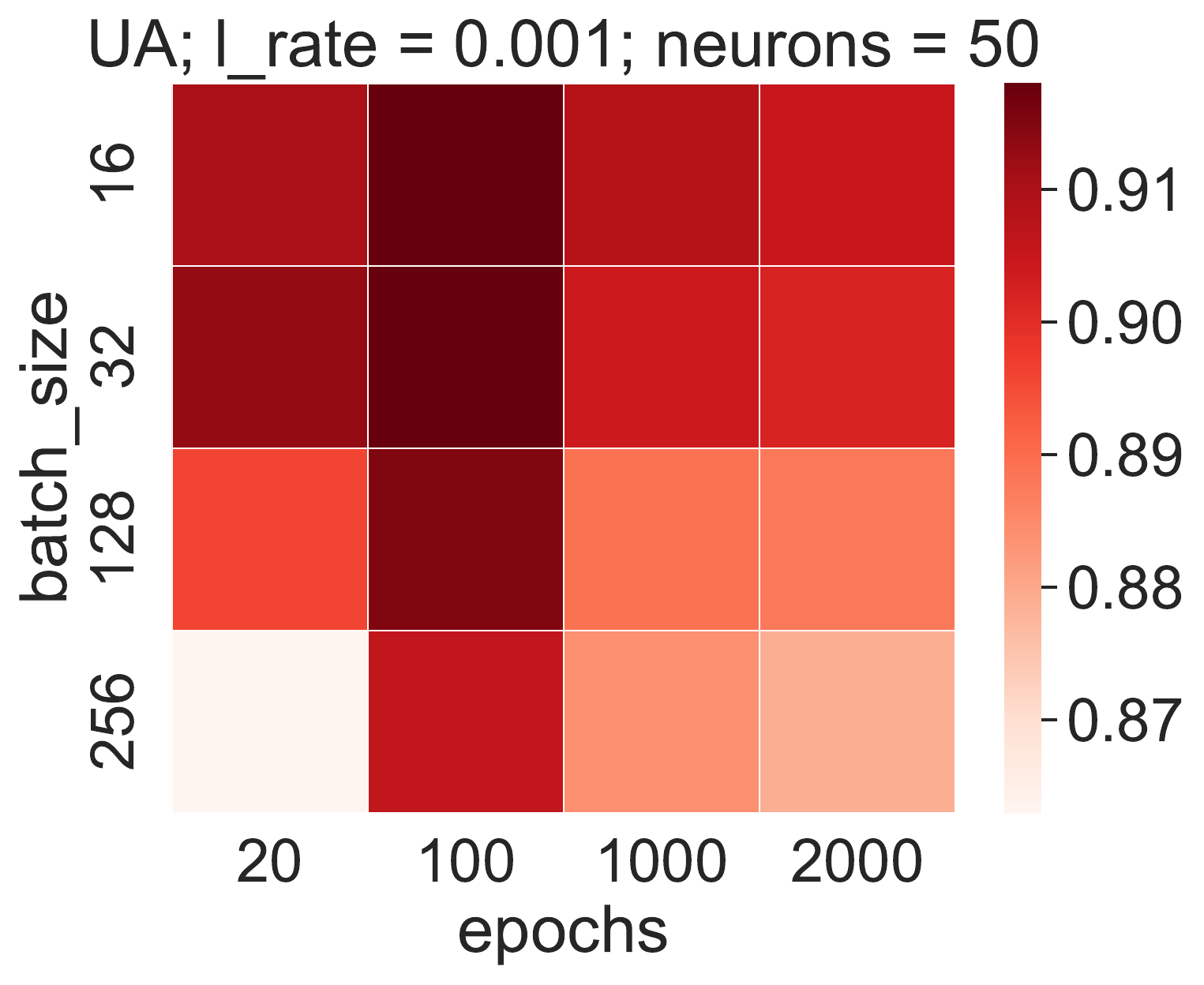}
  \caption{\textbf{Hyperparameter grid search for SOMO energy prediction: UA resolution.}
           Hyperparameters tested via grid search 
           are reported on the heatmap axes, that are colored according to the corresponding
           5-fold cross-validated $R^2$ obtained after training.
          }
  \label{fig:hyperparameter-heatmaps-UA}
\end{figure}
\begin{figure}[H]
  \centering
  \includegraphics[width=0.31\textwidth]{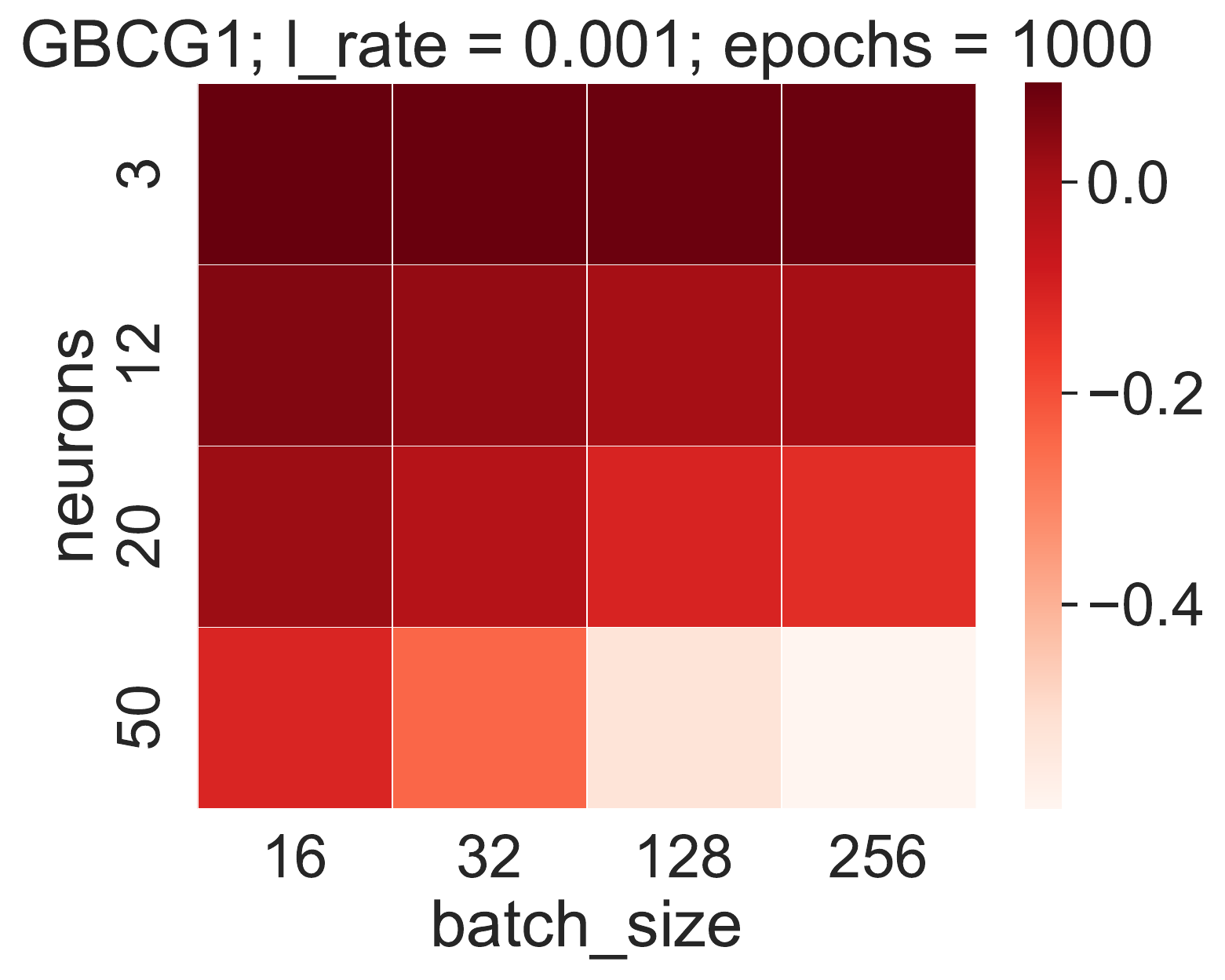}
  \includegraphics[width=0.31\textwidth]{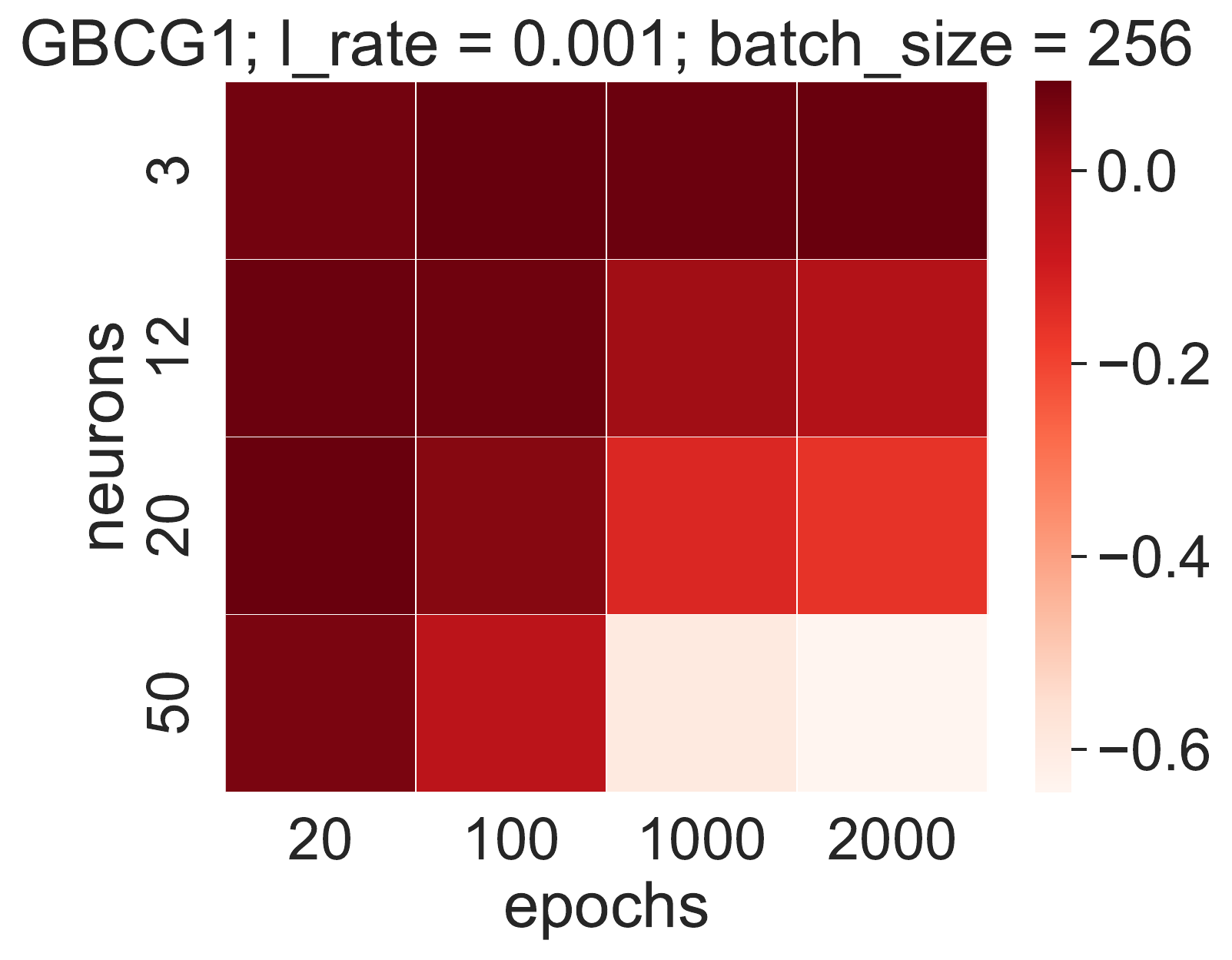}
  \includegraphics[width=0.31\textwidth]{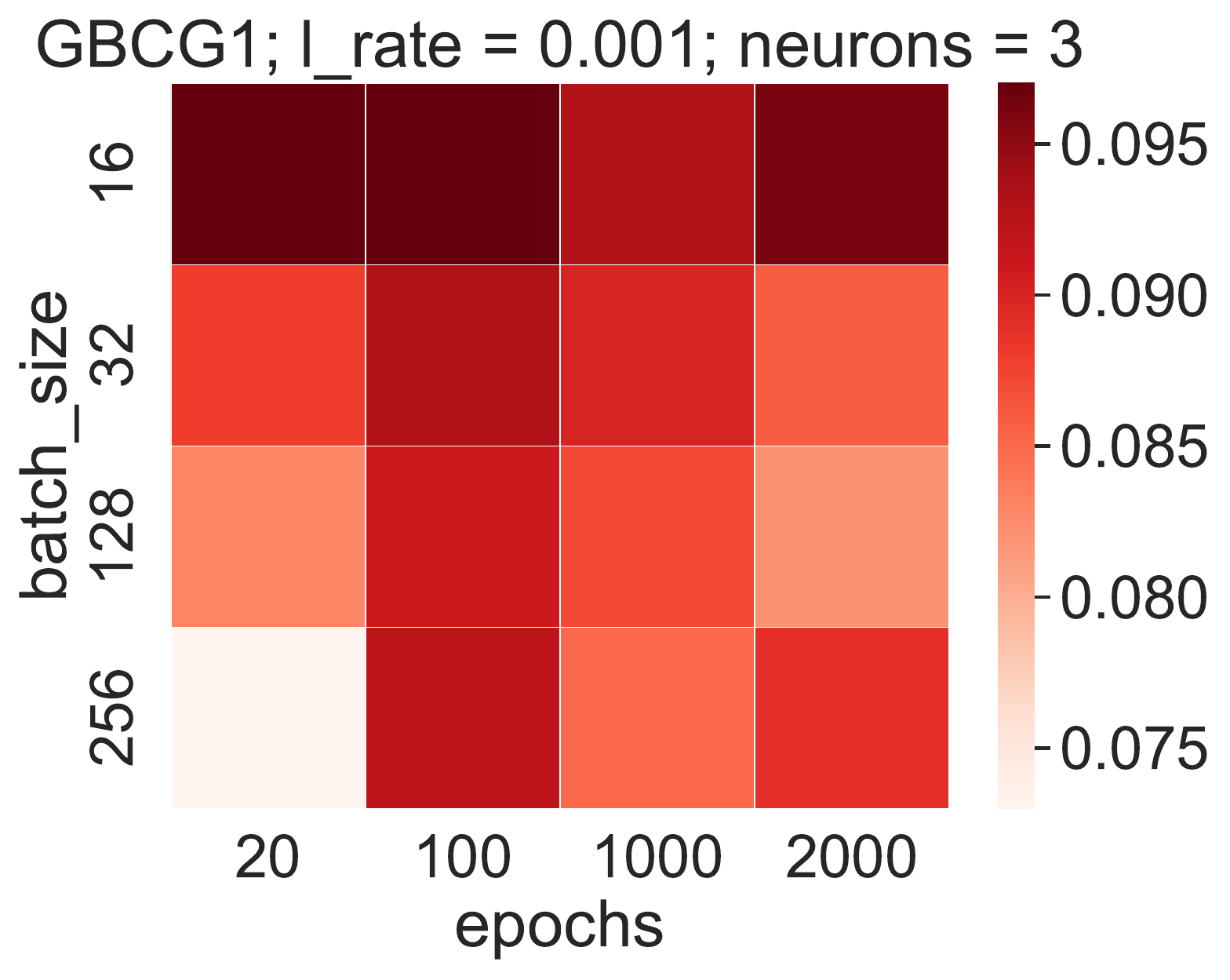}

  \includegraphics[width=0.31\textwidth]{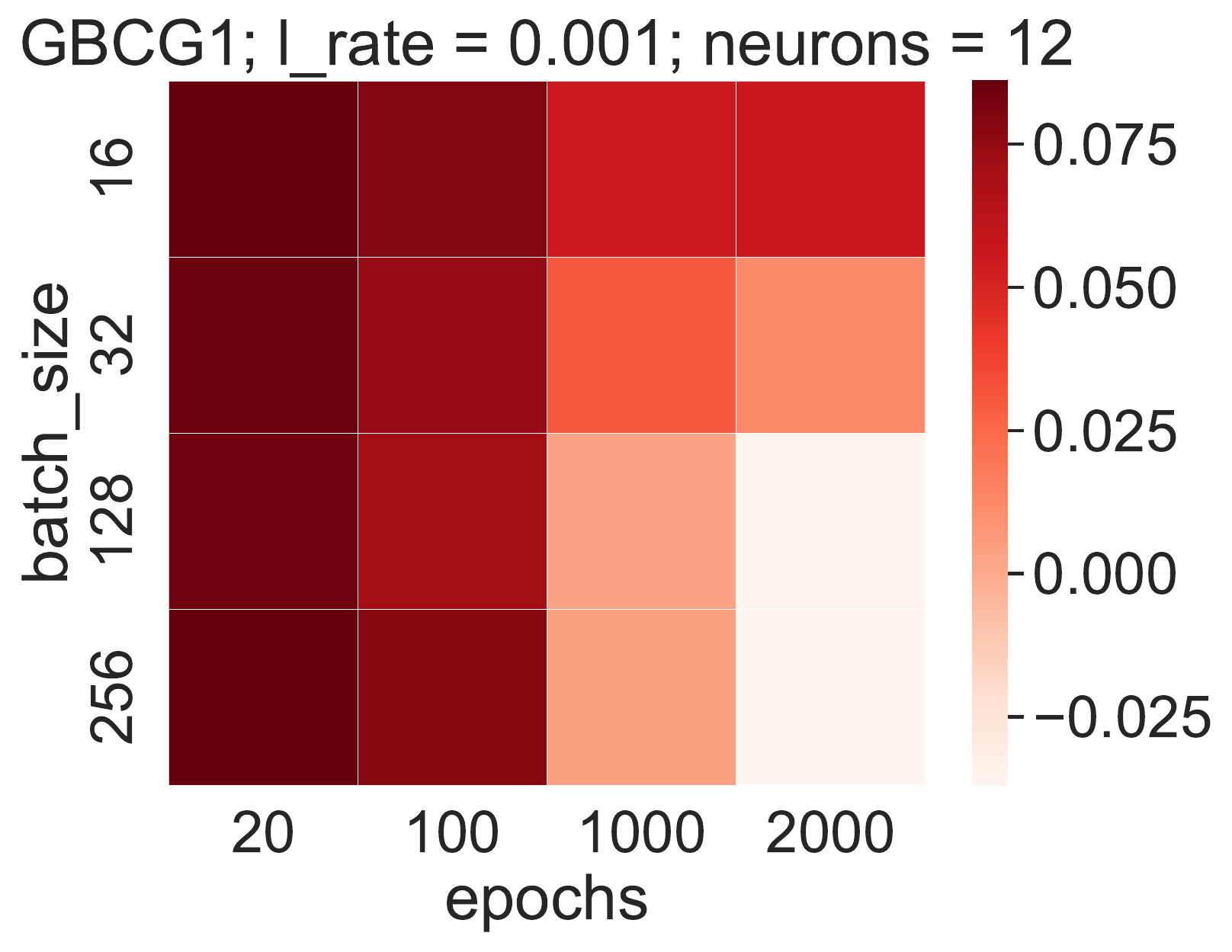}
  \includegraphics[width=0.31\textwidth]{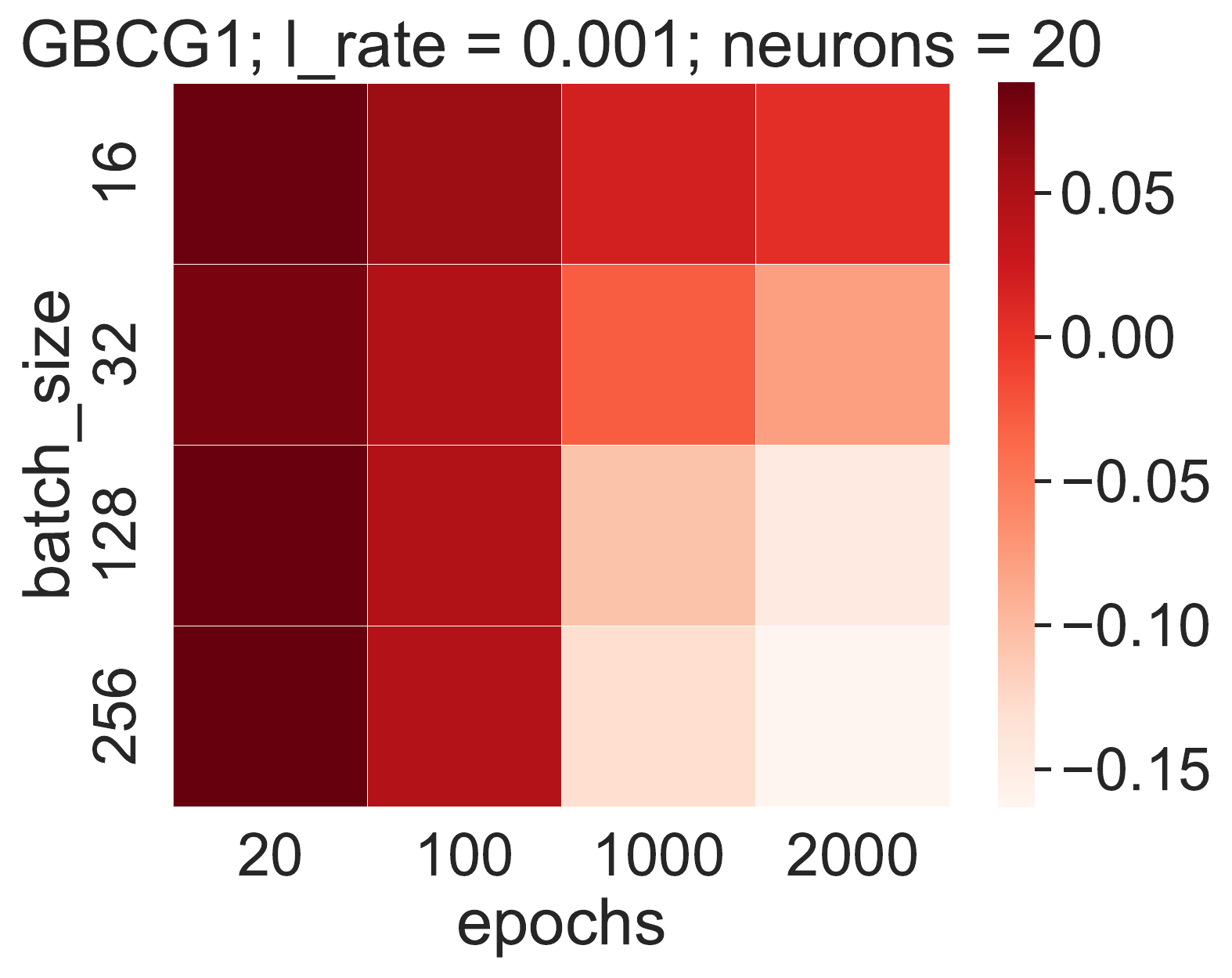}
  \includegraphics[width=0.31\textwidth]{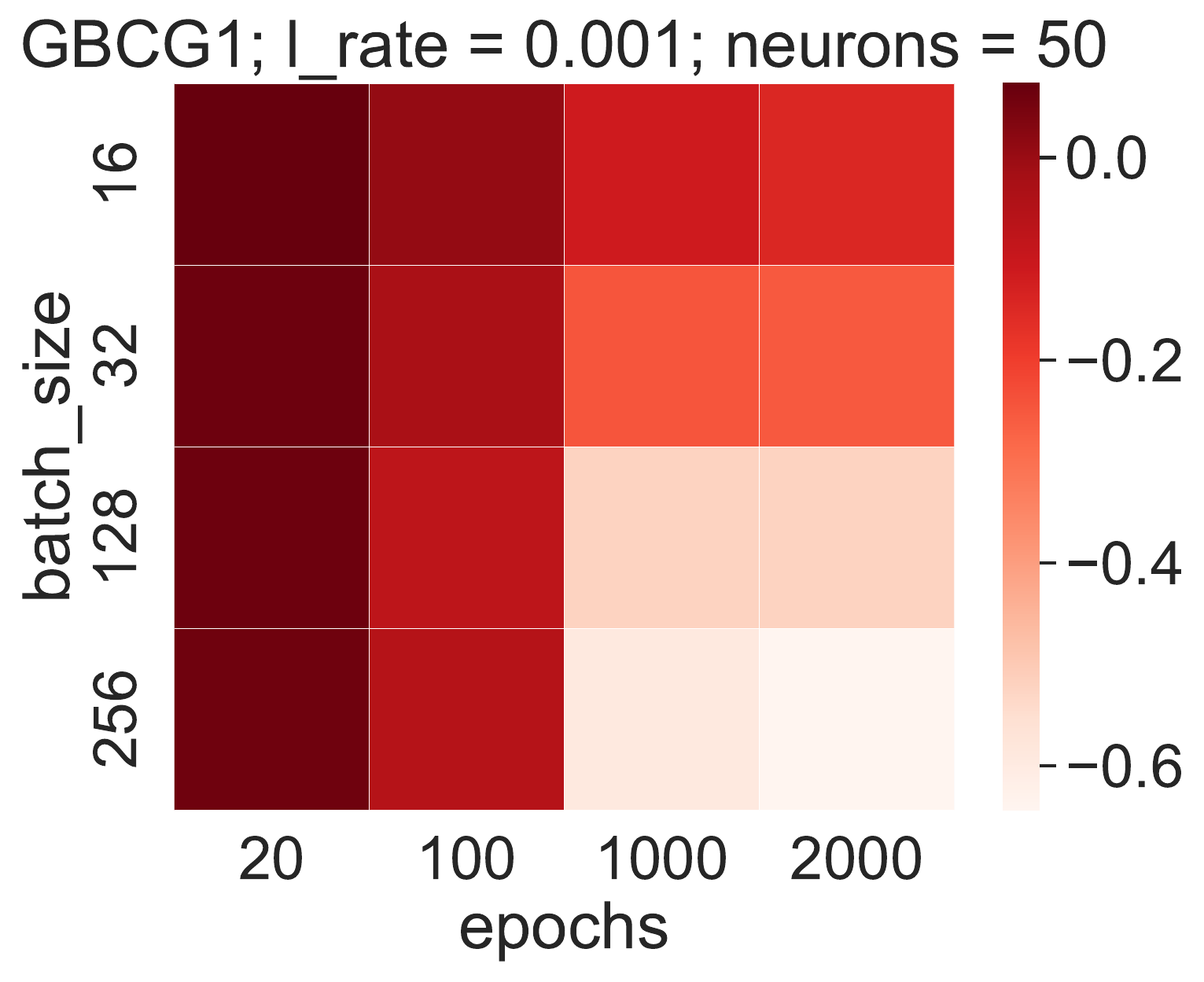}
  \caption{\textbf{Hyperparameter grid search for SOMO energy prediction: GBCG1 resolution.}
           Hyperparameters tested via grid search 
           are reported on the heatmap axes, that are colored according to the corresponding
           fivefold cross-validated $R^2$ obtained after training.
          }
  \label{fig:hyperparameter-heatmaps-GBCG1}
\end{figure}
\begin{figure}[H] 
  \centering
  \includegraphics[width=0.31\textwidth]{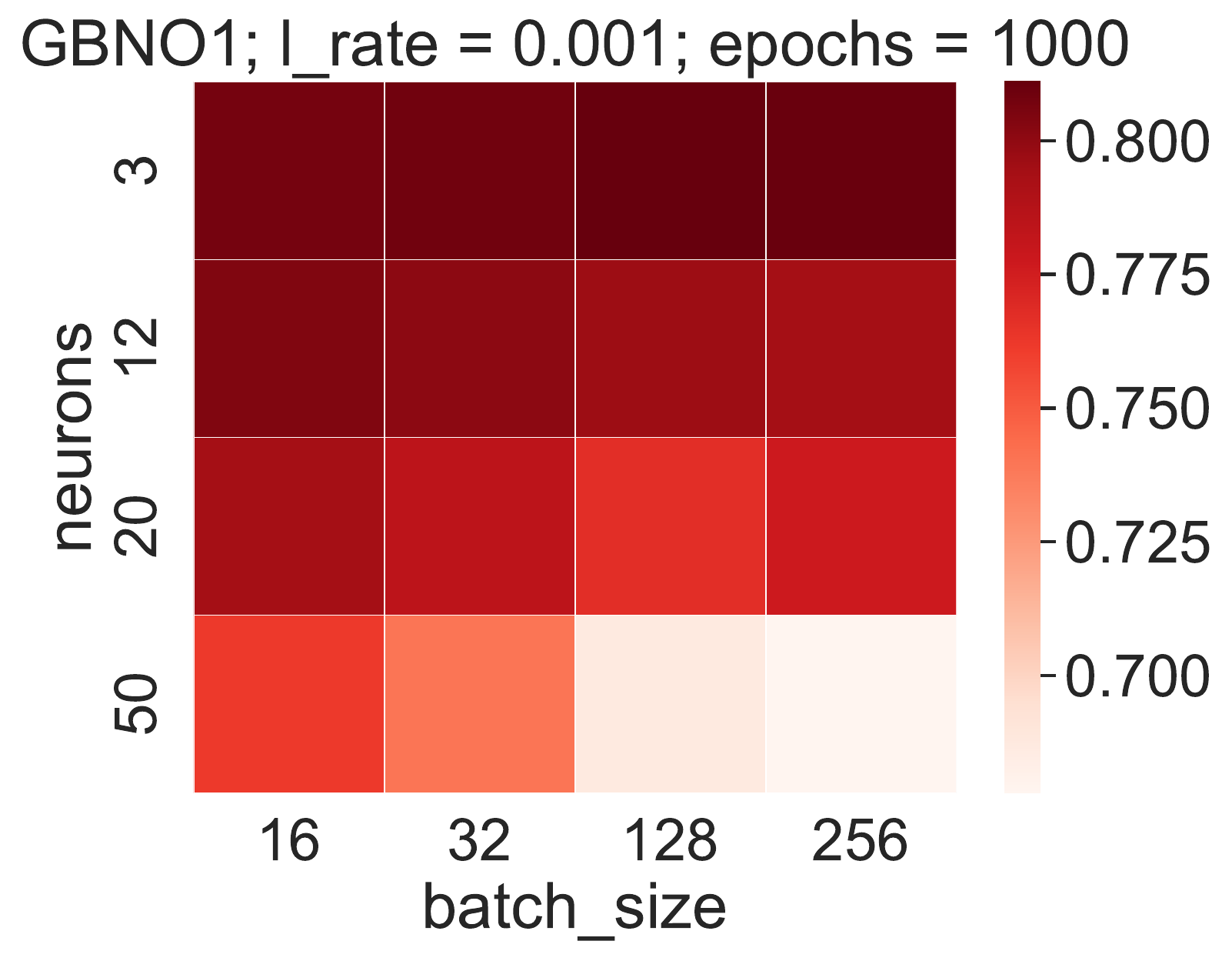}
  \includegraphics[width=0.31\textwidth]{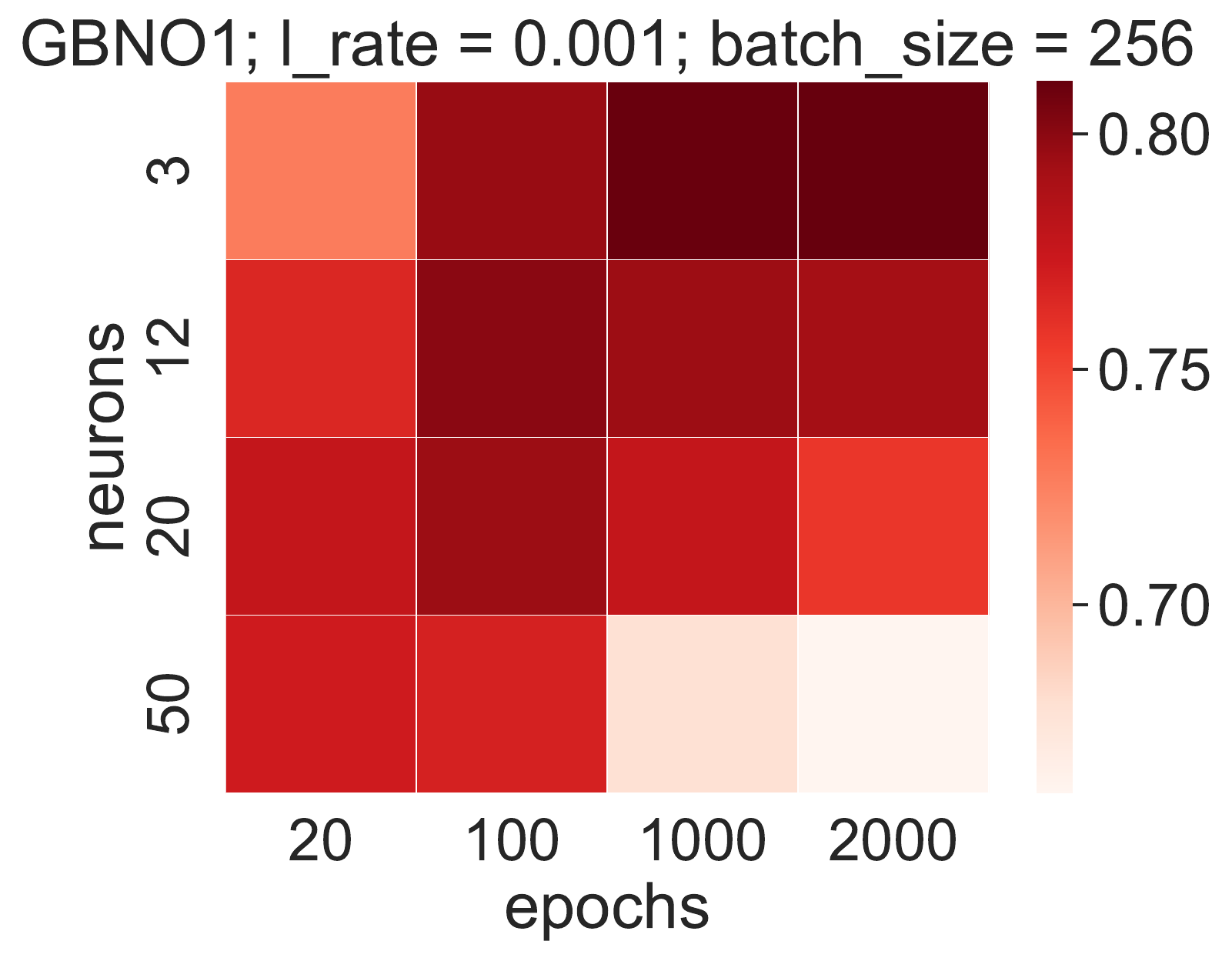}
  \includegraphics[width=0.31\textwidth]{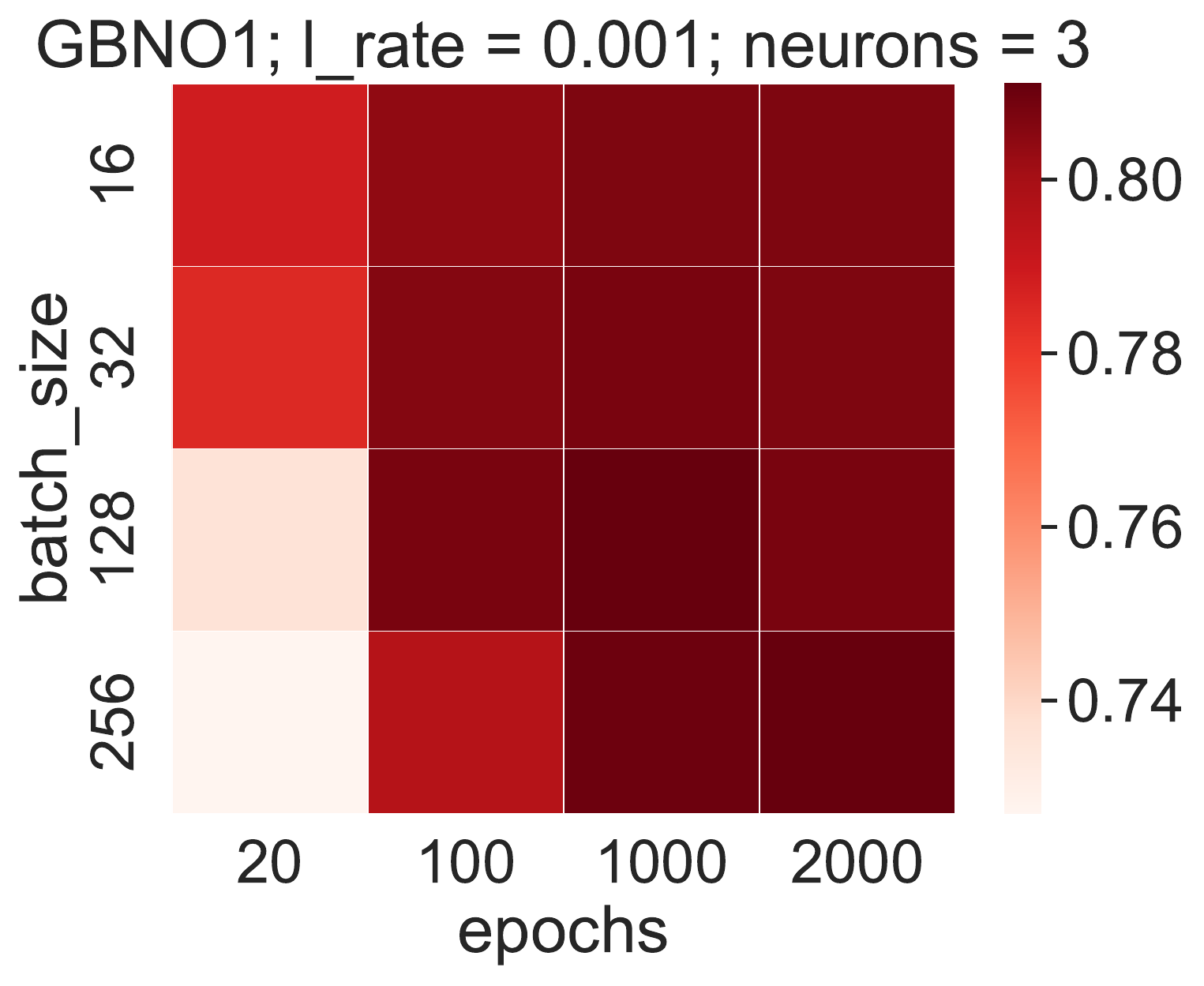}

  \includegraphics[width=0.31\textwidth]{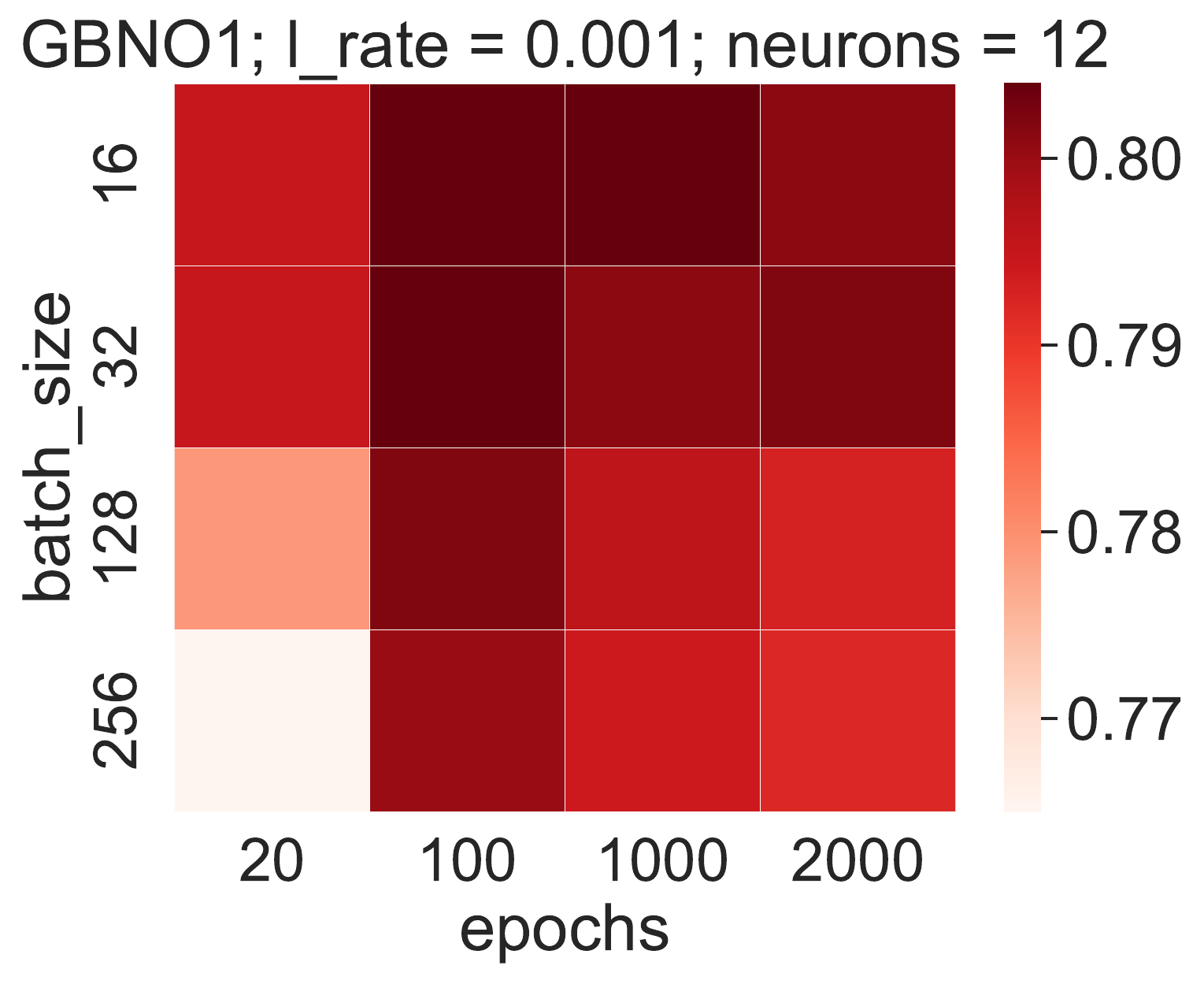}
  \includegraphics[width=0.31\textwidth]{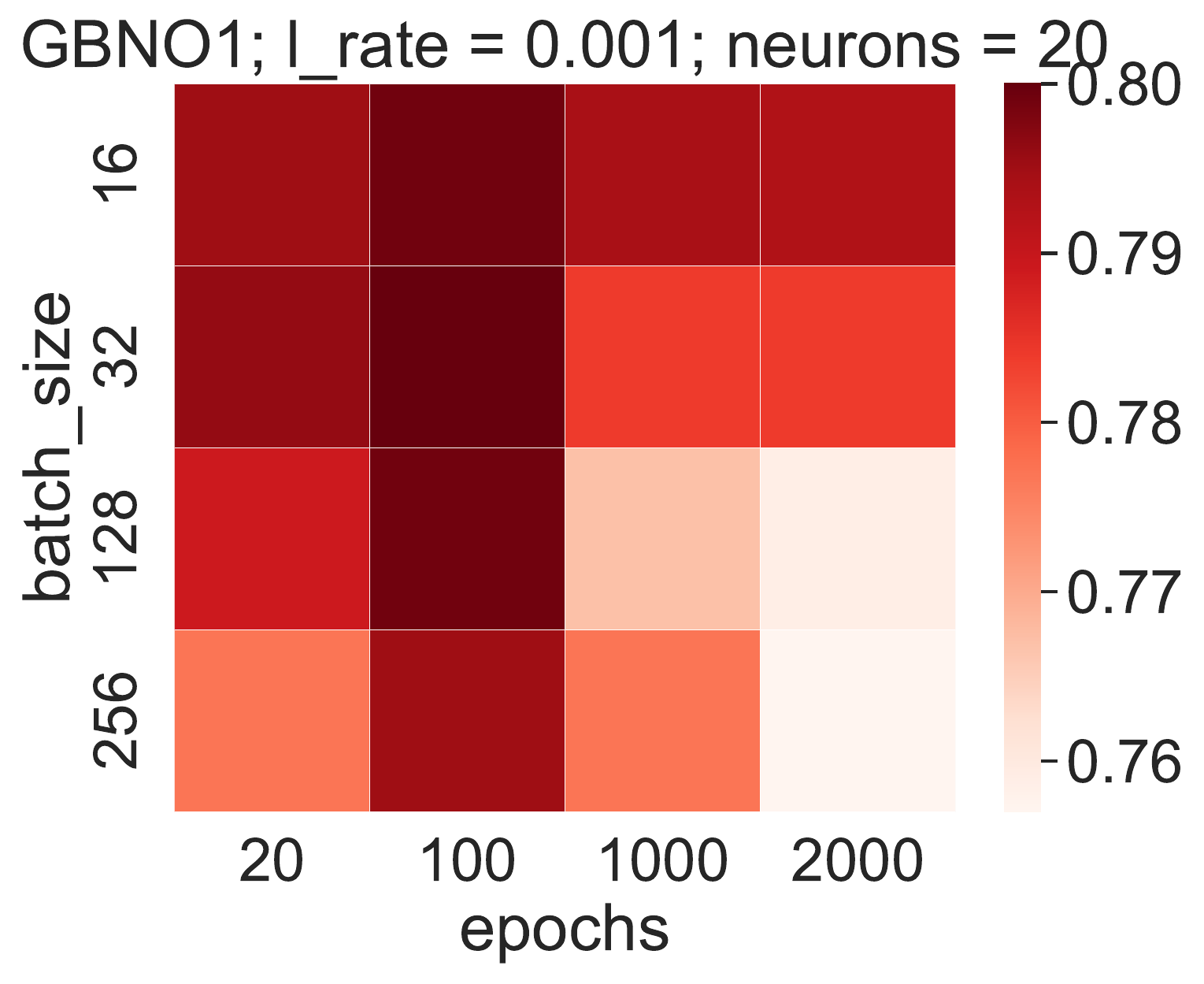}
  \includegraphics[width=0.31\textwidth]{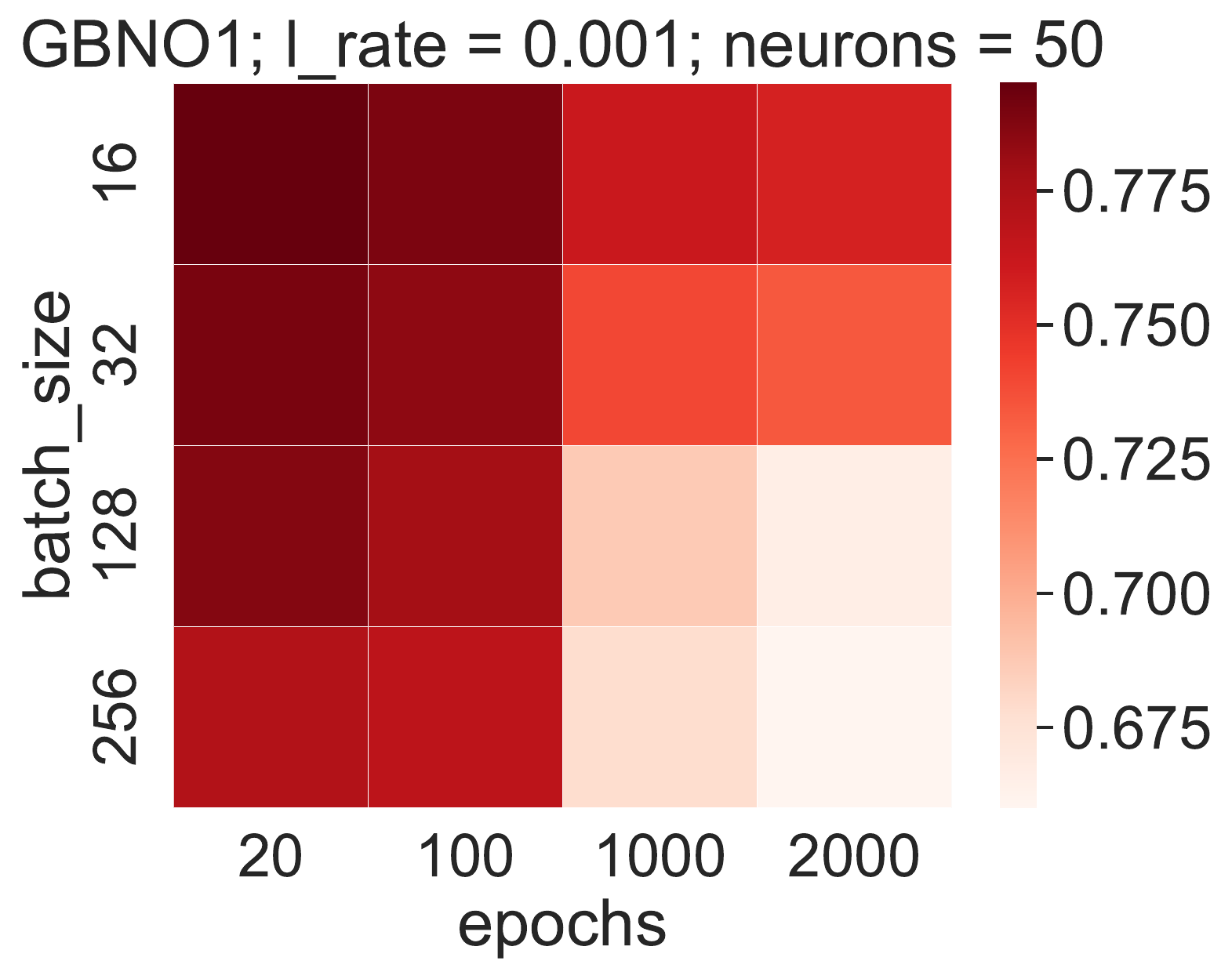}
  \caption{\textbf{Hyperparameter grid search for SOMO energy prediction: GBNO1 resolution.}
           Hyperparameters tested via grid search 
           are reported on the heatmap axes, that are colored according to the corresponding
           fivefold cross-validated $R^2$ obtained after training.
          }
  \label{fig:hyperparameter-heatmaps-GBNO1}
\end{figure}

\newpage
\section{All-atom model}
\label{SI:sec:AA}
\begin{table}[htbp]
  \caption{\textbf{Mass densities: AA model vs. experiments.}
           Mass densities for the fragments constituting PTMA, namely 
           methyl methacrylate (MMA) and TEMPO.
           Densities obtained via AA simulations are reported along with the reference
           experimental values. All values in g cm$^{-3}$. Also reported are the
           error percentages with respect to the experimental values.
          }
  \centering
  \label{tbl:densities}
  \begin{tabular}{lrr}
\hline
                & exp.                                & AA (err \%)                \\
\hline                                        
TEMPO (crystal) & 1.077~\cite{2007YYonekuta-ChemLett} & 1.071$\pm$0.002 ($-0.5$\%) \\
MMA             & 0.94~\cite{2014CRCHandbook}         & 0.963$\pm$0.003 ($+2.4$\%) \\
\hline
  \end{tabular}
\end{table}

\begin{figure}[H] 
  \centering
  \includegraphics[width=0.48\textwidth]{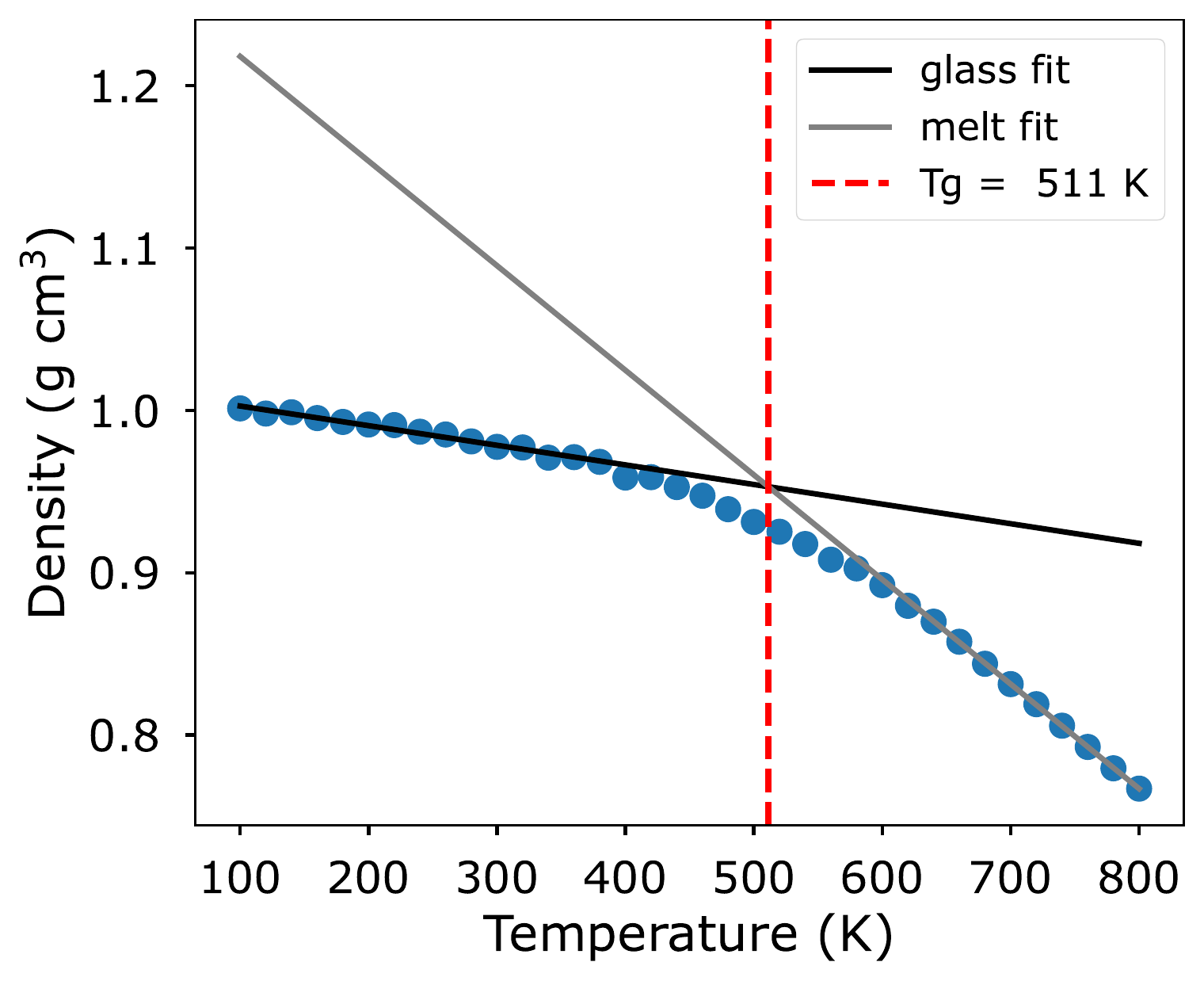}
  \includegraphics[width=0.48\textwidth]{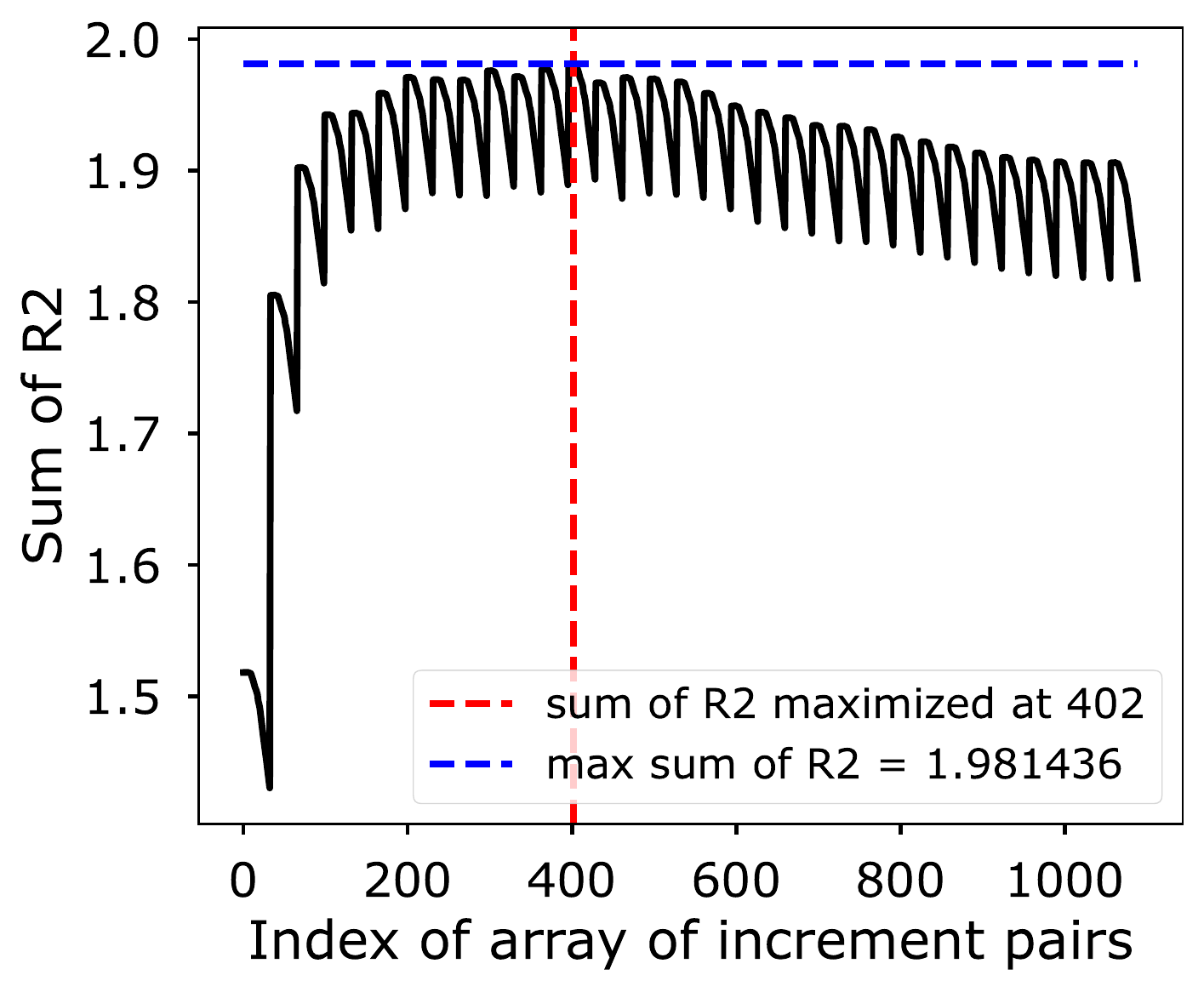}
  \caption{\textbf{Glass transition temperature, $T_g$, estimation.}
           We estimate $T_g$ of the AA model by cooling down a morphology from 800 K
           and equilibrating it at temperatures spaced every 20 K. 
           The simulations were equilibrated at the target temperature for 30 ns
           and the density was extracted as the average density over the subsequent
           20 ns of MD simulation. The data gathered in this way in the range 100-800 K
           are shown on the left-hand side plot.
           Two independent linear fits of the density-temperature data have been performed to
           individuate the glass and melt regions. To choose the best fits, the following
           procedure was employed: 
           starting from the 4th datapoint from the lowest temperature and from
           the highest temperature for the glass and melt fits, respectively, we scanned
           through all combinations of fits by progressively increasing the number of datapoints 
           to be included in the fits; the best fits were selected by choosing the
           ones that maximize the
           sum of the $R^2$ coefficients of the linear fits (right-hand side plot).
           $T_g$ was estimated to be at the point where the two fits cross, i.e., 511 K.
          }
  \label{fig:Tg-AA-model}
\end{figure}

\newpage
\section{Coarse-grained models}
\label{SI:sec:CG}
\begin{figure}[htbp]
  \centering
  \includegraphics[width=1.00\textwidth]{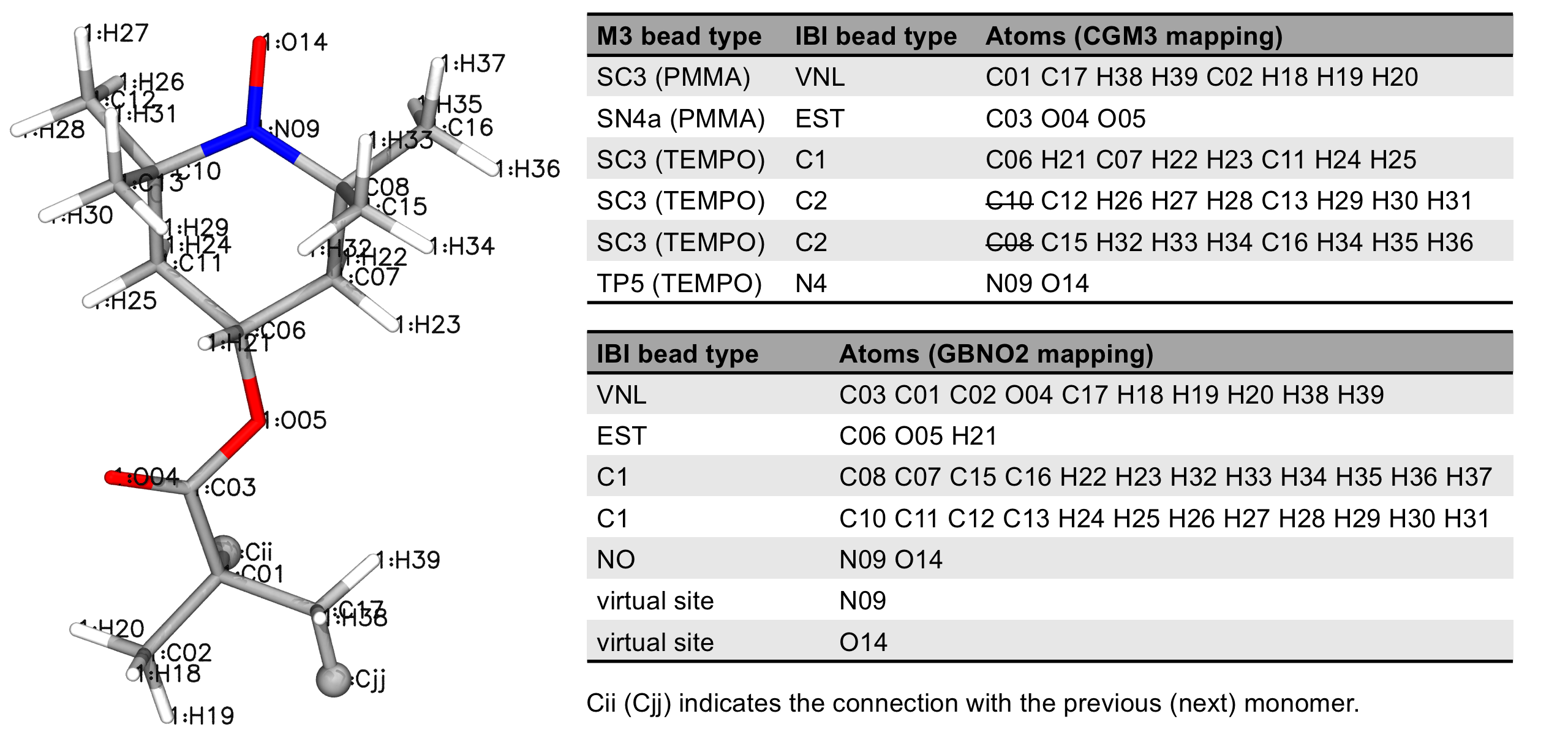}
  \caption{\textbf{Atom labels and mappings for PTMA: CGM3 and GBNO2.}
           (left) Atom labels. (right)
           The atom-to-bead correspondence is given for CGM3 and GBNO2 mappings, 
           along with the Martini 3 (M3) bead types~\cite{2021PCTSouza-NatMethods} 
           and the names for the IBI bead types.
          }
  \label{fig:summary-CG-IBI-PTMA}
\end{figure}

\begin{figure}[htbp]
  \centering
  \includegraphics[width=\textwidth]{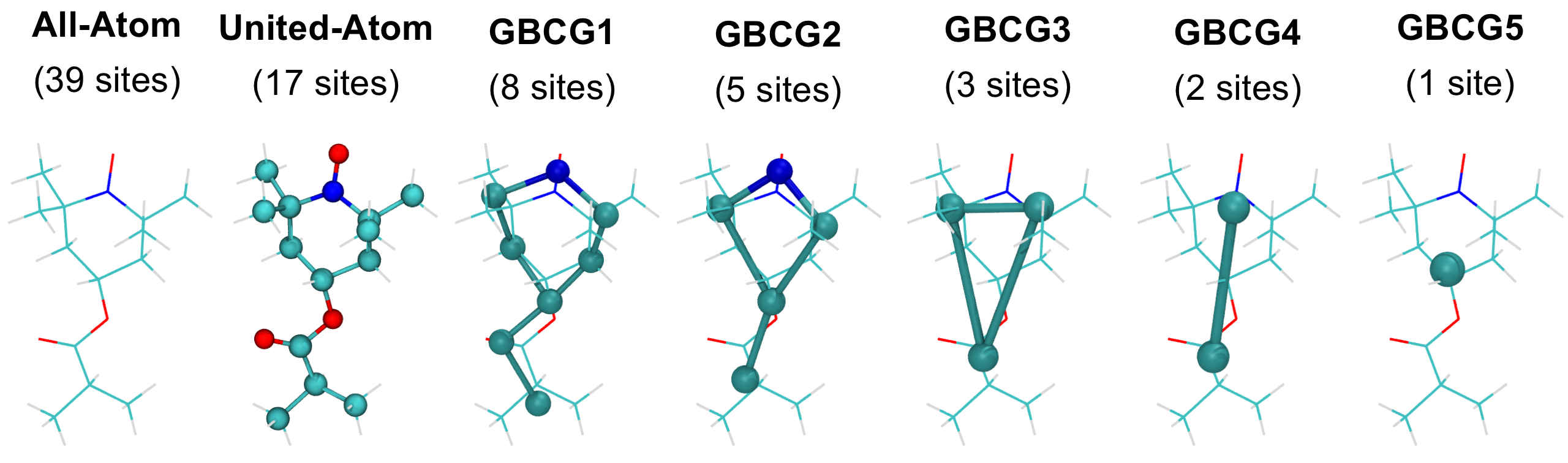}
  \caption{\textbf{Graph-Based CG mappings for a monomer of PTMA.} 
           From (left) all-atom (AA) resolution to (right) a 1-site CG model.
           Figure~\ref{fig:GBCG-mappings-for-PTMA-labels} details the atom-to-bead mapping.
           This Figure is a larger version of Figure~\ref{M-fig:results-mappings-and-NO}B
           of the main text.
           }
  \label{fig:GBCG-mappings-for-PTMA}
\end{figure}

\textbf{Mappings.} The CGM3 and GBNO2 mappings, for which CG model parameters
have been derived, are given in Figure~\ref{fig:summary-CG-IBI-PTMA}. 
GBCG mappings are given in Figure~\ref{fig:GBCG-mappings-for-PTMA} (schematic representation)
and Figure~\ref{fig:GBCG-mappings-for-PTMA-labels} (atom-to-bead correspondence).
GBCG mappings that include an explicit nitroxide group description are given in
Figure~\ref{fig:GBCG-mappings-for-PTMA-labels-NO} (atom-to-bead correspondence).

\begin{figure}
  \caption{\textbf{Graph-Based CG mappings.} Beads are separated by commas.
           Atom labels correspond to
           the ones of Figure~\ref{fig:summary-CG-IBI-PTMA}.}
  \label{fig:GBCG-mappings-for-PTMA-labels}
\begin{lstlisting}
"GBCG1": [C01 C02 C17 H18 H19 H20 H38 H39, C03 O04, C06 O05 H21, C07 H22 H23, C08 C15 C16 H32 H33 H34 H35 H36 H37, N09 O14, C10 C12 C13 H26 H27 H28 H29 H30 H31, C11 H24 H25] 
"GBCG2": [C03 C01 C02 O04 C17 H18 H19 H20 H38 H39, C06 O05 H21, C08 C07 C15 C16 H22 H23 H32 H33 H34 H35 H36 H37, N09 O14, C10 C11 C12 C13 H24 H25 H26 H27 H28 H29 H30 H31] 
"GBCG3": [C06 C01 C02 C03 O04 O05 C17 H18 H19 H20 H21 H38 H39, C08 C07 N09 O14 C15 C16 H22 H23 H32 H33 H34 H35 H36 H37, C10 C11 C12 C13 H24 H25 H26 H27 H28 H29 H30 H31] 
"GBCG4": [C06 C01 C02 C03 O04 O05 C17 H18 H19 H20 H21 H38 H39, C08 C07 N09 C10 C11 C12 C13 O14 C15 C16 H22 H23 H24 H25 H26 H27 H28 H29 H30 H31 H32 H33 H34 H35 H36 H37] 
"GBCG5": [C08 C01 C02 C03 O04 O05 C06 C07 N09 C10 C11 C12 C13 O14 C15 C16 C17 H18 H19 H20 H21 H22 H23 H24 H25 H26 H27 H28 H29 H30 H31 H32 H33 H34 H35 H36 H37 H38 H39]
\end{lstlisting}

\end{figure}
\begin{figure}
  \caption{\textbf{Graph-Based CG mappings with explicit nitroxide group description.}
           Beads are separated by commas.
           Atom labels correspond to
           the ones of Figure~\ref{fig:summary-CG-IBI-PTMA}.}
  \label{fig:GBCG-mappings-for-PTMA-labels-NO}
\begin{lstlisting}
"GBNO1": [C01 C02 C17  H18 H19 H20 H38 H39 , C03 O04, C06 O05 H21, C07 H22 H23, C08 C15 C16 H32 H33 H34 H35 H36 H37, N09, O14, C10 C12 C13 H26 H27 H28 H29 H30 H31, C11 H24 H25]
"GBNO2": [C03 C01 C02 O04 C17  H18 H19 H20 H38 H39 , C06 O05 H21, C08 C07 C15 C16 H22 H23 H32 H33 H34 H35 H36 H37, N09, O14, C10 C11 C12 C13 H24 H25 H26 H27 H28 H29 H30 H31]
"GBNO3": [C06 C01 C02 C03 O04 O05 C17  H18 H19 H20 H21 H38 H39 , N09, O14, C08 C07 C15 C16 H22 H23 H32 H33 H34 H35 H36 H37, C10 C11 C12 C13 H24 H25 H26 H27 H28 H29 H30 H31]
"GBNO4": [C06 C01 C02 C03 O04 O05 C17  H18 H19 H20 H21 H38 H39 , N09, O14, C08 C07 C10 C11 C12 C13 C15 C16 H22 H23 H24 H25 H26 H27 H28 H29 H30 H31 H32 H33 H34 H35 H36 H37] 
"GBNO5": [N09, O14, C08 C01 C02 C03 O04 O05 C06 C07 C10 C11 C12 C13 C15 C16 C17 H18 H19 H20 H21 H22 H23 H24 H25 H26 H27 H28 H29 H30 H31 H32 H33 H34 H35 H36 H37 H38 H39]
\end{lstlisting}

\end{figure}

\begin{figure}[htbp]
  \centering
  \includegraphics[width=\textwidth]{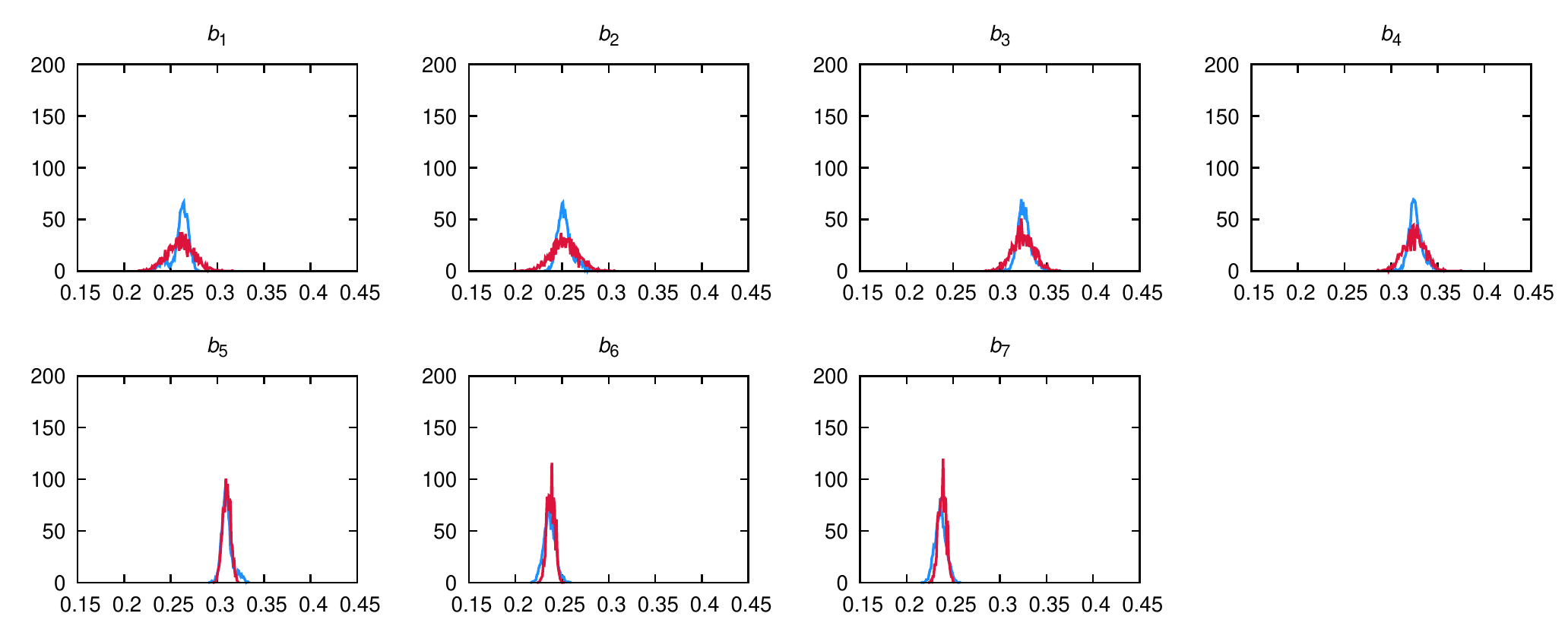}
  \includegraphics[width=\textwidth]{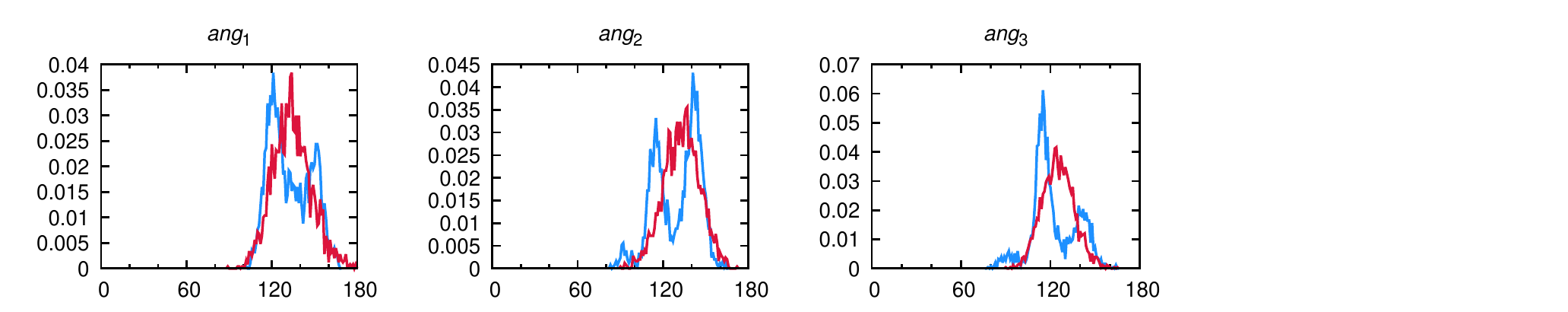}
  \includegraphics[width=\textwidth]{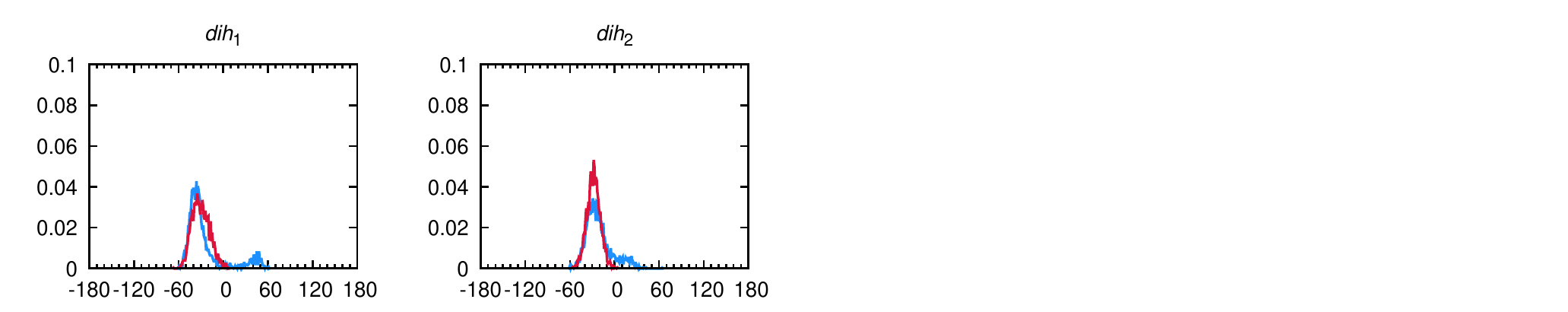}
  \caption{\textbf{Bond, angle, and torsion distributions for a PTMA monomer (CGM3 mapping).}
           Atomistic reference distributions are in blue, CG distributions are in red.
           Compare header to labels in Figure~\ref{fig:CG-PTMA-topology-CGM3-IBI} 
           for information on the particular degree of freedom shown. 
          }
  \label{fig:TMA-distributions-CGM3}
\end{figure}

\textbf{Bonded parameters.} 
Bonded parameters were obtained by extracting
bond, angle, and torsion distributions from mapped atomistic simulations and fitting 
them to standard harmonic bond, angle, and dihedral (i.e., improper dihedral) potentials.
CG sites are mapped to the center of geometry (COG) of the group of atoms they represent
(see \textbf{Mappings.}).
The resulting CG parameters are given in Figure~\ref{fig:CG-PTMA-topology-CGM3-IBI}
and the distributions are shown in Figure~\ref{fig:TMA-distributions-CGM3}.
Similarly, CG parameters and distributions are shown in Figure~\ref{fig:CG-PTMA-topology-GBNO2}
and Figure~\ref{fig:TMA-distributions-GBNO2}, respectively, for the GBNO2 mapping.
The two virtual dummy sites used to describe the nitrogen and oxygen atoms
of the nitroxide group are ``virtual'' sites because their displacement 
is not calculated by the integrator algorithm,
but their position is instead recomputed 
from the new positions of the constructing CG particles
(i.e., a number of CG sites that completely define the virtual sites' position) 
after each integration step.
They are ``dummy'' sites because they do not interact via nonbonded interactions 
with any of the other CG sites. See Figure~\ref{fig:CG-PTMA-topology-GBNO2}
for their definition.

The bonded parameters for the PMMA-like backbone were taken from the 
Martini 3 PMMA model~\cite{2022FGrunewald-NatCommun} (and are shown in 
Figure~\ref{fig:CG-PMMA-topology-CGM3}) in the case of the CGM3 mapping.
For the GBNO2 mapping, new parameters were obtained and are shown in
Figure~\ref{fig:CG-PMMA-topology-GBNO2} (and corresponding distributions in
Figure~\ref{fig:PMMA-distributions-GBNO2}).

\textbf{Nonbonded parameters: Iterative Boltzmann inversion.} 
For the IBI procedure, reference atomistic RDFs are needed. 
Bonded parameters are taken from the previous step, and only nonbonded pair interactions
are optimized using IBI as implemented in VOTCA~\cite{2009VRuhle-JChemTheoryComput}. 
The same nonbonded exclusions used in the calculations of the reference atomistic RDFs
should be used in the IBI iteration steps and hence in the final CG model.
Typically, one excludes intramolecular pairs of CG beads which interact via bonded interactions.
In the case of PTMA, we exclude all the \textit{intramonomer} bead interactions and 
the 1-2 and 1-3 interactions between all the other beads. 
In GROMACS, the former is achieved by explicitly excluding the intramonomer pairs via a
\texttt{[exclusions]} section, and the latter by setting \texttt{nrexcl = 2}, 
as shown in the topology file in Figure~\ref{fig:CG-PTMA-topology-CGM3-IBI}
and Figure~\ref{fig:CG-PTMA-topology-GBNO2}
(the files are Polyply \textit{block} files in \texttt{.ff} format; they use
GROMACS-like syntax to define the repeating unit of a polymer needed by 
Polyply~\cite{2022FGrunewald-NatCommun}).

\begin{figure}[htbp]
  \centering

  \includegraphics[width=0.28\textwidth]{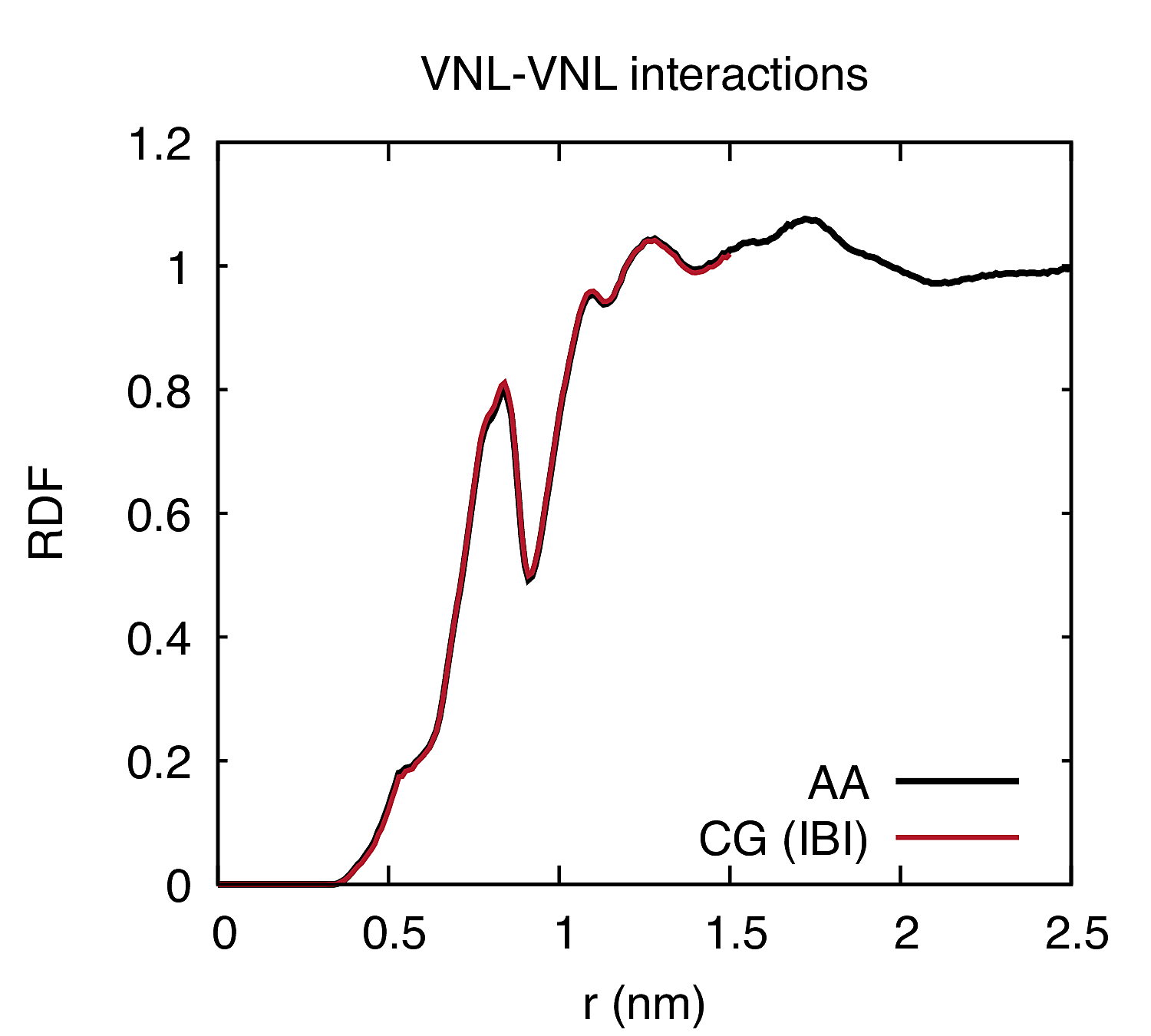}
  \includegraphics[width=0.28\textwidth]{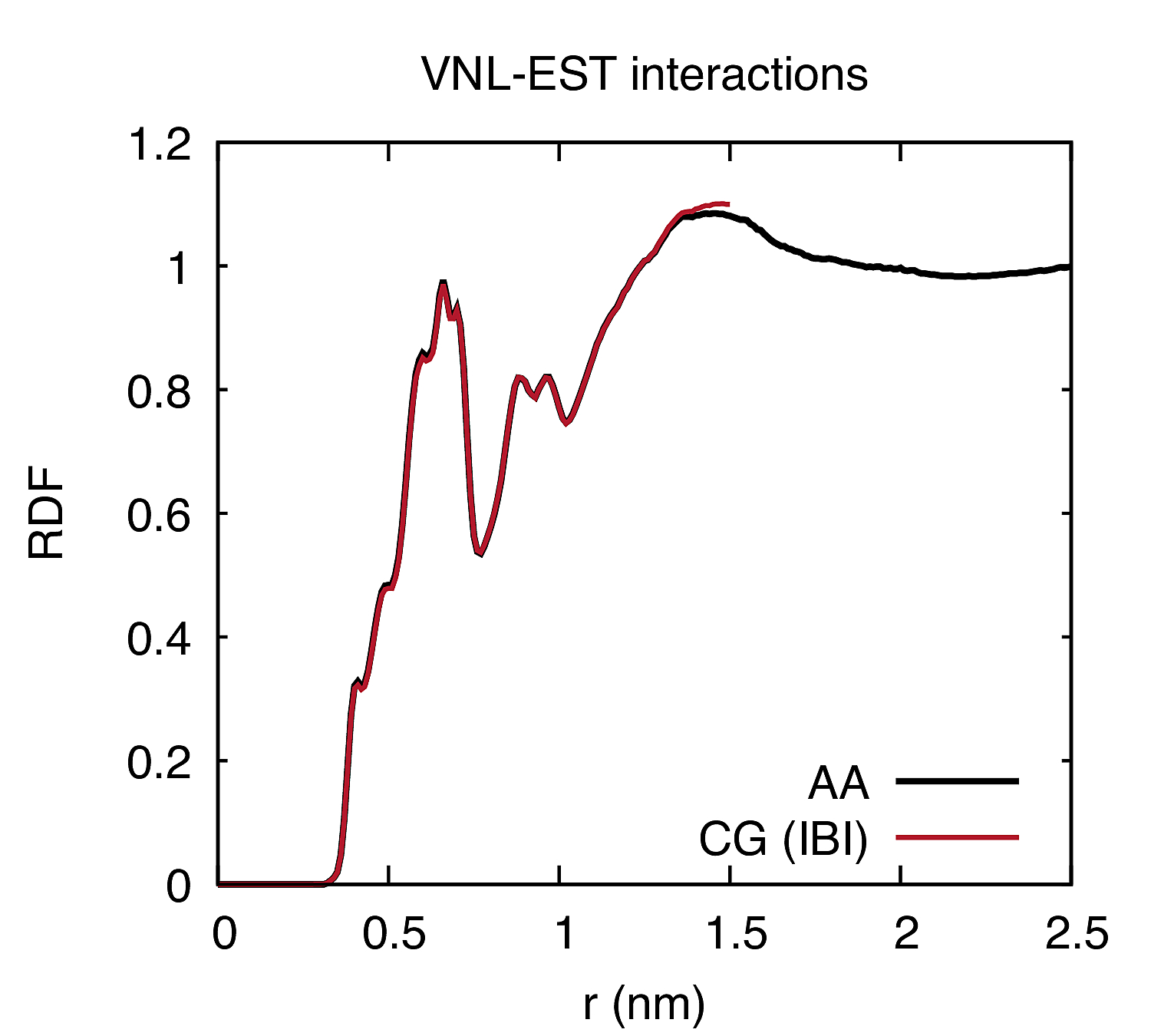}
  \includegraphics[width=0.28\textwidth]{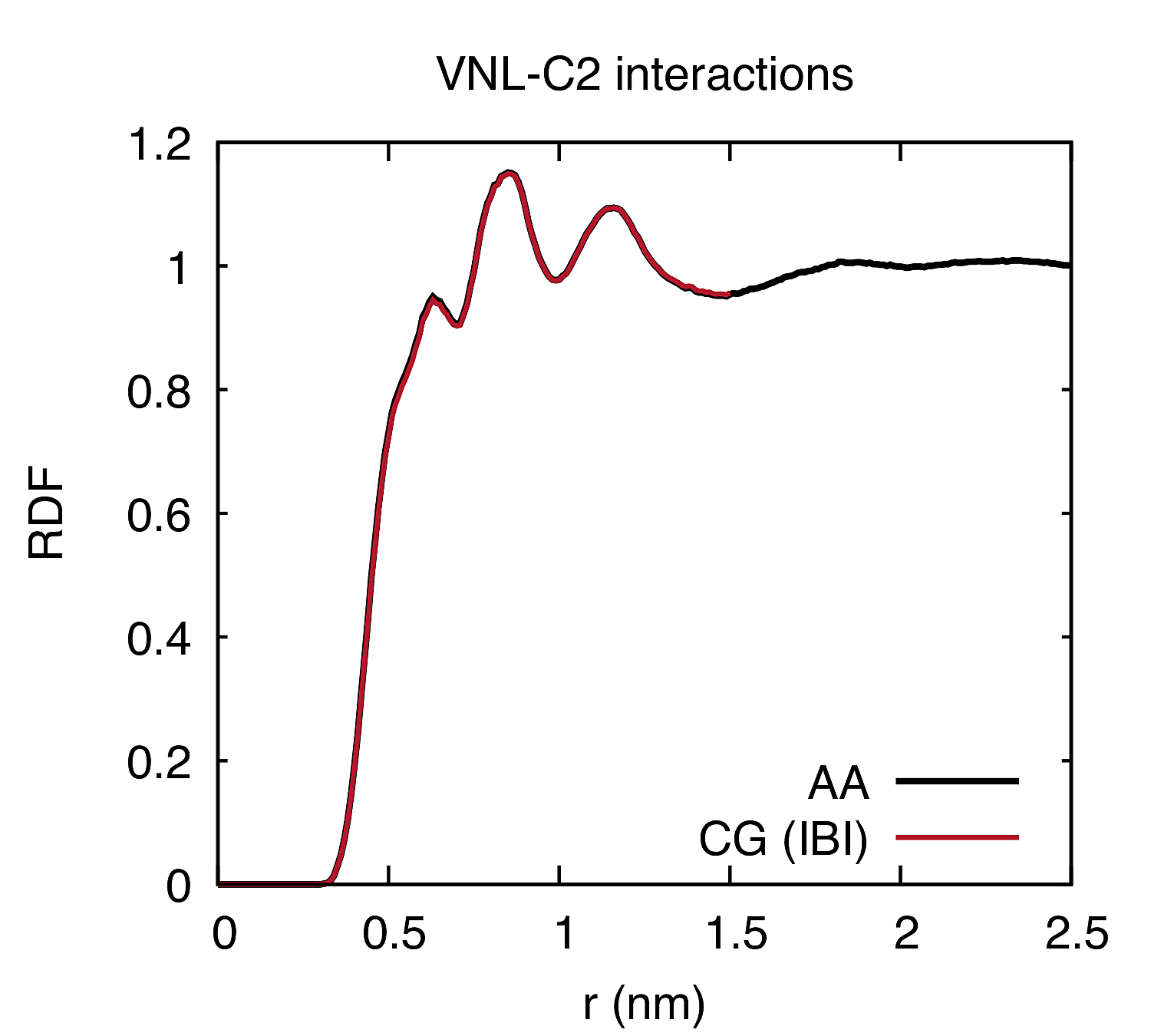}

  \includegraphics[width=0.28\textwidth]{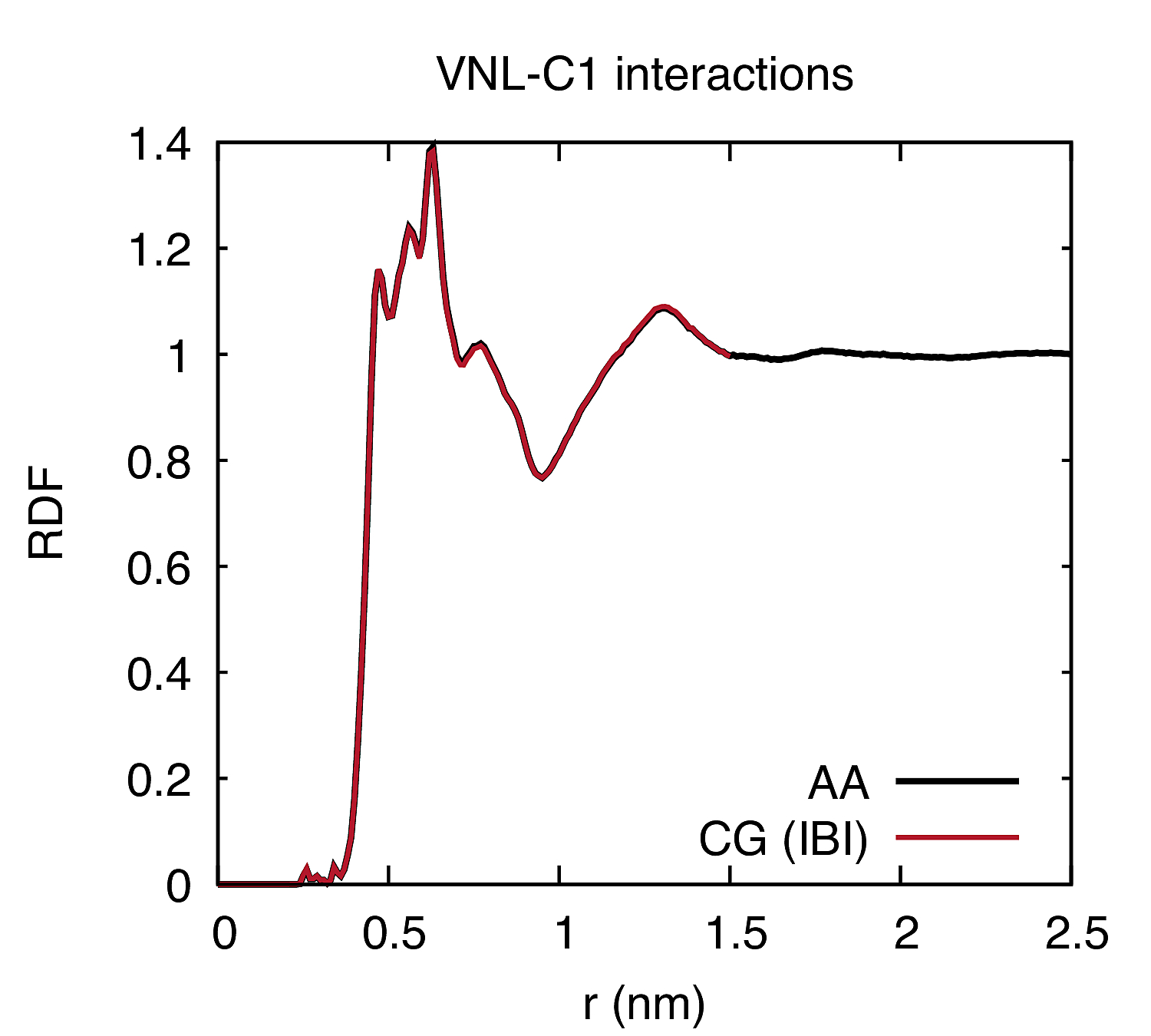}
  \includegraphics[width=0.28\textwidth]{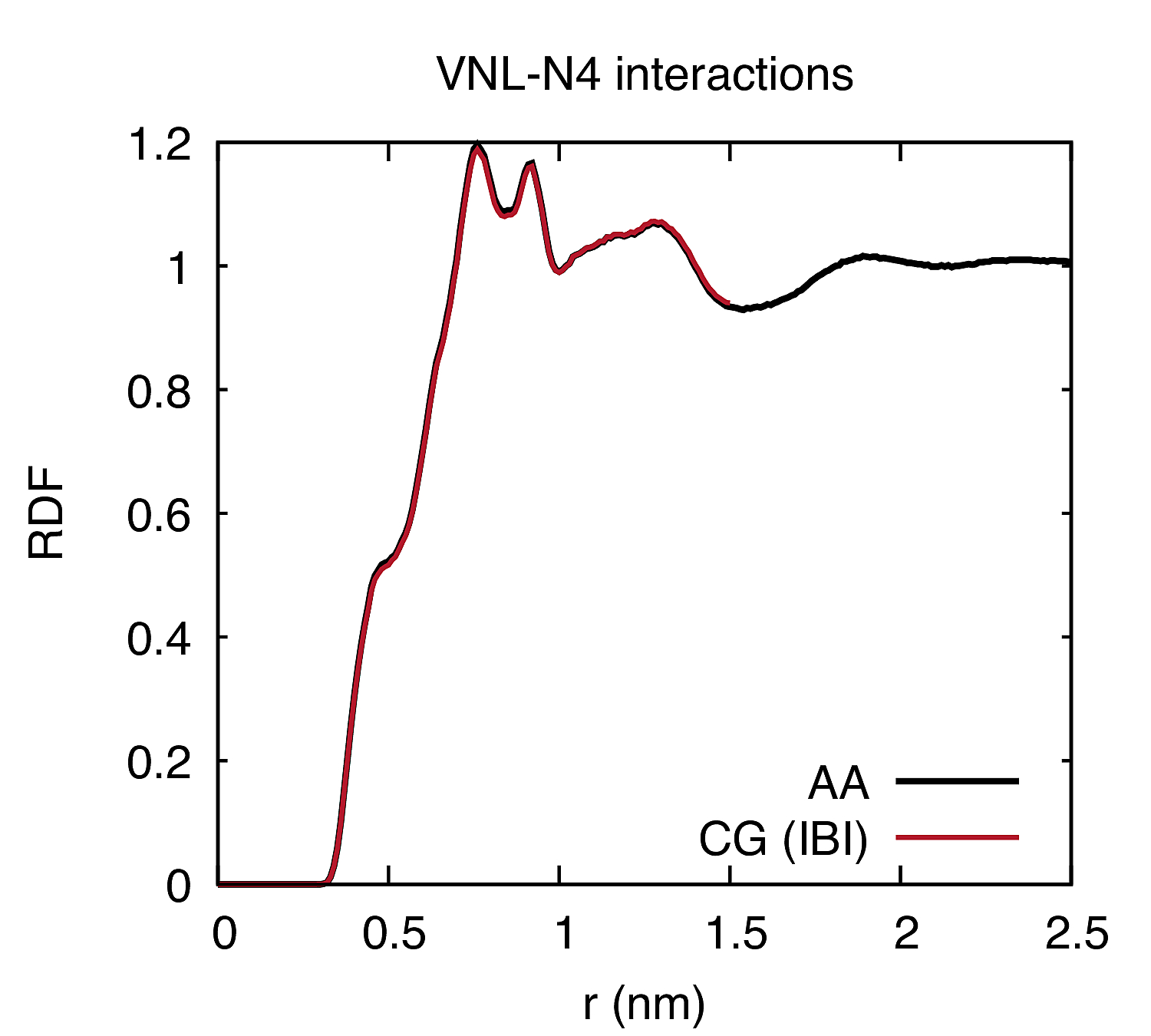}
  \includegraphics[width=0.28\textwidth]{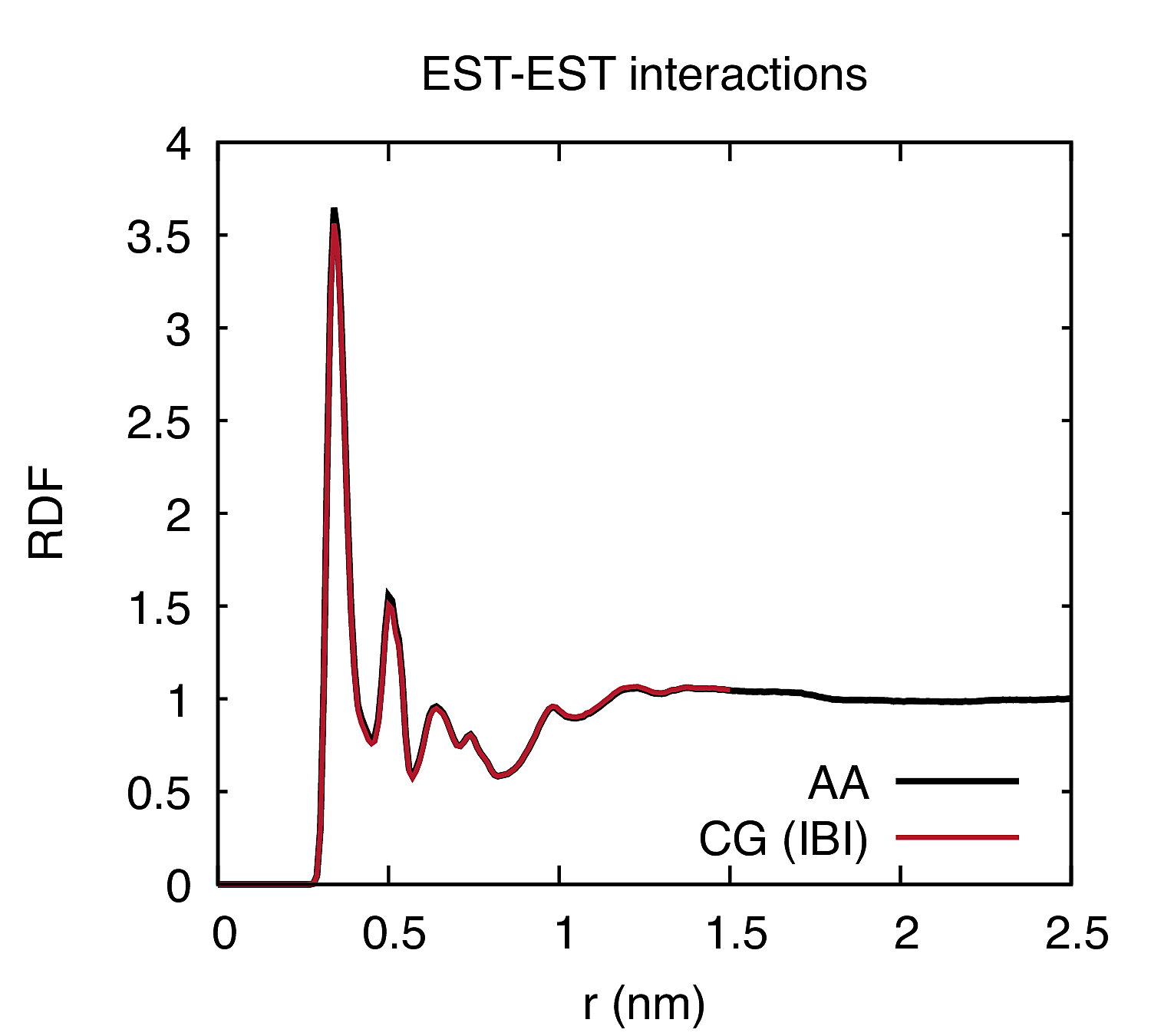}

  \includegraphics[width=0.28\textwidth]{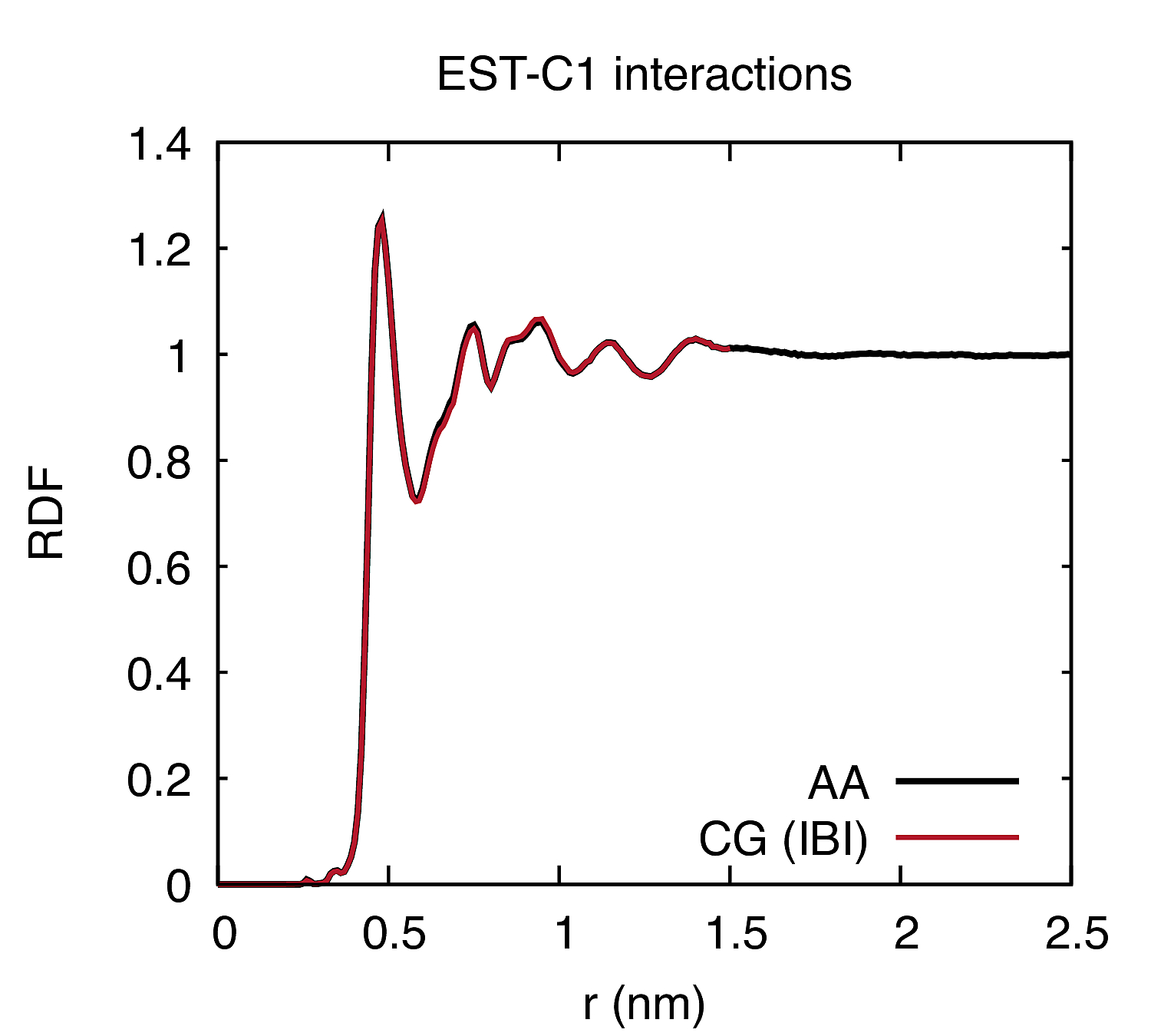}
  \includegraphics[width=0.28\textwidth]{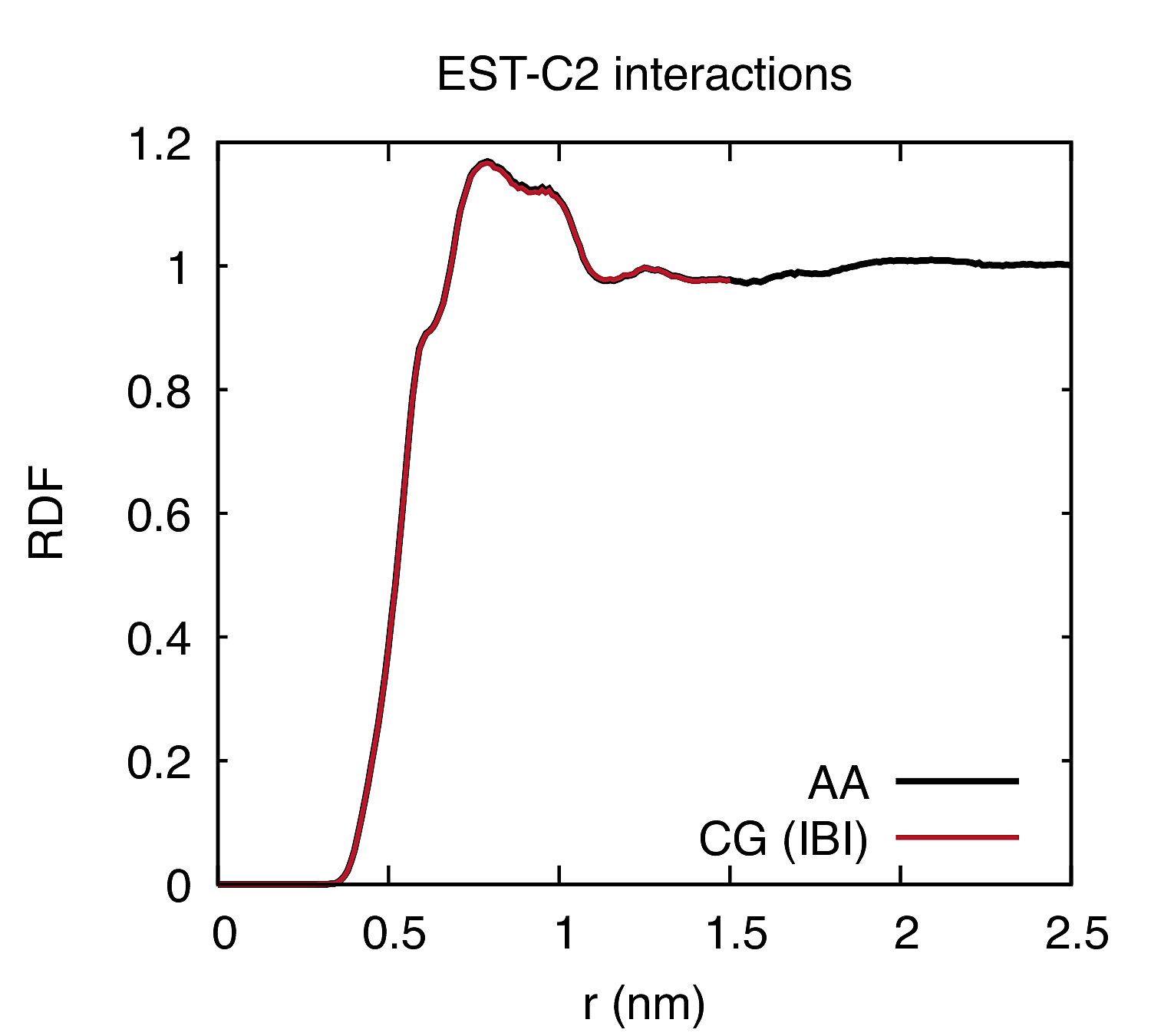}
  \includegraphics[width=0.28\textwidth]{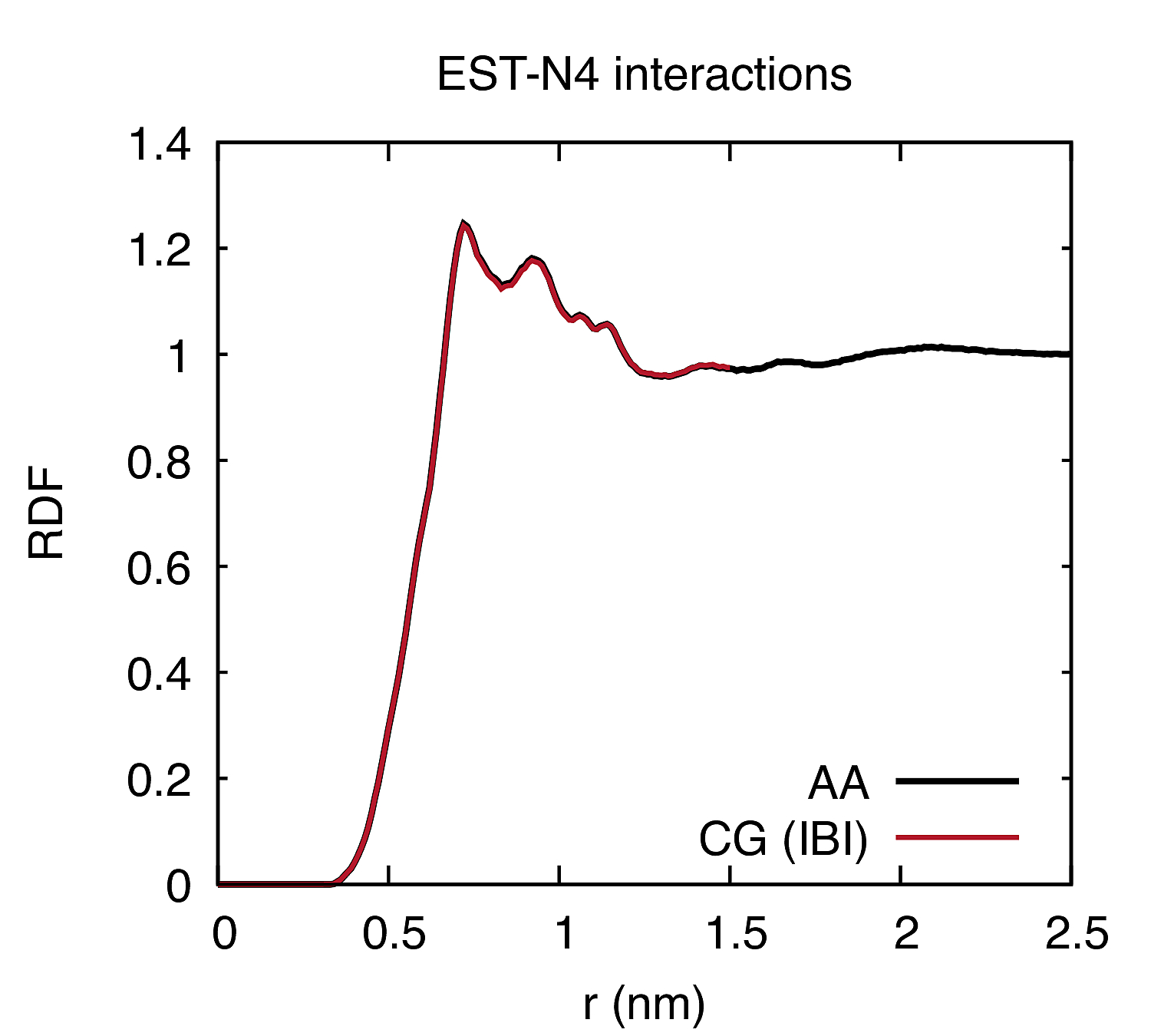}

  \includegraphics[width=0.28\textwidth]{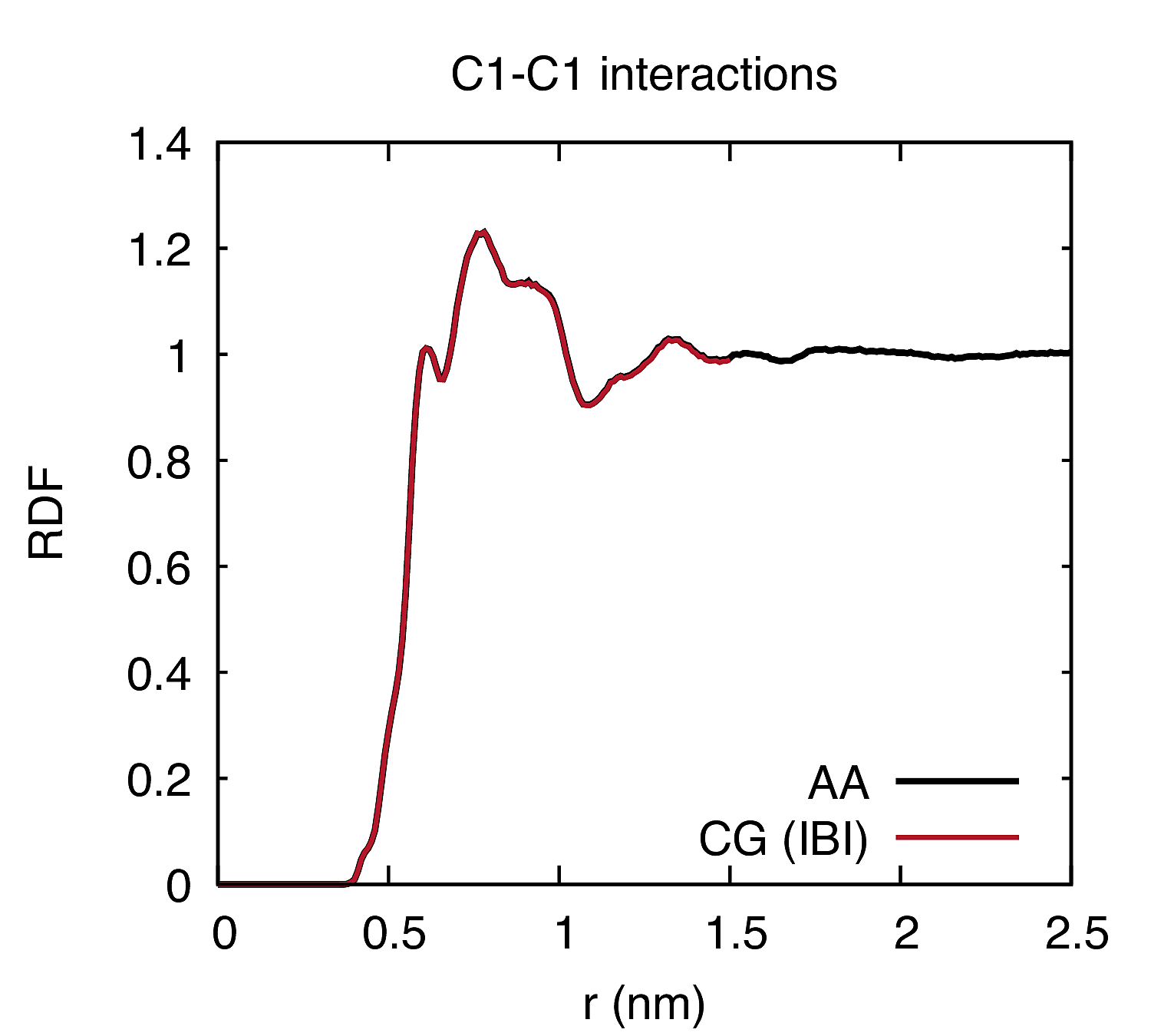}
  \includegraphics[width=0.28\textwidth]{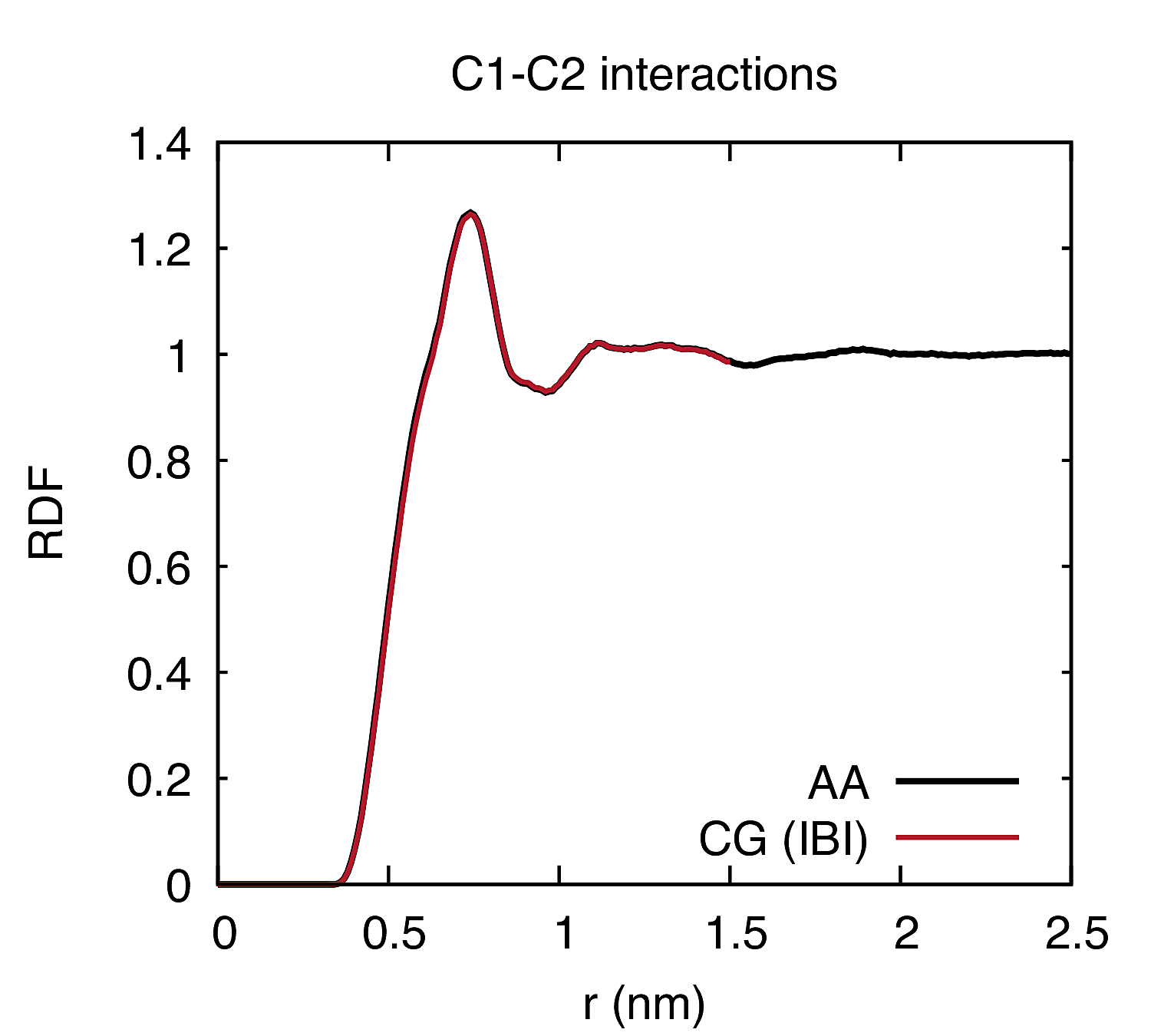}
  \includegraphics[width=0.28\textwidth]{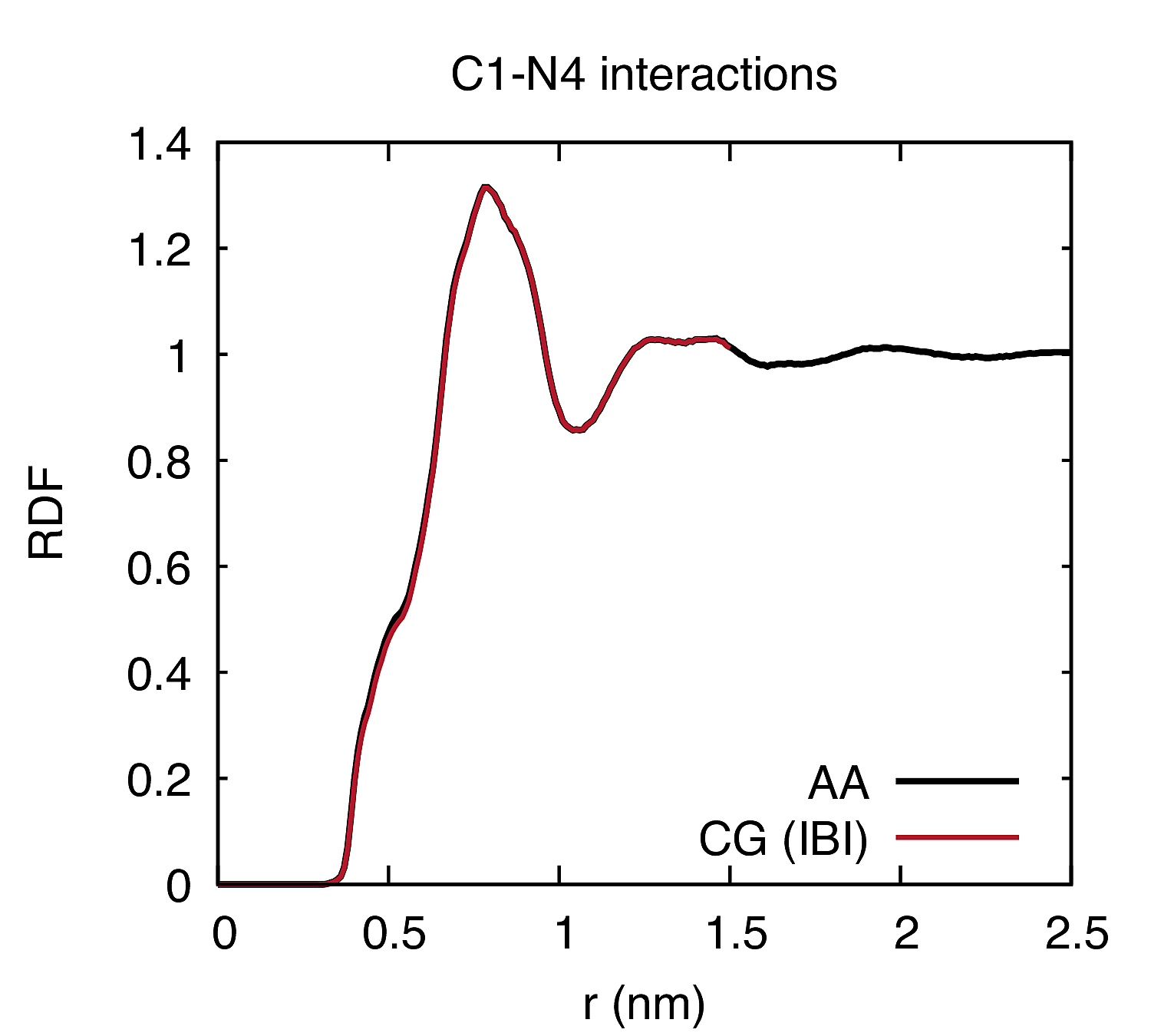}

  \includegraphics[width=0.28\textwidth]{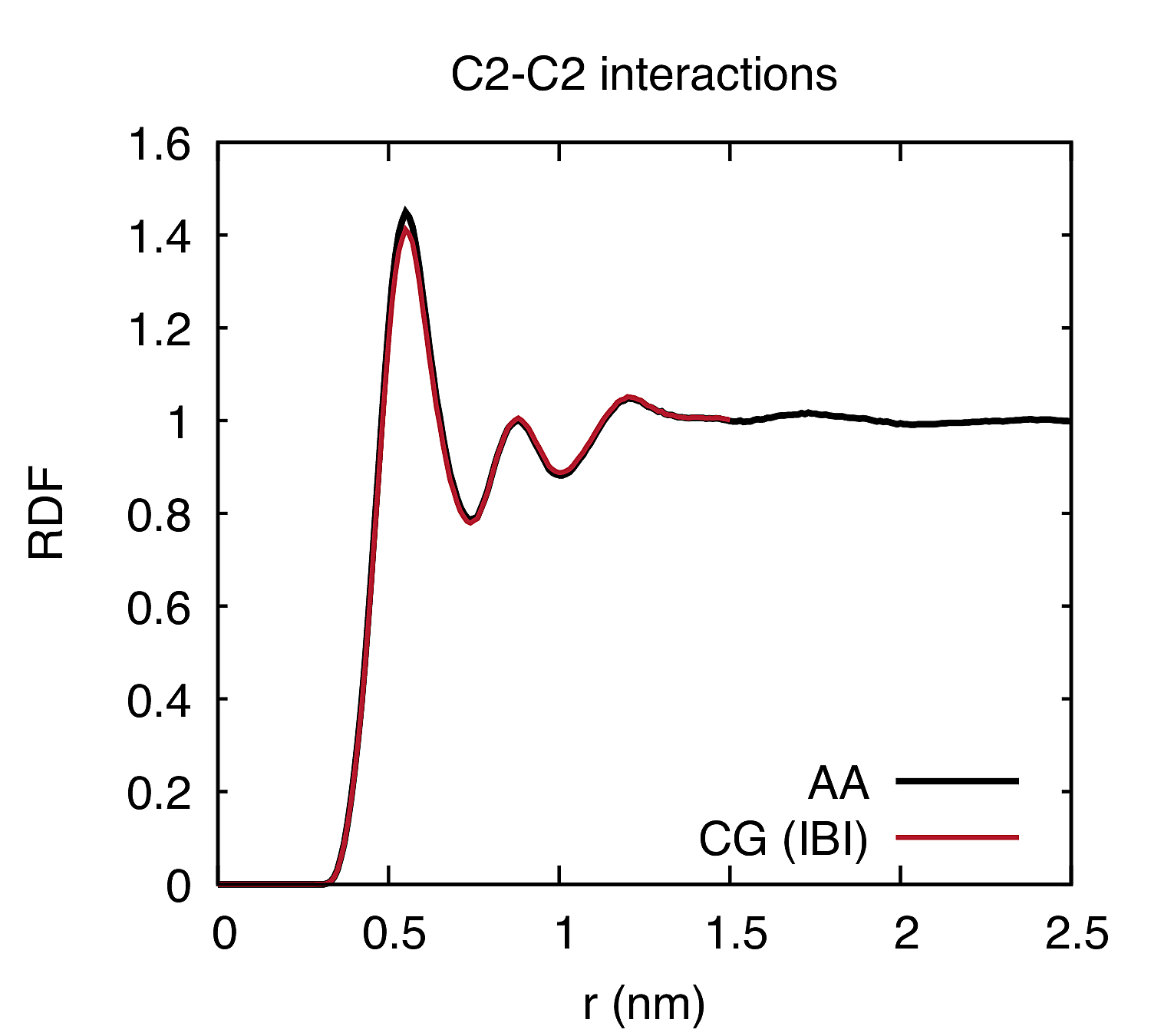}
  \includegraphics[width=0.28\textwidth]{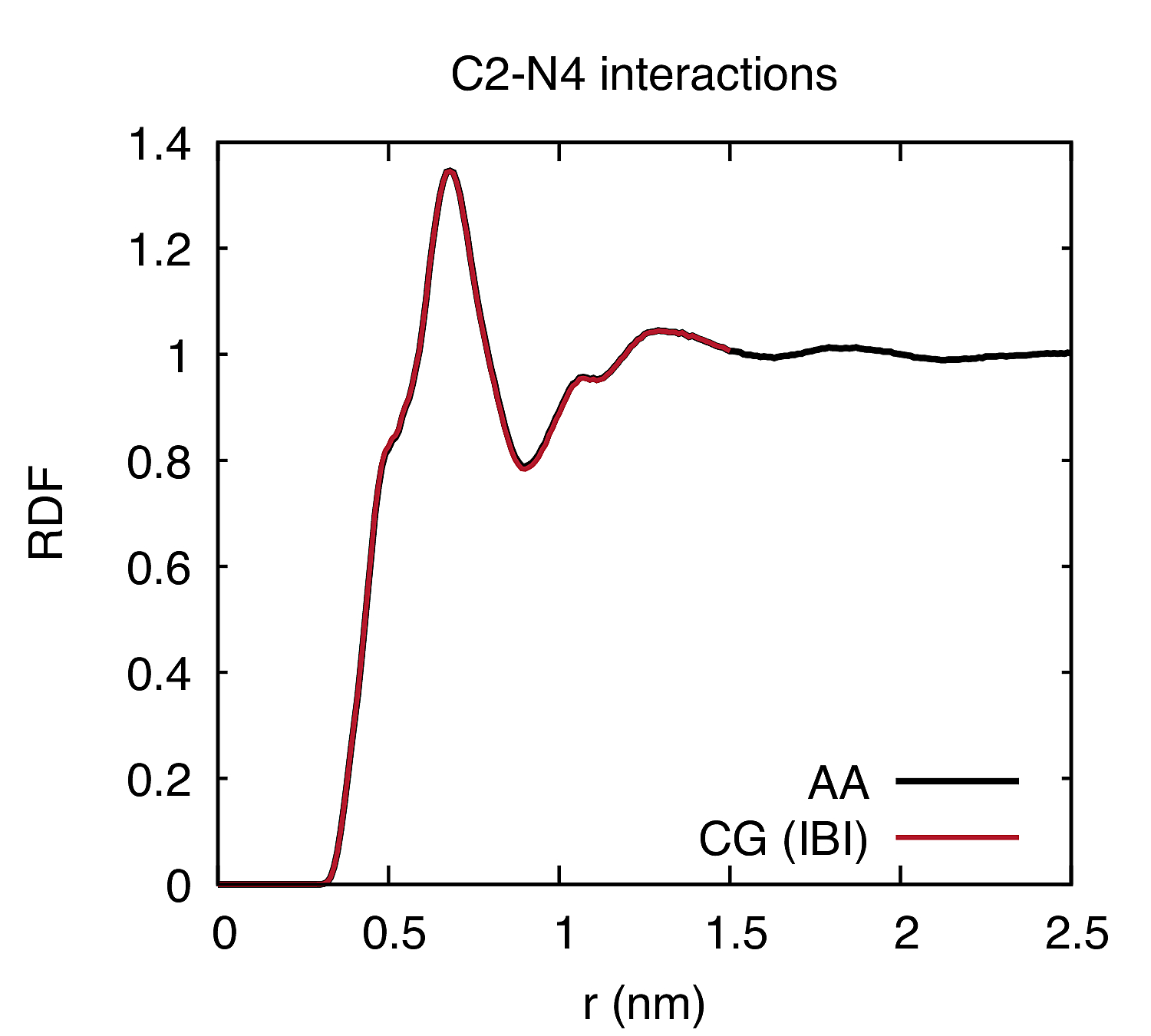}
  \includegraphics[width=0.28\textwidth]{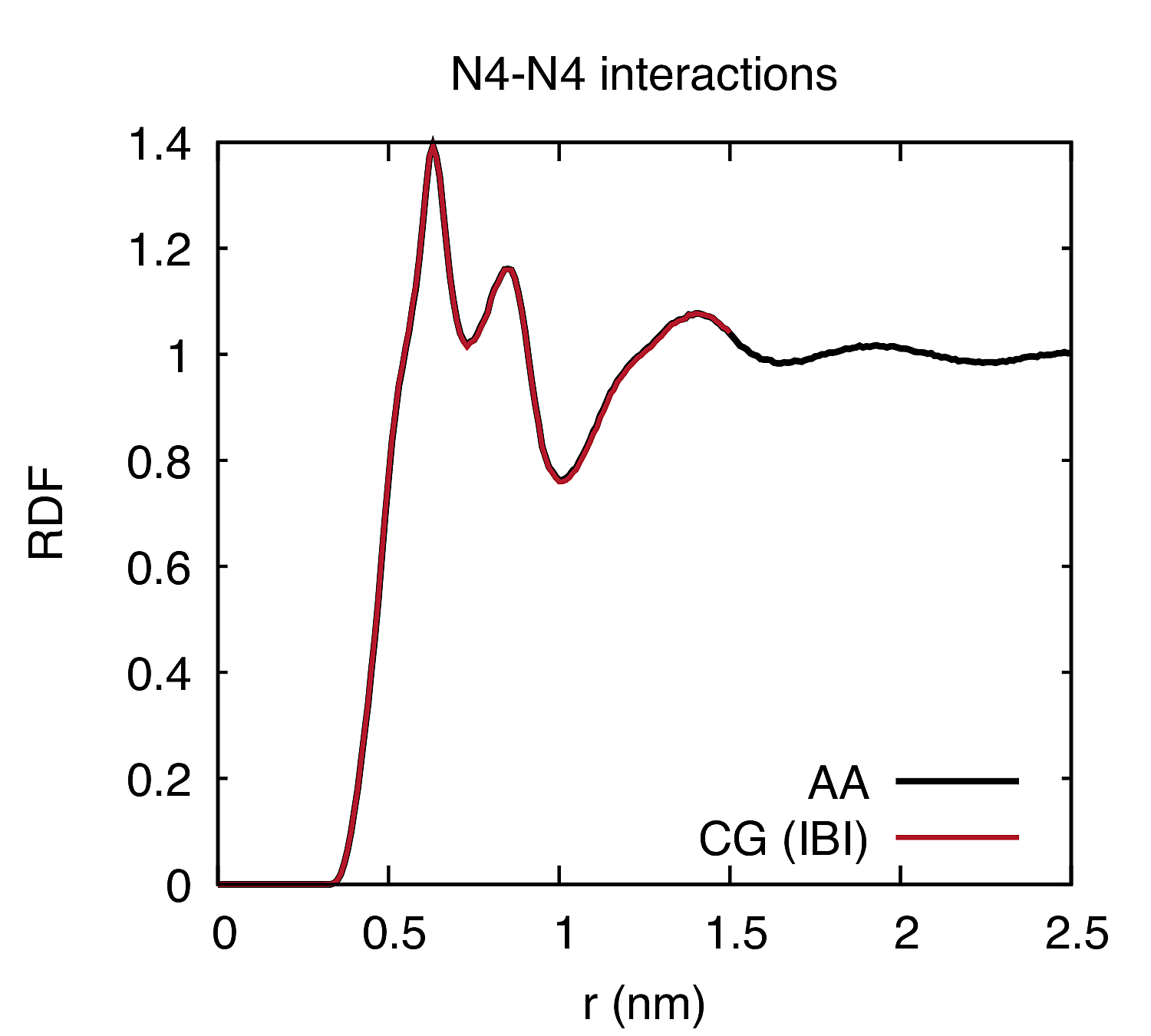}
  \caption{\textbf{Pair radial distribution functions (RDFs) between CG bead types 
           for the CGM3 mapping.}
           In each panel, results using the CGM3 representation are
           obtained from mapped all-atom (AA) simulations (black) and CG simulations 
           with pair potentials computed from iterative Boltzmann inversion (IBI) (red).
           }
  \label{fig:PTMA-CG-IBI-CGM3}
\end{figure}

\begin{figure}[htbp]
  \centering

  \includegraphics[width=0.28\textwidth]{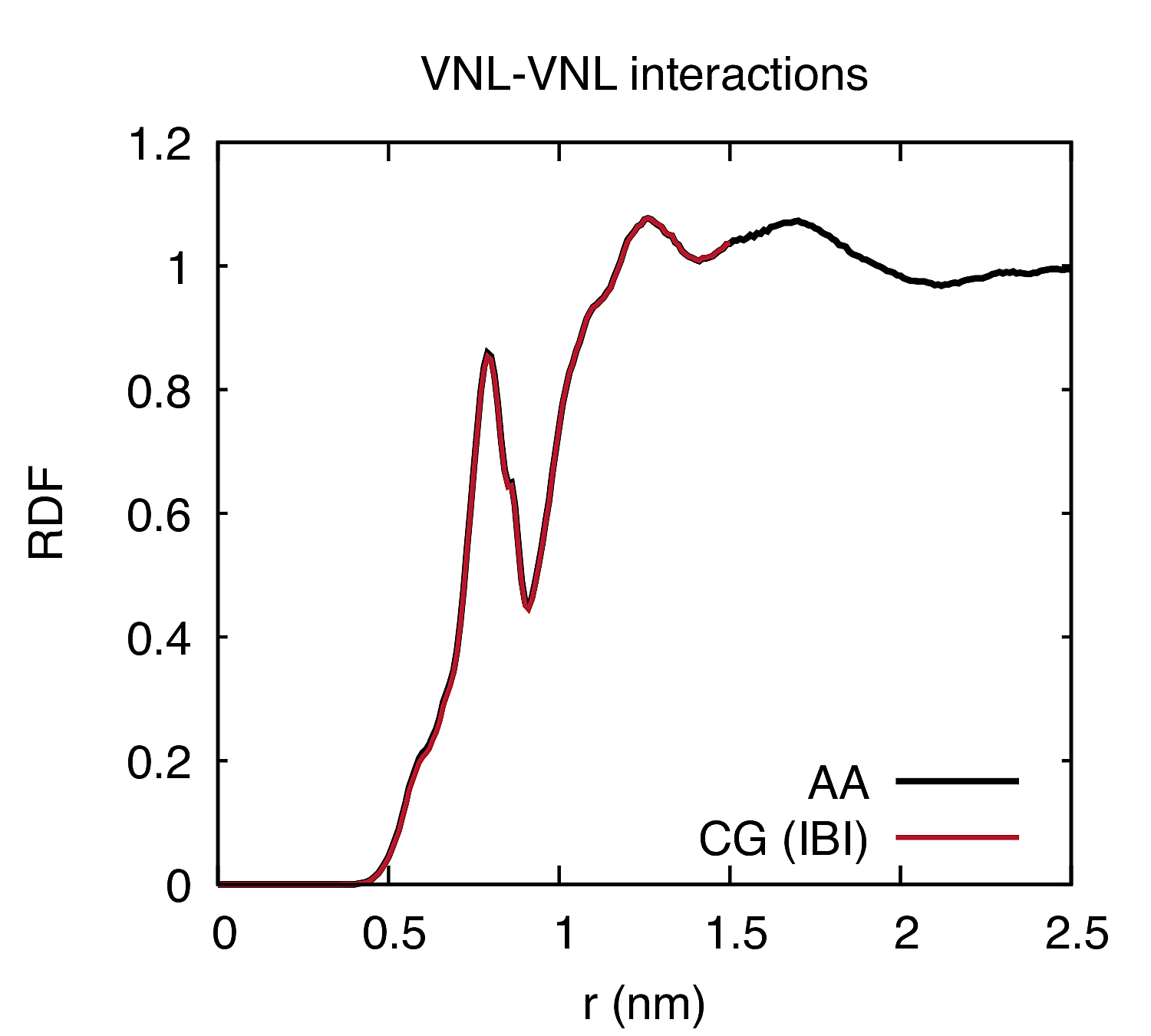}
  \includegraphics[width=0.28\textwidth]{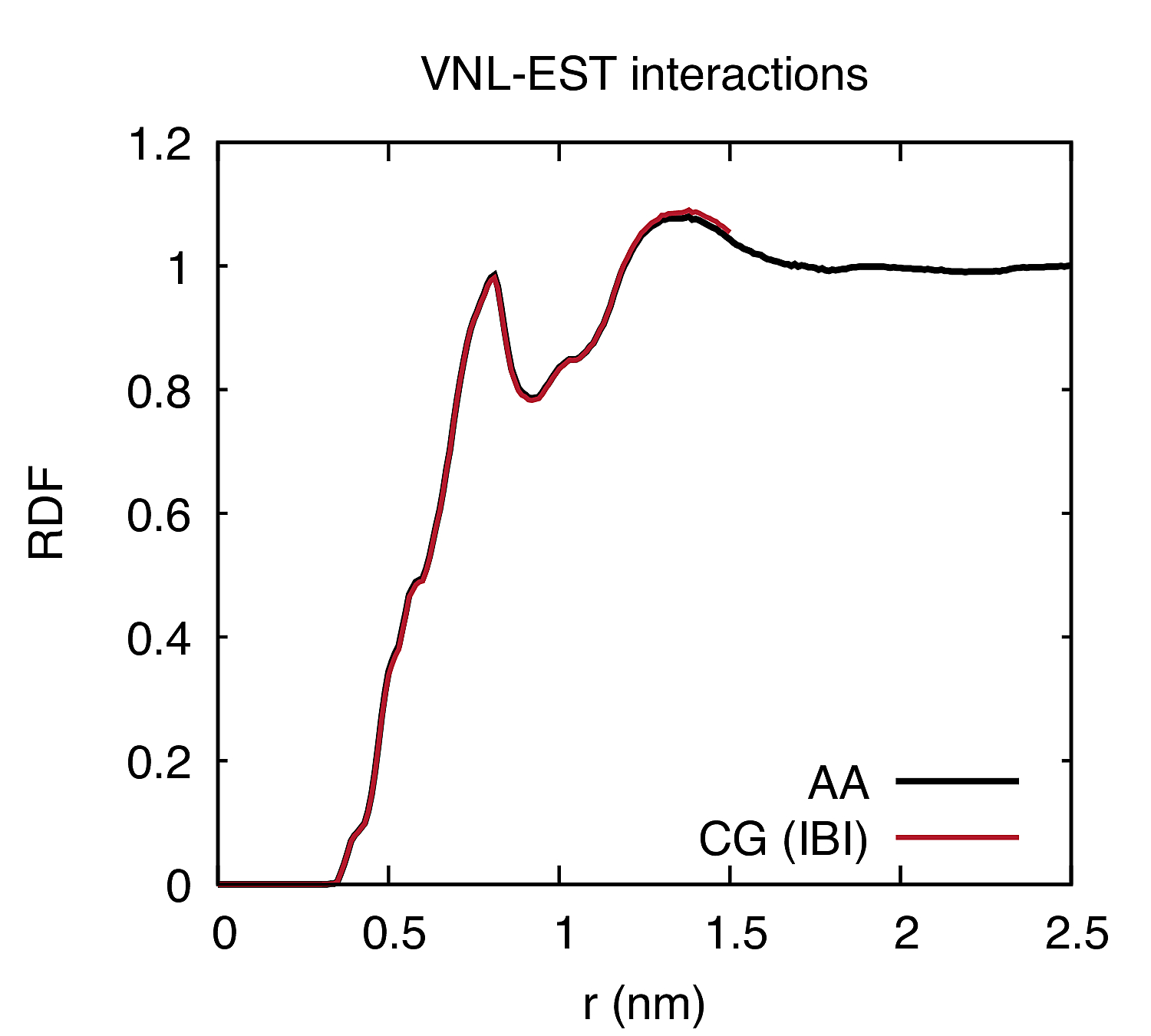}
  \includegraphics[width=0.28\textwidth]{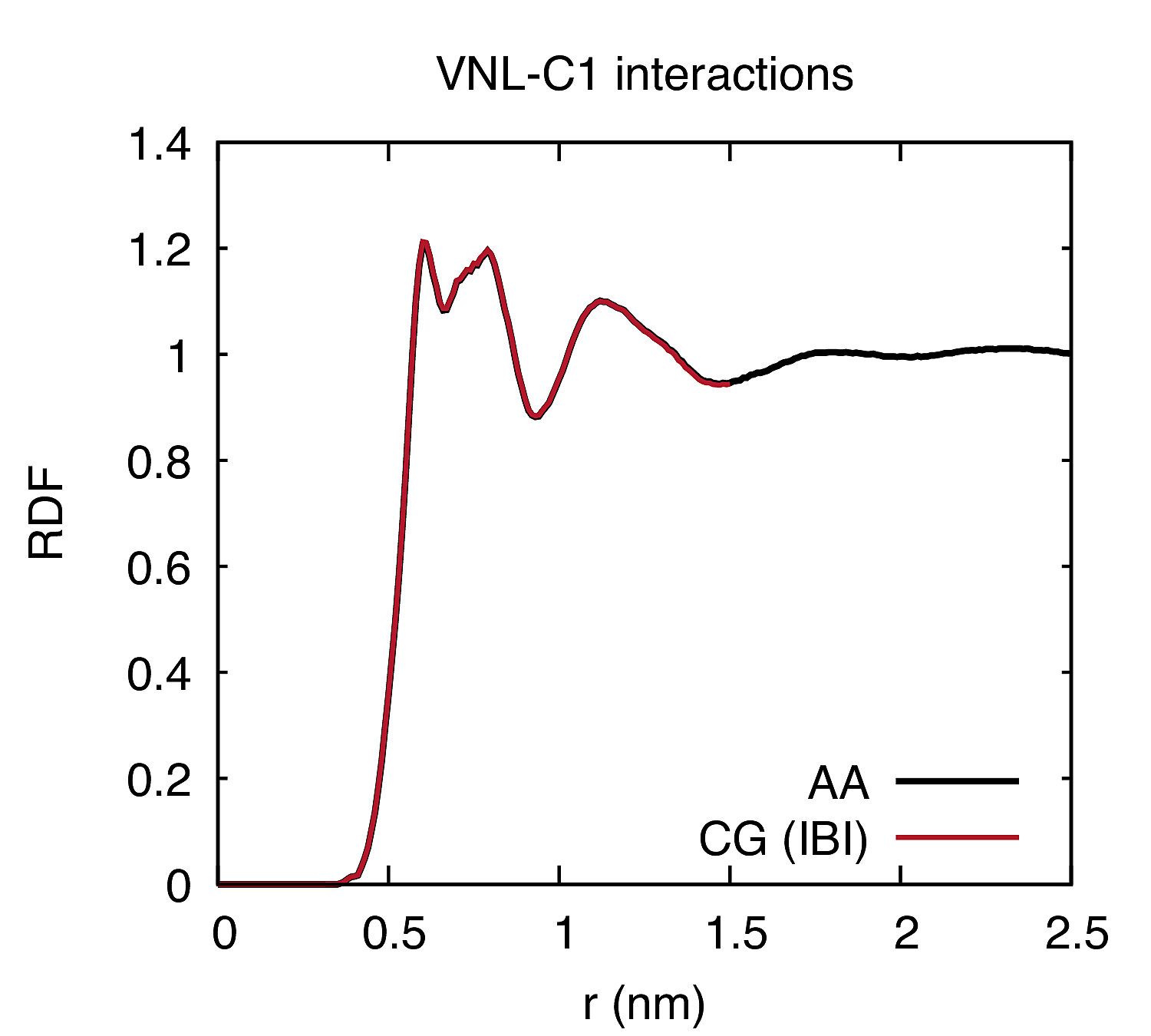}

  \includegraphics[width=0.28\textwidth]{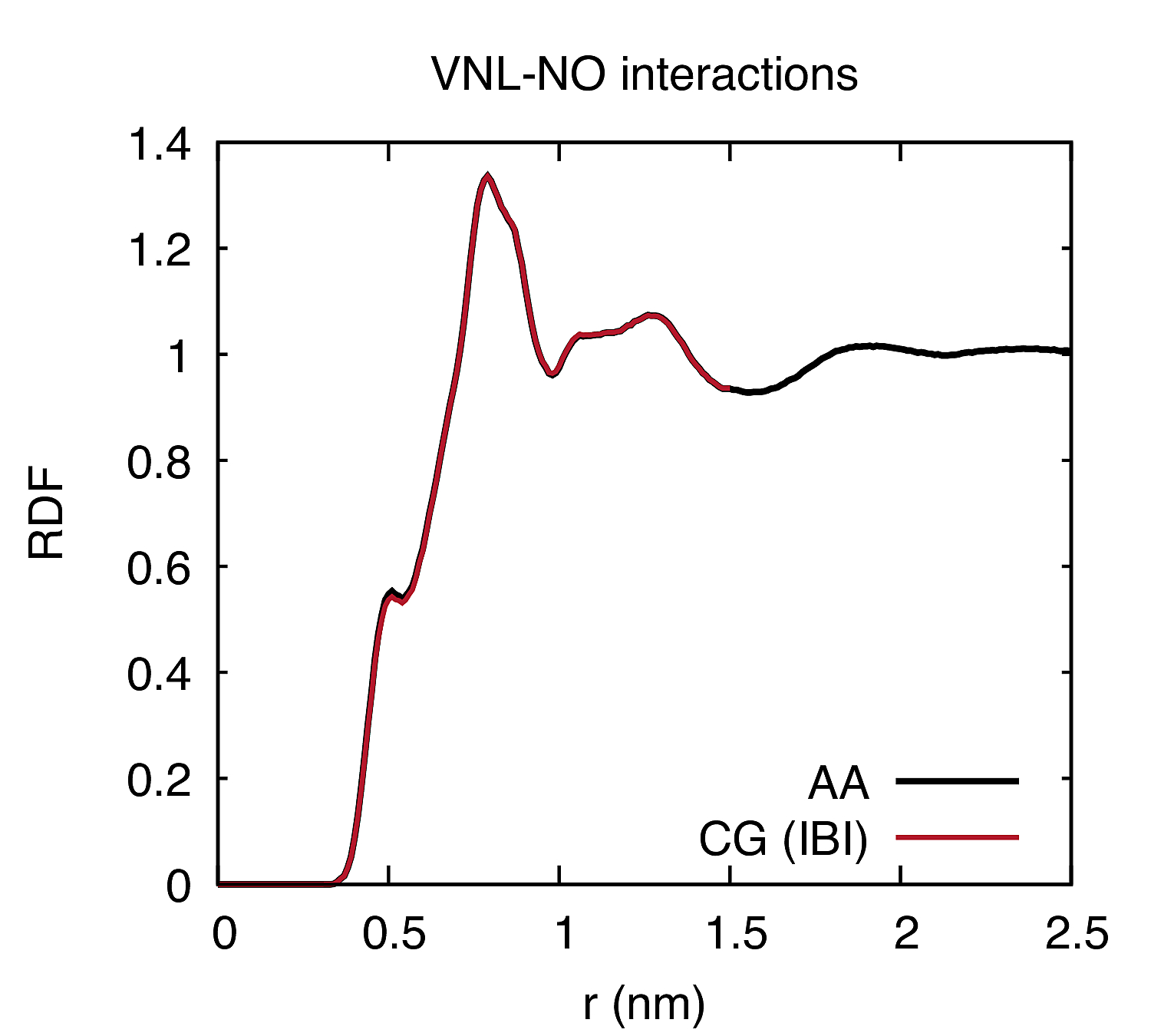}
  \includegraphics[width=0.28\textwidth]{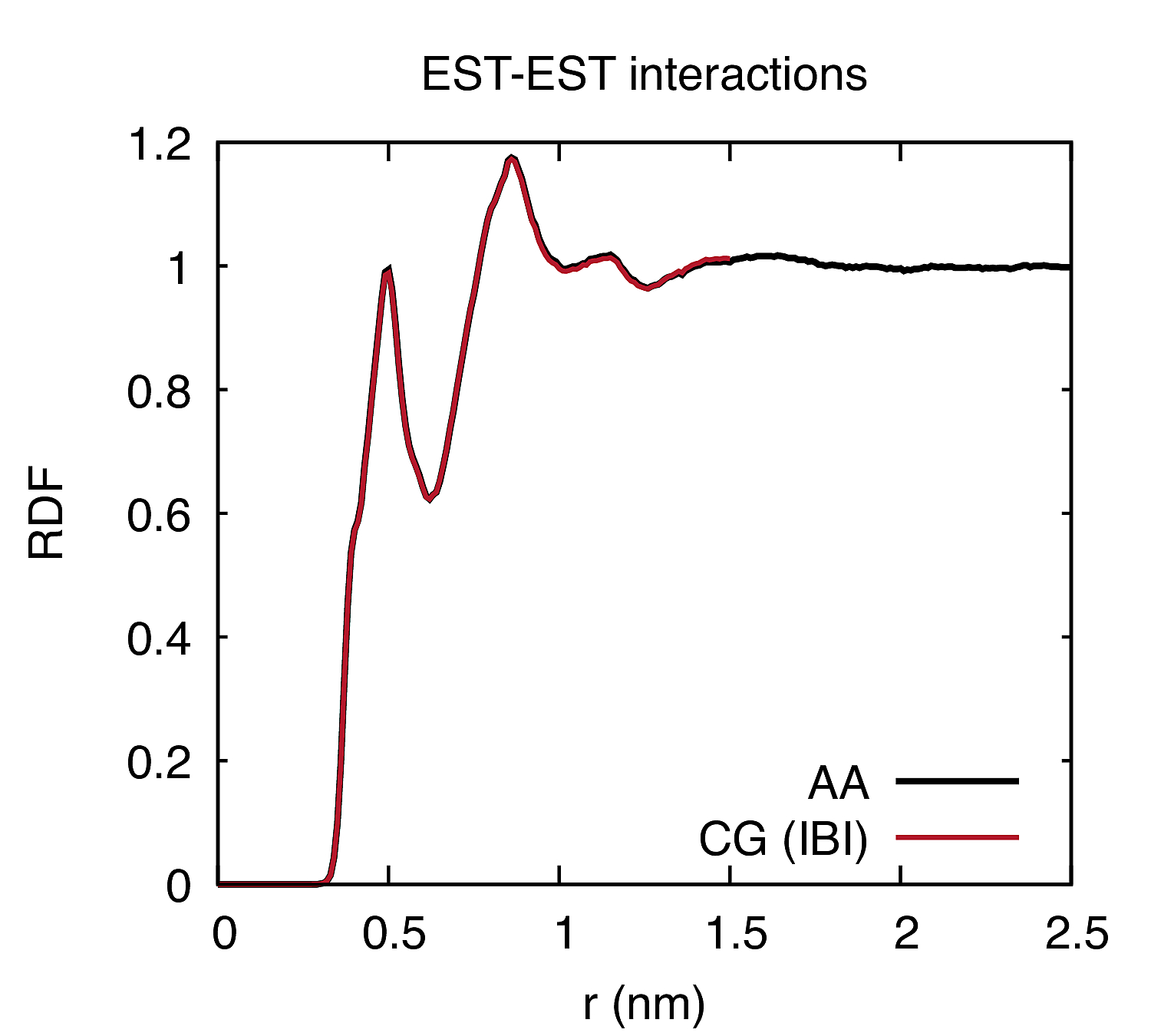}
  \includegraphics[width=0.28\textwidth]{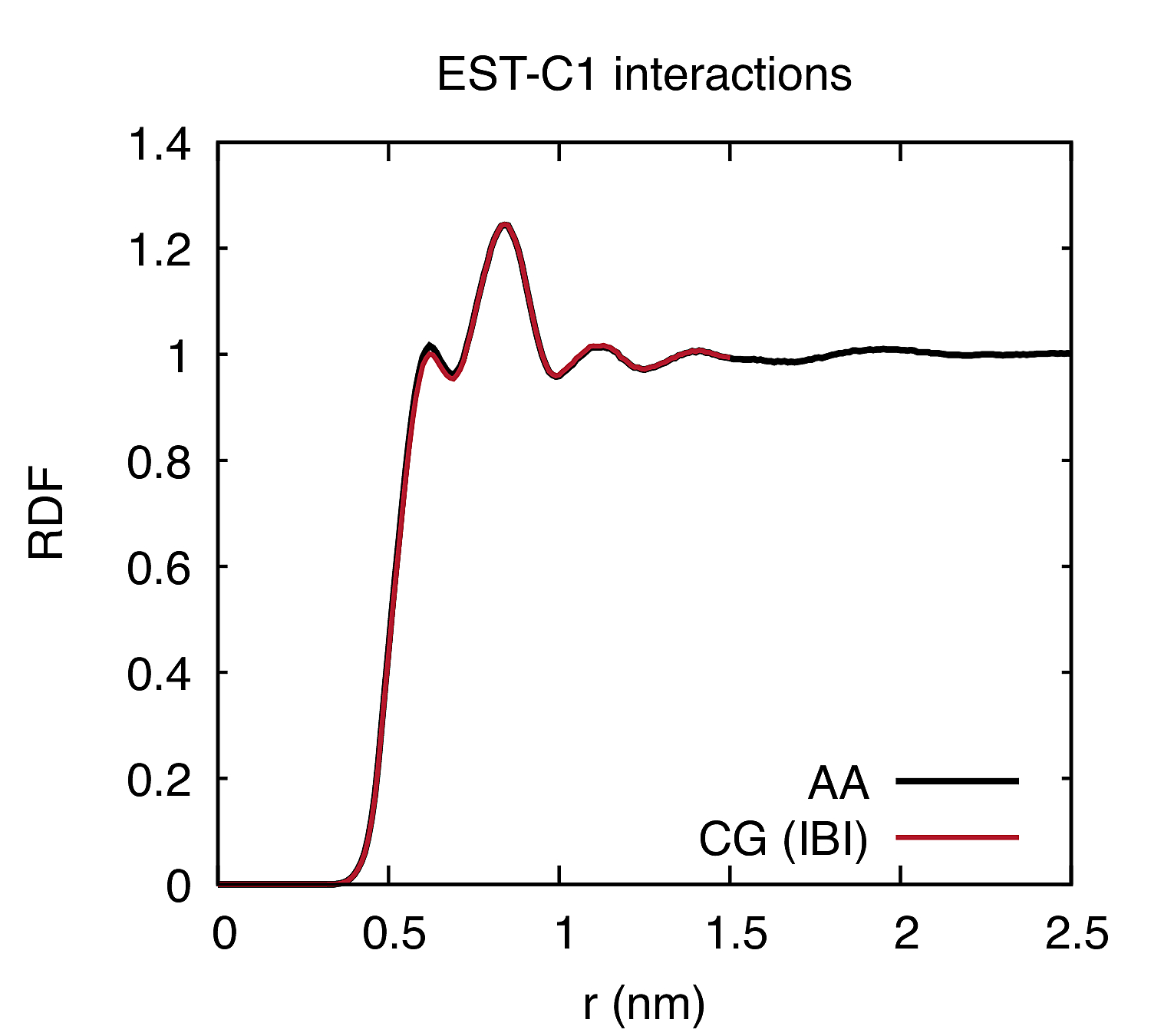}

  \includegraphics[width=0.28\textwidth]{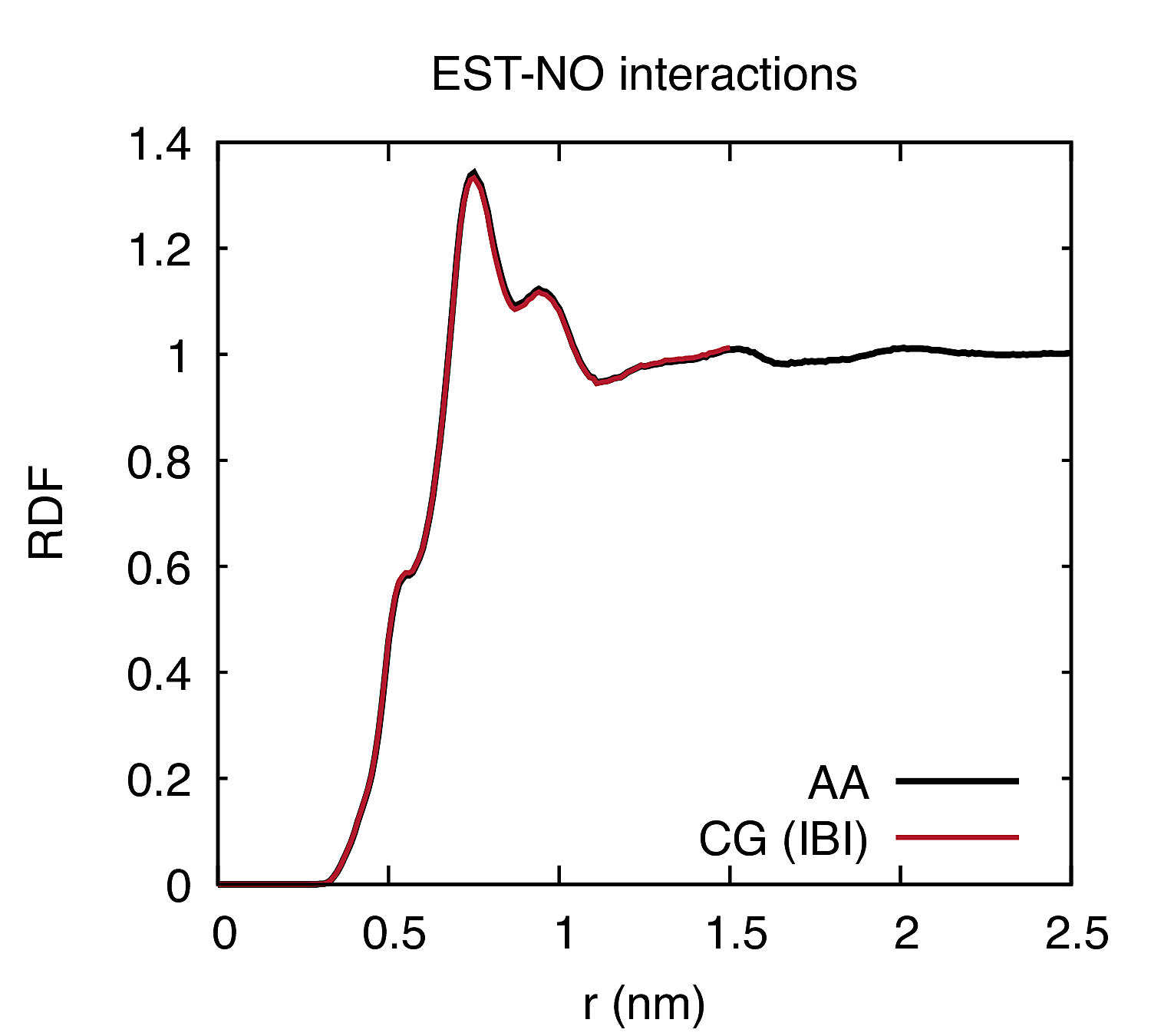}
  \includegraphics[width=0.28\textwidth]{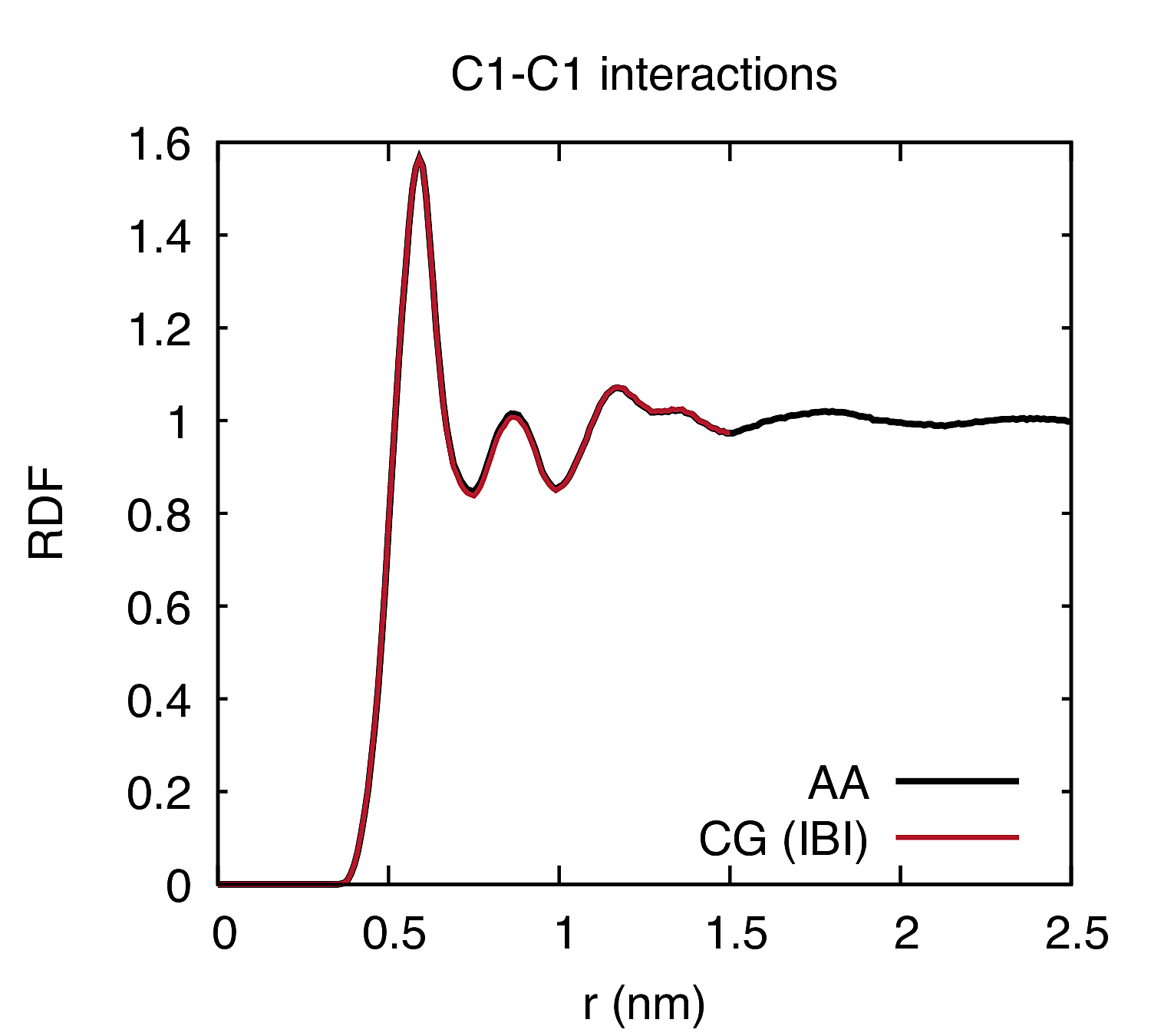}
  \includegraphics[width=0.28\textwidth]{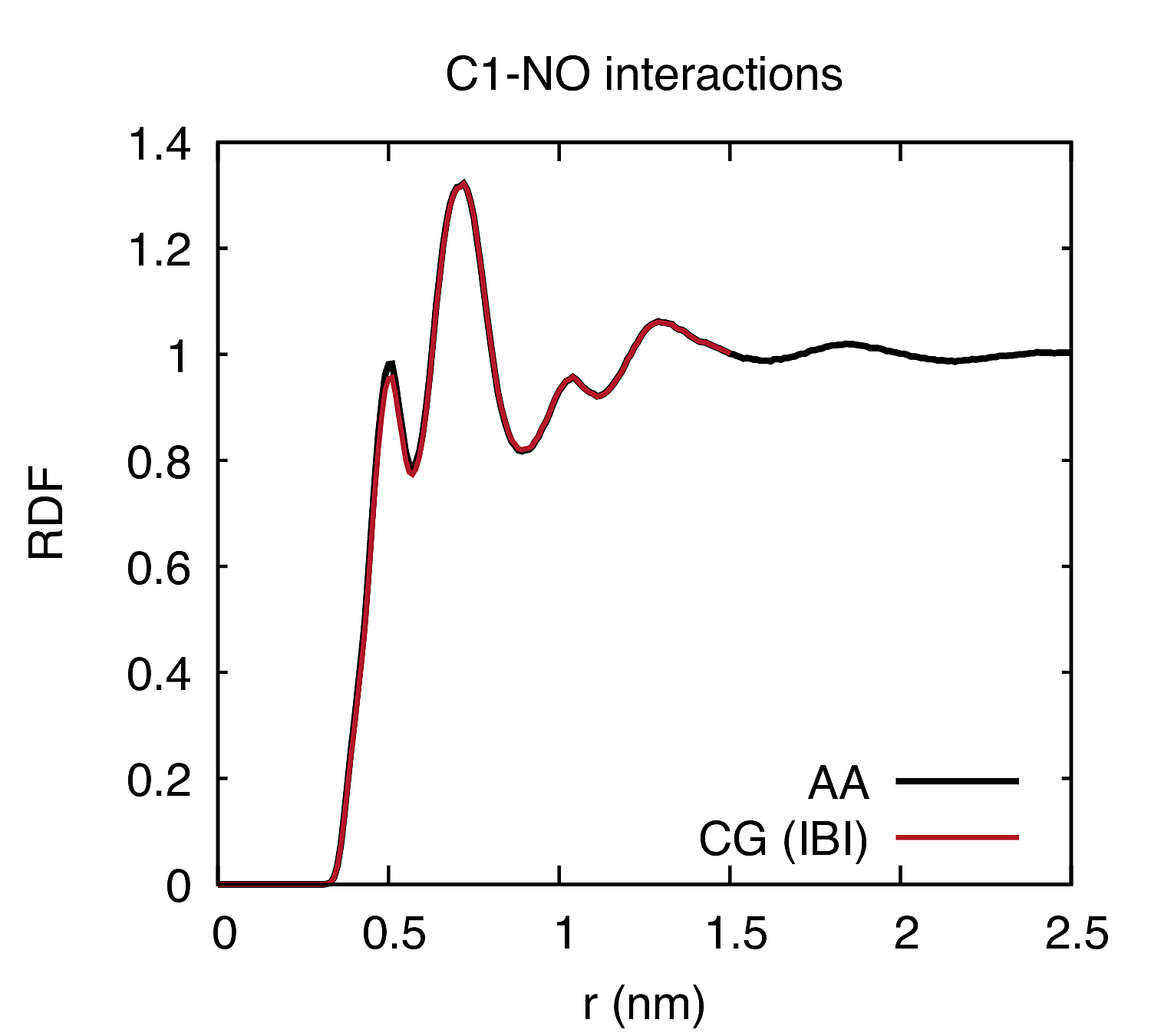}

  \includegraphics[width=0.28\textwidth]{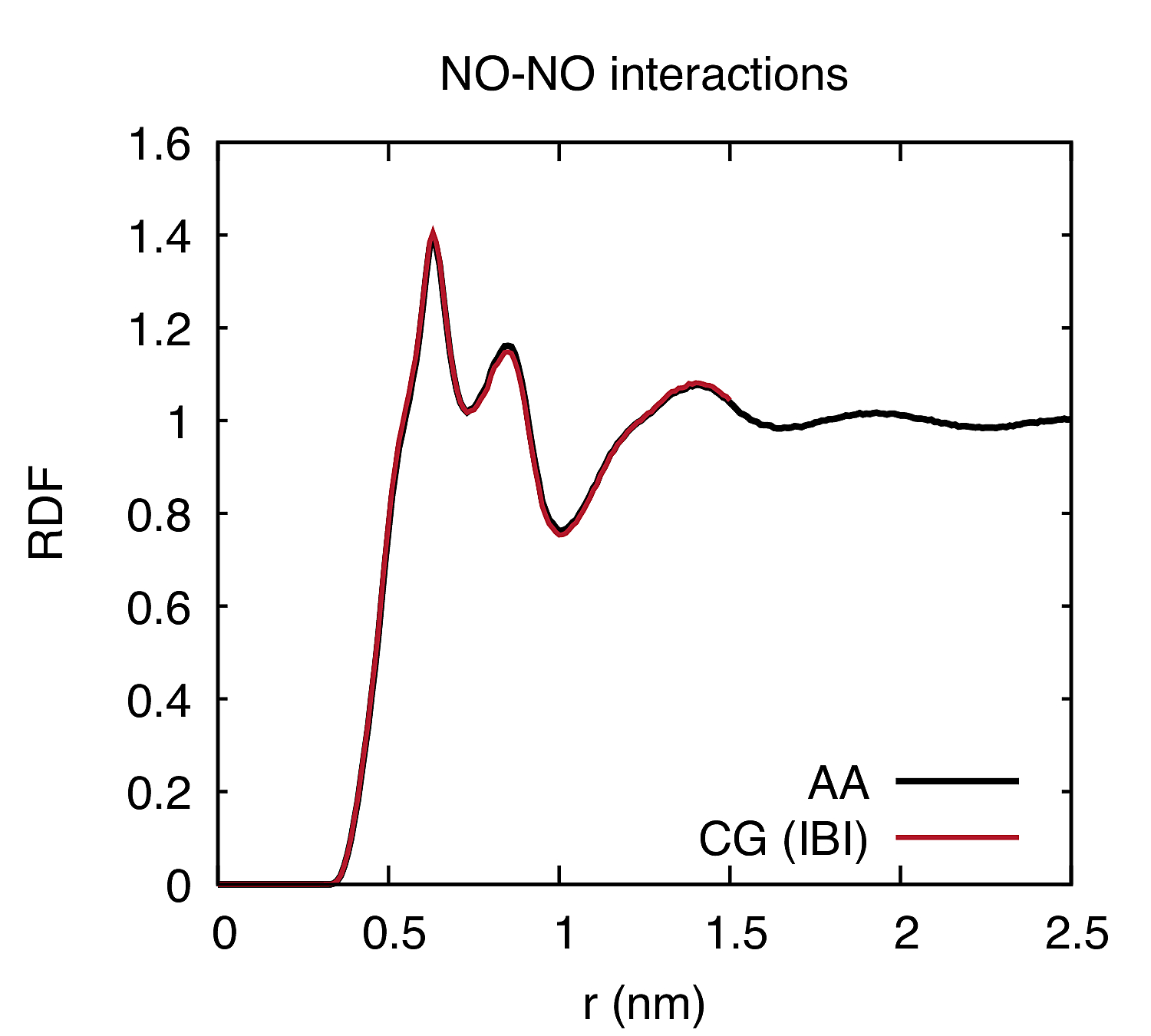}
  \caption{\textbf{Pair radial distribution functions (RDFs) between CG bead types 
           for the GBNO2 mapping.}
           In each panel, results using the GBNO2 representation are
           obtained from mapped all-atom (AA) simulations (black) and CG simulations 
           with pair potentials computed from iterative Boltzmann inversion (IBI) (red).
           }
  \label{fig:PTMA-CG-IBI-GBNO2}
\end{figure}

A comparison between the AA reference RDFs and the RDFs obtained with
the final IBI-derived CG potentials for PTMA is shown in Figure~\ref{fig:PTMA-CG-IBI-CGM3} (CGM3)
and Figure~\ref{fig:PTMA-CG-IBI-GBNO2} (GBNO2). 

\begin{figure}
  \caption{\textbf{Topology of the PTMA CG model (CGM3 mapping).}
           Shown is the Polyply \textit{block} \texttt{.ff} file (that uses GROMACS syntax).
           The exclusion settings apply to the CG IBI model.
           In the CG M3 model, the \texttt{[exclusions]} are omitted and
           \texttt{nrexcl = 1} (the Martini default).
           Units for the bond, angle, and dihedral parameters are indicated
           in the respective sections.}
  \label{fig:CG-PTMA-topology-CGM3-IBI}
\begin{lstlisting}
[ moleculetype ]
; name  nrexcl
PTMA     2

[ atoms ]
; id  type  resnr residue atom  cgnr  charge  mass
   1   VNL    1    PTMA   VNL    1     0.0     54
   2   EST    1    PTMA   EST    2     0.0     54
   3    C1    1    PTMA    C1    3     0.0     54
   4    C2    1    PTMA    C2    4     0.0     54
   5    C2    1    PTMA    C3    5     0.0     54
   6    N4    1    PTMA    N4    6     0.0     36

[ bonds ]
; i  j   funct   length force_k
; harmonic        (nm)  (kJ mol-1 nm-2)
  1  2      1     0.262   15000 ; b_1
  2  3      1     0.253   15000 ; b_2
  3  4      1     0.326   25000 ; b_3
  3  5      1     0.326   25000 ; b_4

[ constraints ]
; i  j   funct   length
;                 (nm)
  3  6      1     0.310 ; b_5
  4  6      1     0.238 ; b_6
  5  6      1     0.238 ; b_7

[ angles ]
; i  j  k  funct  ref.angle force_k
; harmonic          (deg)   (kJ mol-1 rad-2)
  1  2  3    2      132.9    70 ; ang_1
  2  3  4    2      129.0    60 ; ang_2
  2  3  5    2      121.9    95 ; ang_3

[dihedrals]
; i j k l  funct  ref.angle   force_k
; harmonic          (deg)   (kJ mol-1 rad-2)
  3 4 5 6    2      140.00      20 ; dih_1
  2 4 5 6    2      150.00      80 ; dih_2

[exclusions]
  1 2 3 4 5 6
  2 3 4 5 6
  3 4 5 6
  4 5 6
  5 6
\end{lstlisting}

\end{figure}

\begin{figure}
  \caption{\textbf{Topology of the PTMA CG model (GBNO2 mapping).}
           Shown is the Polyply \textit{block} \texttt{.ff} file (that uses GROMACS syntax).
           Units for the bond, angle, and dihedral parameters are indicated
           in the respective sections.}
  \label{fig:CG-PTMA-topology-GBNO2}
\begin{lstlisting}
[ moleculetype ]
; name  nrexcl
PTMA     2

[ atoms ]
; id  type  resnr  residu atom  cgnr  charge  mass
   1   VNL    1     PTMA  VNL    1     0     71.099
   2   EST    1     PTMA  EST    2     0     29.018
   3    C1    1     PTMA   C1    3     0     56.108
   4     U    1     PTMA   NN    4     0      0.000
   5     U    1     PTMA   OO    5     0      0.000
   6    C1    1     PTMA   C2    6     0     56.108
   7    NO    1     PTMA   NO    7     0     30.006

[ bonds ]
; i  j   funct   length force_k
; harmonic        (nm)  (kJ mol-1 nm-2)
  1  2      1     0.35337 12500 ; b_1

[ constraints ]
; i  j   funct   length
;                 (nm)
  2  3      1     0.32325 b_2
  2  6      1     0.32369 b_3
  3  6      1     0.38838 b_4
  3  7      1     0.23794 b_5
  6  7      1     0.23793 b_6

[ angles ]
; i  j  k  funct  ref.angle force_k
; harmonic          (deg)   (kJ mol-1 rad-2)
  1  2  3    2      130.000  90 ; ang_1
  1  2  6    2      130.000  75 ; ang_2

[dihedrals]
; i j k l  funct  ref.angle   force_k
; harmonic          (deg)   (kJ mol-1 rad-2)
  1 3 6 7    2      165.98      50 ; dih_2
  2 3 6 7    2      179.20      80 ; dih_3

[virtual_sites3]
; peripheral virtual sites
; site positioned as a linear combination of 3 atoms (same plane)
; site  from      funct    a      b
   4    7  3  6     1    0.230  0.230
   5    7  3  6     1   -0.230 -0.230

[exclusions]
  1  2  3  4  5  6  7
  2  3  4  5  6  7
  3  4  5  6  7
  4  5  6  7
  5  6  7
  6  7
\end{lstlisting}

\end{figure}

\begin{figure}[htbp]
  \centering
  \includegraphics[width=\textwidth]{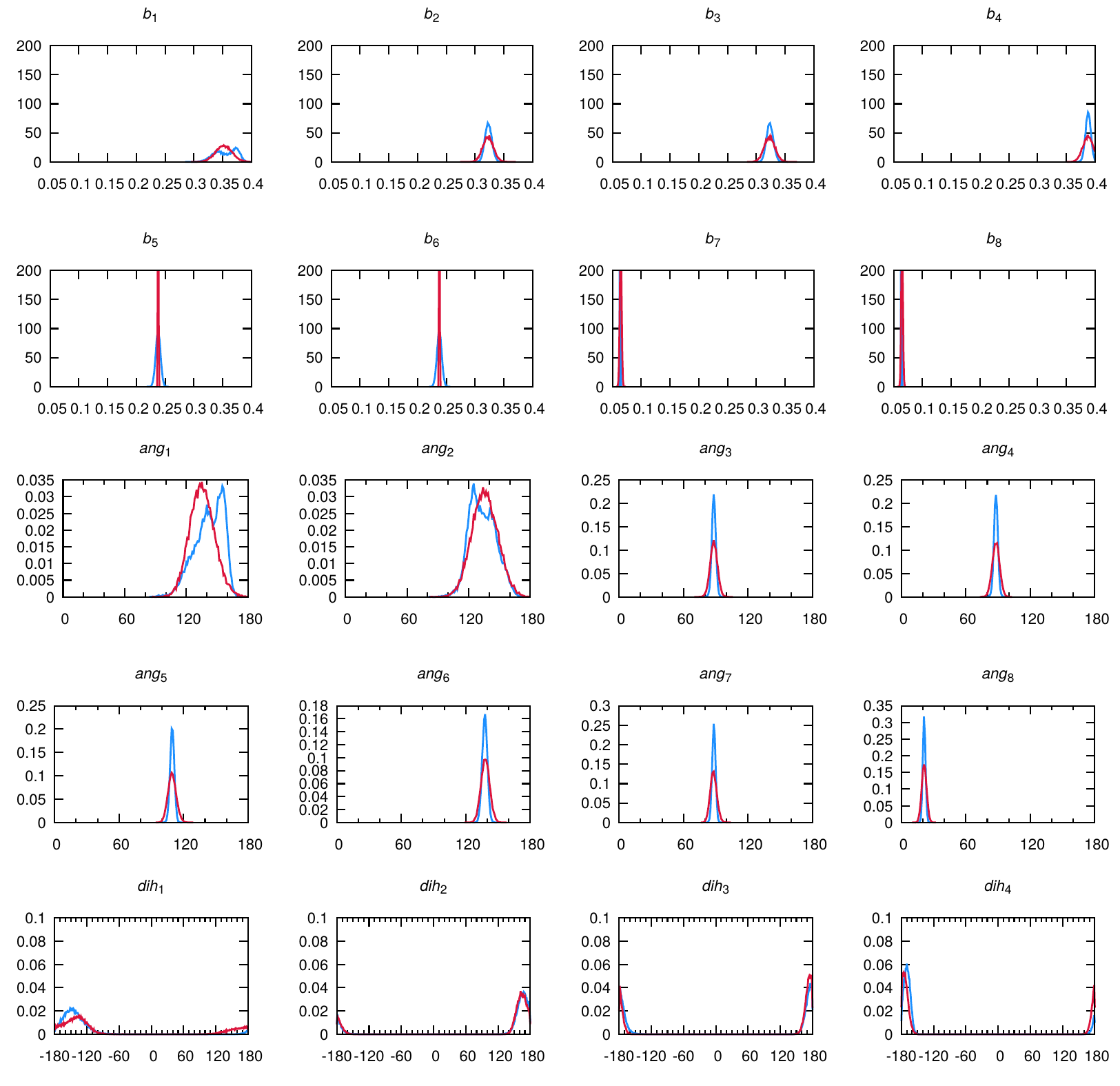}
  \caption{\textbf{Bond, angle, and torsion distributions for a PTMA monomer (GBNO2 mapping).}
           Atomistic reference distributions are in blue, CG distributions are in red.
           Compare header to labels in Figure~\ref{fig:CG-PTMA-topology-GBNO2} 
           for information on the particular degree of freedom shown. 
           Degrees of freedom $b_7-b_8$, $ang_3-ang_8$, and $dih_1$ and $dih_4$ (which have no explicit 
           potential in the topology, see Figure~\ref{fig:CG-PTMA-topology-GBNO2}) correspond to those
           that are either kept in place by the virtual sites or by the other potentials.
          }
  \label{fig:TMA-distributions-GBNO2}
\end{figure}

\begin{figure}
  \caption{\textbf{Topology of the PMMA-like backbone model (CGM3 mapping).}
           Shown is the Polyply \textit{link} \texttt{.ff} file (that uses GROMACS syntax).
           Units for the bond, angle, and dihedral parameters are indicated
           in the respective sections. From Ref.\citenum{2022FGrunewald-NatCommun}.}
  \label{fig:CG-PMMA-topology-CGM3}
\begin{lstlisting}
[ link ]
resname "PTMA"
[ bonds ]
; i  j   funct   length force_k
; harmonic        (nm)  (kJ mol-1 nm-2)
VNL +VNL  1      0.315  4000  {"group": "vinyl backbone"}

[ link ]
resname "PTMA"
[ angles ]
; i   j    k  funct ref.angle force_k
; harmonic          (deg)   (kJ mol-1 rad-2)
VNL +VNL ++VNL  2   115        35  {"group": "vinyl backbone"}

[ link ]
resname "PTMA"
[ angles ]
; i   j    k  funct ref.angle force_k
; harmonic          (deg)   (kJ mol-1 rad-2)
EST  VNL  +VNL  2    70        20
\end{lstlisting}

\end{figure}
\begin{figure}
  \caption{\textbf{Topology of the PMMA-like backbone model (CGM3 mapping).}
           Shown is the Polyply \textit{link} \texttt{.ff} file (that uses GROMACS syntax).
           Units for the bond, angle, and dihedral parameters are indicated
           in the respective sections.}
  \label{fig:CG-PMMA-topology-GBNO2}
\begin{lstlisting}
[ link ]
resname "PTMA"
[ bonds ]
; i  j   funct   length force_k
; harmonic        (nm)  (kJ mol-1 nm-2)
VNL +VNL  1      0.305 12000  {"group": "vinyl backbone"} ; b_1

[ link ]
resname "PTMA"
[ angles ]
; i   j    k  funct ref.angle force_k
; harmonic          (deg)   (kJ mol-1 rad-2)
VNL +VNL ++VNL  2   126        40  {"group": "vinyl backbone"} ; ang_1

[ link ]
resname "PTMA"
[ angles ]
; i   j    k  funct ref.angle force_k
; harmonic          (deg)   (kJ mol-1 rad-2)
EST  VNL  +VNL  2    90        20 ; ang_2
\end{lstlisting}

\end{figure}
\begin{figure}[htbp]
  \centering
  \includegraphics[width=\textwidth]{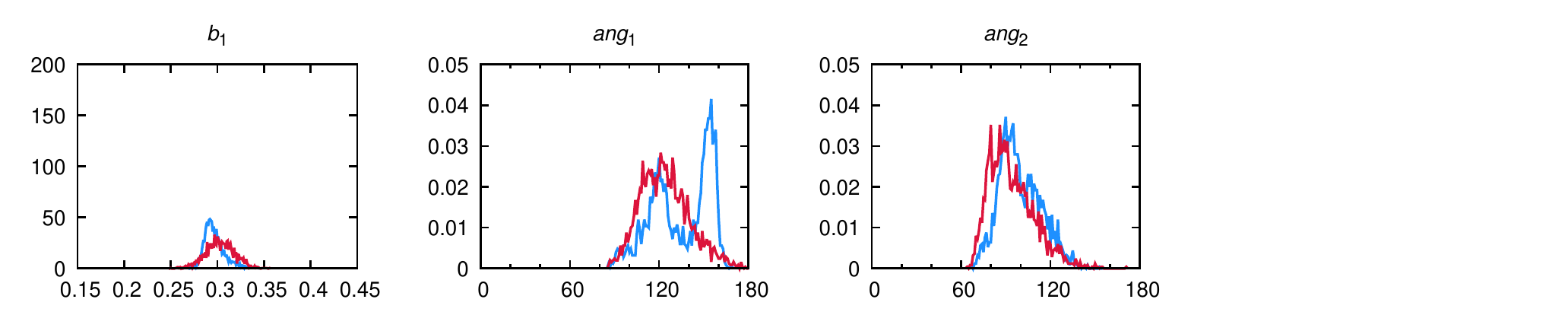}
  \caption{\textbf{Bond, angle, and torsion distributions for a PMMA-like backbone (GBNO2 mapping).}
           Atomistic reference distributions are in blue, CG distributions are in red.
           Compare header to labels in Figure~\ref{fig:CG-PMMA-topology-GBNO2} 
           for information on the particular degree of freedom shown. 
          }
  \label{fig:PMMA-distributions-GBNO2}
\end{figure}

\newpage 

\textbf{Nonbonded interactions: Martini 3.} 
Bead types for the backbone beads are taken from the Martini 3~\cite{2021PCTSouza-NatMethods} 
PMMA model (available in the Polyply library)~\cite{2022FGrunewald-NatCommun} 
and are SC3 and SN4a (see also Figure~\ref{fig:CG-PTMA-topology-CGM3-M3}). 
Martini 3 bead types for the 2,2,6,6-Tetramethyl-1-piperidineolate (TEMPO) 
unit have been chosen following recommendations
based on the Martini 3 small molecule dataset~\cite{2022RAlessandri-AdvTheorySimul},
when available. When not, such as in the case of the nitroxide group, the bead type 
was chosen by computing the free energies of transfer and comparing to experimental data
(see Table~\ref{tbl:transfer-free-energies}). Accordingly, we found that a TP5 bead
gives a $\Delta G$ in fair agreement with the available experimental 
logP of 1.85~\cite{2013ONAleksandrova-JGeochemExplor--TEMPOlogP}.

\begin{figure}[htbp]
  \centering
  \includegraphics[width=0.32\textwidth]{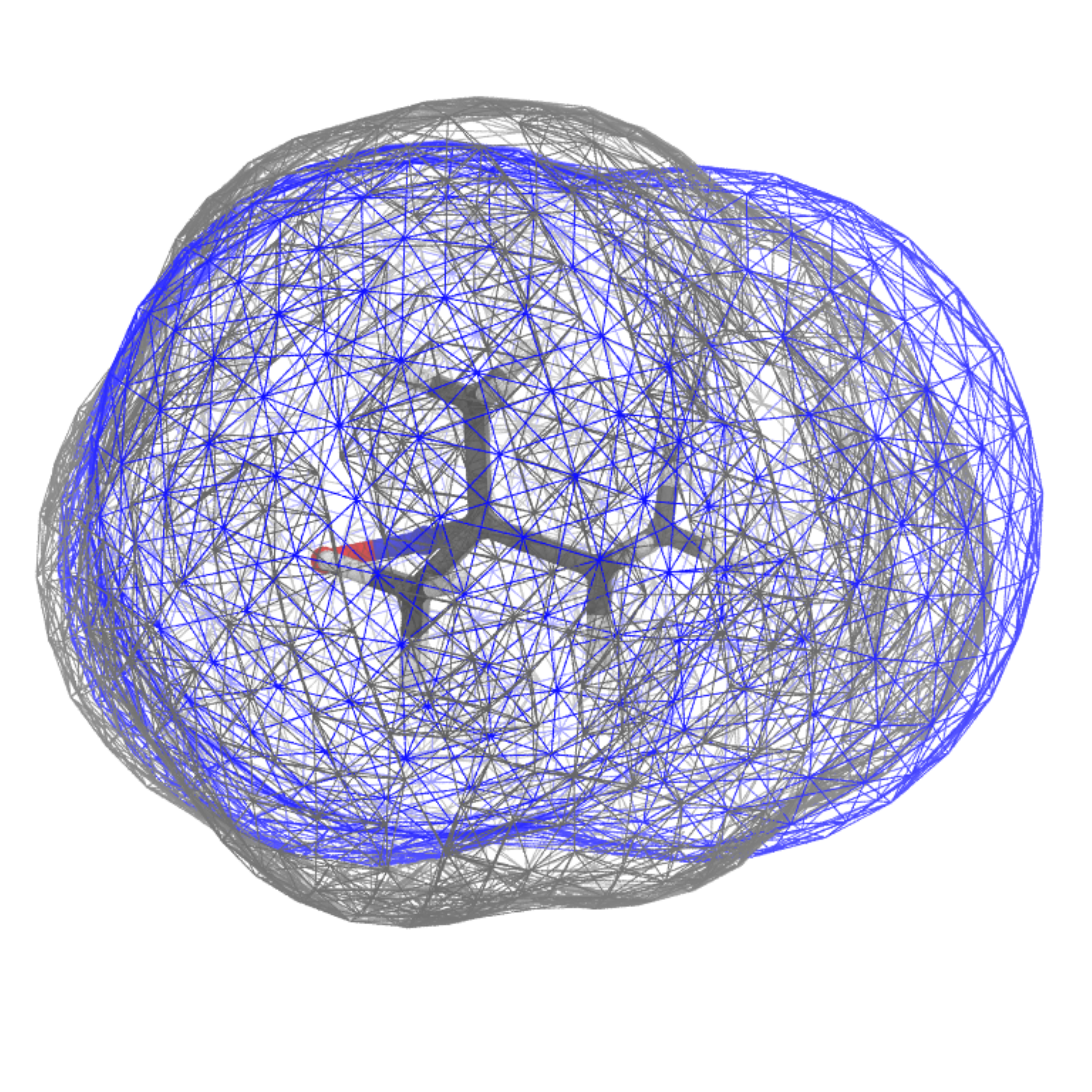}
  \includegraphics[width=0.32\textwidth]{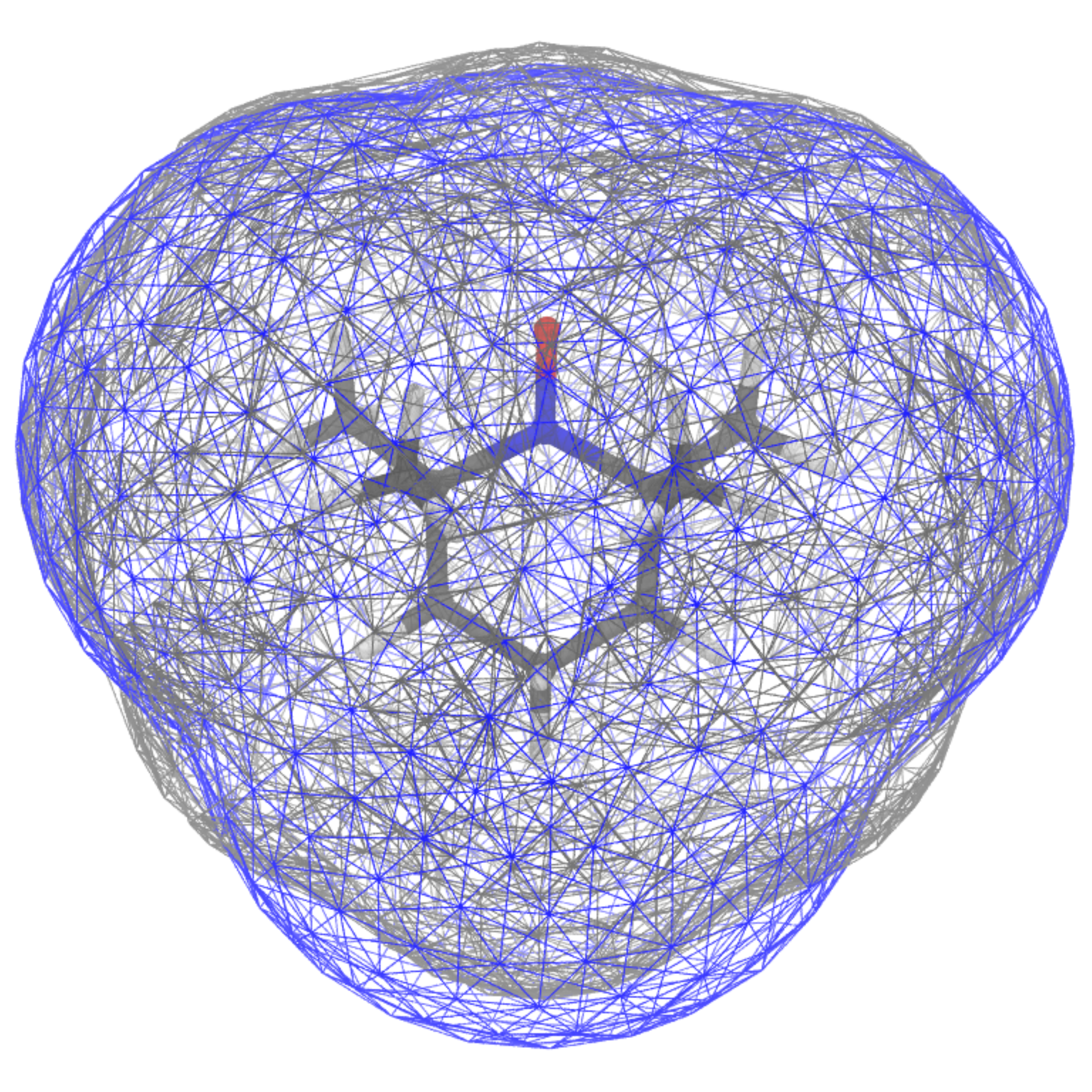}
  \includegraphics[width=0.32\textwidth]{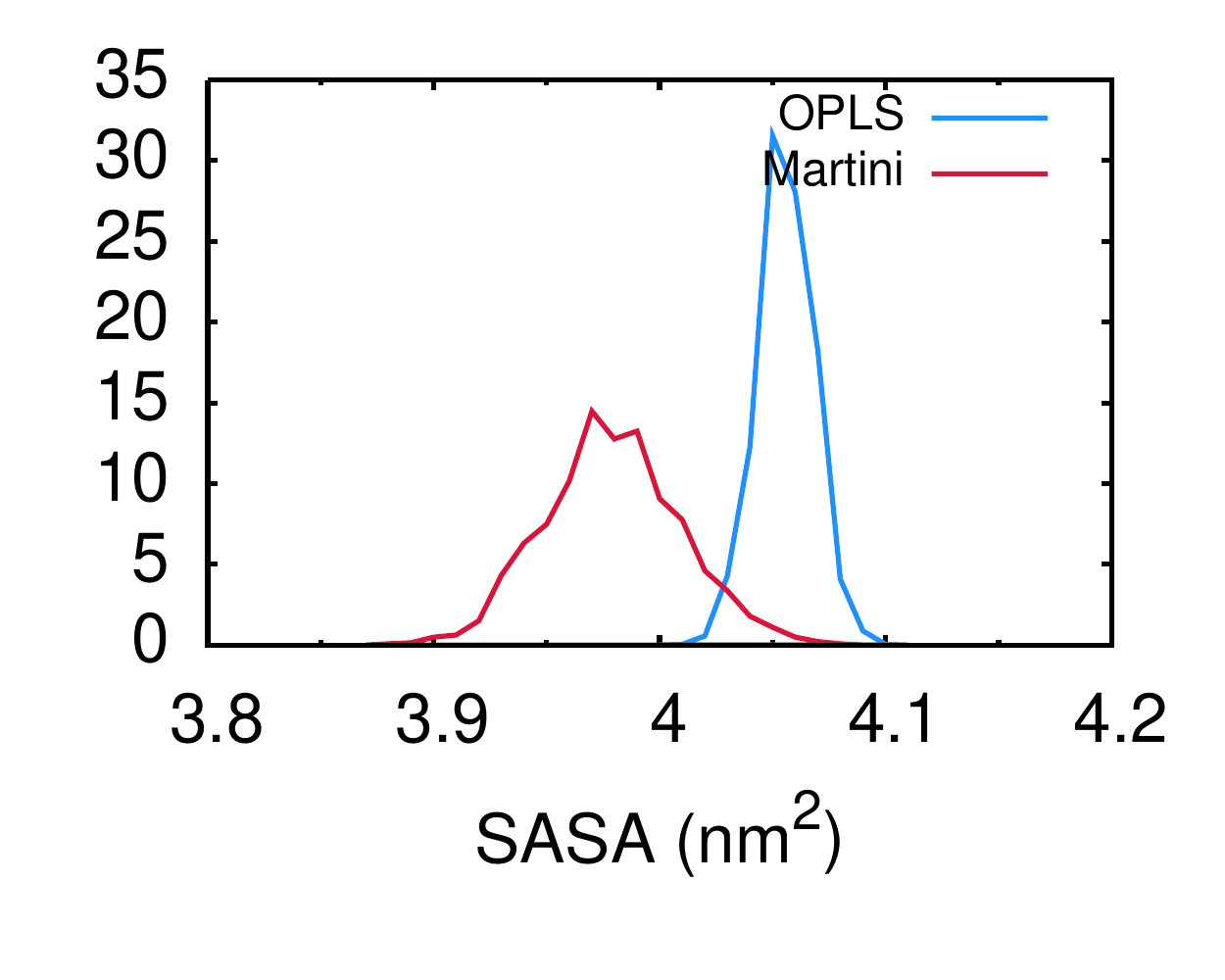}
  \caption{\textbf{Solvent accessible surface areas and Connolly surfaces for TEMPO: AA vs. CG models.}
           The CG surface is depicted in blue while the AA surface in gray.
           Mean and standard deviation of the SASA distributions are as follows:
           SASA$_{\mathrm{AA}}=4.06\pm0.01$ nm$^2$
           SASA$_{\mathrm{CG}}=3.98\pm0.03$ nm$^2$ (the CG SASA is $1.9$\% lower than the AA one).
          }
  \label{fig:SASA-and-Connolly-TEMPO}
\end{figure}

We moreover analyzed the shape and molecular volume of the Martini CG model
for TEMPO with respect to its corresponding AA model by computing solvent accessible
surface areas (SASAs) and Connolly surfaces, according to the Martini 3 
protocol described elsewhere~\cite{2021PCTSouza-NatMethods, 2022RAlessandri-AdvTheorySimul}. 
Results are shown in Figure~\ref{fig:SASA-and-Connolly-TEMPO} and show that
although the SASA average values are in good agreement (within 2\%),
the Connolly surface comparison shows mismatches between the AA and CG models.
In particular, the lower thickness of the CG Connolly surface is likely 
responsible for the RDF discrepancies that can be observed in Figure~\ref{fig:CGvsNN-results-M3}:
namely, as a consequence of the too low thickness of the Martini 3 model,
the TEMPO structures can come closer in space, likely giving rise to
the extra peak at low distances in the nitroxide-nitroxide RDF (Figure~\ref{fig:CGvsNN-results-M3}B).
\begin{figure}
  \caption{\textbf{Martini 3 bead types for the PTMA CG model.}
           Shown is the \texttt{[atoms]} section of the Polyply
           Polyply \textit{block} \texttt{.ff} file (that uses GROMACS syntax).
           Martini 3 bead types are shown in the second column.}
  \label{fig:CG-PTMA-topology-CGM3-M3}
\begin{lstlisting}
[ moleculetype ]
; name nrexcl.
PTMA     1

[ atoms ]
; id  type  resnr  residu atom  cgnr  charge  mass
   1   SC3    1    PTMA   VNL    1     0.0     54 ; Martini 3 PMMA model
   2   SN4a   1    PTMA   EST    2     0.0     54 ; Martini 3 PMMA model
   3   SC3    1    PTMA    C1    3     0.0     54
   4   SC3    1    PTMA    C2    4     0.0     54
   5   SC3    1    PTMA    C3    5     0.0     54
   6   TP5    1    PTMA    N4    6     0.0     36
\end{lstlisting}

\end{figure}

\begin{table}[htbp]
\caption{\textbf{Transfer free energies for TEMPO.} 
         The free energy relative to the transfer of the solute molecule
         from solvent S$_1$ to S$_2$ 
         ($ \Delta G_{\mathrm{S_1\rightarrow S_2}} $)
         obtained from experiments. These are shown along with the experimentally determined
         partitioning coefficient logP representing the same
         transfer (logP$_\mathrm{S_1 \rightarrow S_2}$) 
         Solvents are $n$-octanol (OCO) and water (W).
         All the free energies are in kJ mol$^{-1}$.
         Statistical uncertainty for the computed 
         $\Delta G$ is below 0.3 kJ mol$^{-1}$ in
         all cases. 
         Experimental data from Refs.~\citenum{2013ONAleksandrova-JGeochemExplor--TEMPOlogP}.
        }
\begin{center}
\begin{tabular}{l|c|rcc}
\hline 
  \multicolumn{2}{c}{}
  & \multicolumn{3}{|c}{$\Delta G_{\mathrm{OCO\rightarrow W}}$}\\
\hline 
                        & CG model 
                        & (logP$_\mathrm{OCO \rightarrow W}$) exp
                        & CG 
                        & AA \\   
\hline
TEMPO      & (SC3)$_3$-TP5
                        & (1.85) $\:$ 10.6
                        & 13.6    
                        & ---   \\ 
\hline
\end{tabular}
\end{center}
\label{tbl:transfer-free-energies}
\end{table}

\newpage

\section{Random dimer structure generation for Boltzmann-weighting}
\label{SI:sec:dimer}
\begin{figure}[H]
  \centering
  \includegraphics[width=0.48\textwidth]{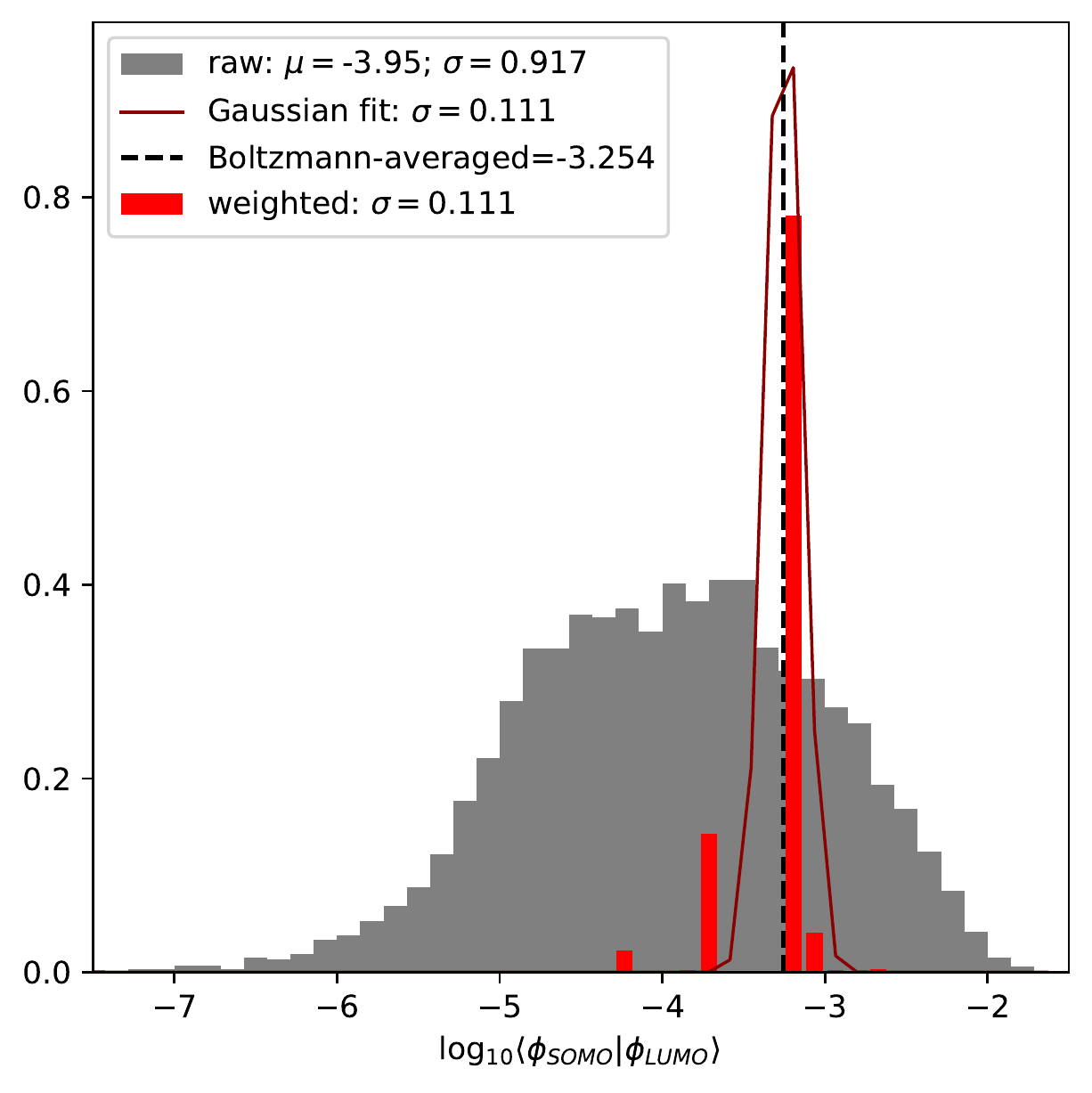}
  \caption{\textbf{Coupling distributions obtained via the gas-phase random dimer structure generation.}
           The grey distribution represents the histogram of the un-weighted overlaps;
           the red distribution represents the histogram of the Boltzmann-weighted overlaps,
           that is, the $y$ axis represents the Boltzmann factor 
           $ \frac{\exp(\frac{-E_i}{k_B T}}{ \sum_{i=1}^{N} \exp(\frac{-E_i}{k_B T}) } $;
           the dark red is the Guassian fit of the Boltzmann-weighted distribution.
           The Boltzmann-averaged (Eq.~\ref{M-eq:Boltzmann-weighting} of the main text) is
           also reported.
           The disorder, $\sigma$, obtained from the Boltzmann-weighted distribution 
           ($\sigma \approx 0.11$) underestimates greatly the disorder found in the 
           condensed phase simulations ($\sigma \approx 0.85$) shown in the main text.
          }
  \label{fig:Boltzmann-dimers}
\end{figure}

To randomly generate the dimer configurations to be Boltzmann-weighted,
we use the algorithm by \citeauthor{2022CHLi-JMaterChemA} and
available at \url{https://github.com/Tabor-Research-Group/redox_mol_screening}~\cite{2022CHLi-JMaterChemA}.
Briefly, the algorithm places two DFT-optimized monomers at an overlapping
position at the origin. From this initial configuration, one of the two monomers
is randomly translated and rotated. To avoid exploring conformational space
where the coupling is zero, the maximum translation allowed was such that
the closest heavy atom distance between two monomers was within 1.5 and 3.0 {\AA}.
The raw and Boltzmann-weighted overlaps are shown in Figure~\ref{fig:Boltzmann-dimers}.

\newpage
\section{Computational performance}
\label{SI:sec:performance}
\begin{table}[htbp]
\caption{\textbf{Benchmark of methods: backmapping-based approach vs. proposed ML-based method}.
         For the training benchmarks, 
         ``datapoints'' refers to the number of different training
         runs that were executed (different hyperparameter combinations,
         different mappings).
        }
\label{tbl:performance-methods}
\begin{tabular}{llll}
\hline
                                          & CPU hours         & datapoints & total CPU hours  \\
                                          & /datapoint        &            &                  \\
\hline                                                                     
dataset generation (SOMOs)                & $\approx$0.91     &  10,778    & $\approx$9,808   \\
dataset generation (couplings)            & $\approx$1.65     & 221,406    & $\approx$365,320 \\
training (SOMOs)                          & $\approx$1.06     & 1,152      & $\approx$1,221   \\
training (couplings)                      & $\approx$13.86    & 216        & $\approx$2,994   \\
prediction single snapshot (SOMOs)        & $\approx$0.000009 &   3,000    & $\approx$0.03    \\
prediction single snapshot (couplings)    & $\approx$0.000009 &  13,164    & $\approx$0.12    \\
\hline                                                                
\textbf{Total}                            &                   &            & $\approx$379,343 \\
\hline                                                                
\hline                                                                
backmapping (incl. relaxation)            & N/A               &  N/A       & $\approx$2,600   \\
QC calculations single snapshot (SOMOs)   & $\approx$0.91     &   3,000    & $\approx$2,730   \\
QC calculations single snapshot (overlaps)& $\approx$1.65     &  13,164    & $\approx$21,721  \\
\hline                                                                
\textbf{Total}                            &                   &            & $\approx$27,051  \\
\hline
\end{tabular}
\end{table}

\begin{figure}[H]
  \centering
  \includegraphics[width=0.48\textwidth]{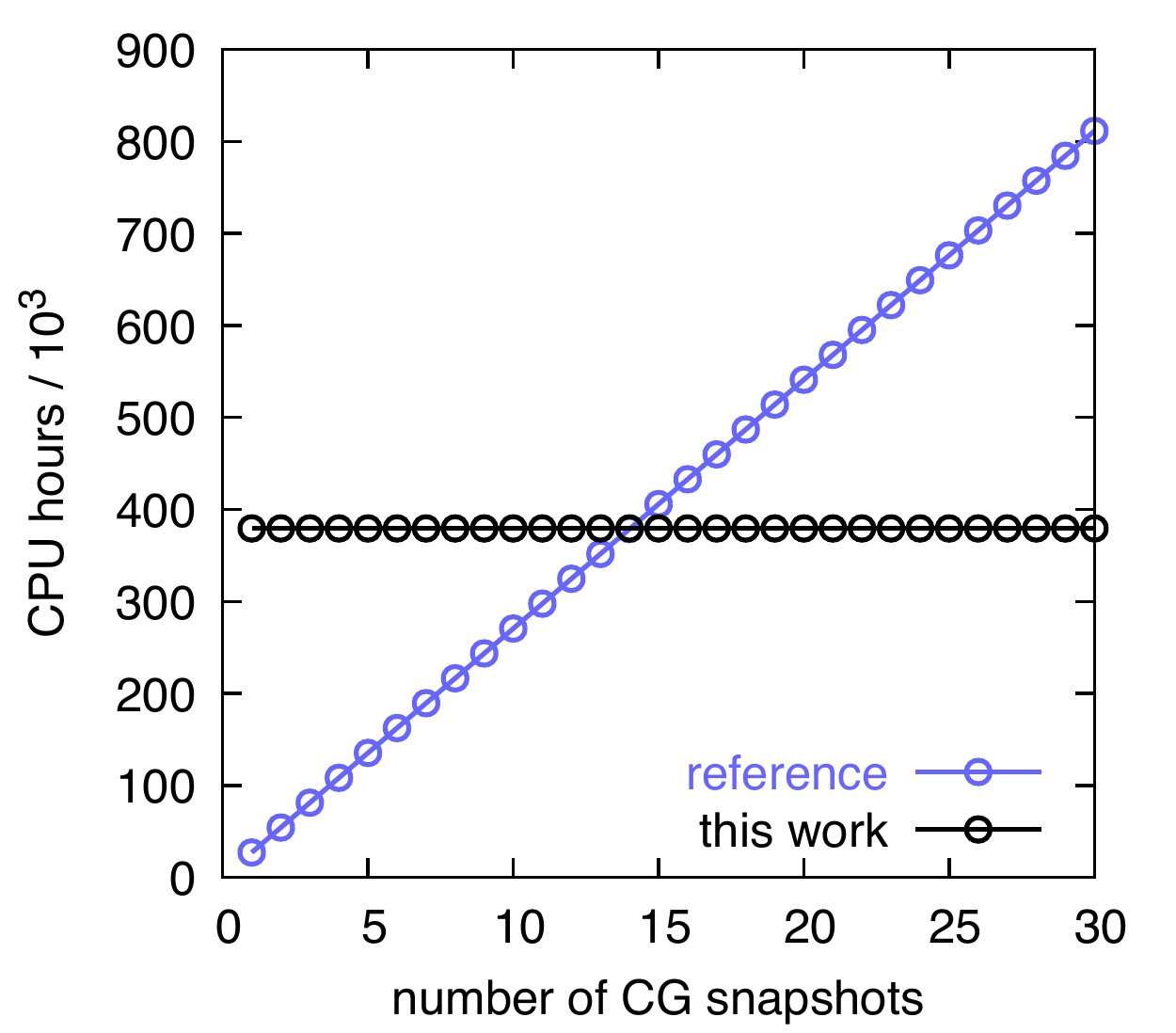}
  \caption{\textbf{Computational cost of the reference method vs. the proposed method.}
           The estimated computational cost (in CPU hours) of the backmapping-based
           reference multiscale approach (blue, ``reference'') is plotted along with the estimated
           computational cost (in CPU hours) of the proposed ML-based method (black, ``this work'')
           as a function of the number of CG snapshots for which the electronic properties are
           predicted.
          }
  \label{fig:comput-performance-methods}
\end{figure}

\begin{table}[H] 
\caption{\textbf{Benchmark of atomistic (AA) and coarse-grained (CG) models computational performance}.
         The performance is reported and compared also between the CG models 
         parametrized using Martini 3 (M3) and IBI.
         The performance refers to MD simulations carried out on 10 CPUs and 
         with GROMACS version 2019.5.
        }
\label{tbl:performance-CG}
\begin{tabular}{llll}
\hline
         & Performance     & speed-up             & speed-up                \\
         & (ns/day)        &          (AA vs. CG) &         (M3 vs. IBI)    \\
\hline
AA       & $\approx$17     &          1           & N/A                     \\
CG IBI   & $\approx$230    & $\approx$13.5        & 1                       \\
CG M3    & $\approx$3900   & $\approx$230         & $\approx$17             \\
\hline
\end{tabular}
\end{table}


\providecommand{\latin}[1]{#1}
\makeatletter
\providecommand{\doi}
  {\begingroup\let\do\@makeother\dospecials
  \catcode`\{=1 \catcode`\}=2 \doi@aux}
\providecommand{\doi@aux}[1]{\endgroup\texttt{#1}}
\makeatother
\providecommand*\mcitethebibliography{\thebibliography}
\csname @ifundefined\endcsname{endmcitethebibliography}
  {\let\endmcitethebibliography\endthebibliography}{}